\documentclass[useAMS,usenatbib]{mn2e}
\usepackage{natbib}
\usepackage{supertabular}
\usepackage{lscape}
\usepackage{graphicx}
\usepackage{epstopdf}
\usepackage{morefloats}

\def\othree{[O\,{\sc{iii}}]}

\def\ca2T{Ca\,{\sc{ii}} Triplet}

\def\kms{km\,s$^{-1}$}
\def\msun{M$_\odot$}
\def\msunyr{M$_\odot$yr$^{-1}$}
\title[Investigating AGN Black Hole Masses and the M-$\sigma_e$ relation for Low Surface Brightness Galaxies]
{Investigating AGN Black Hole Masses and the M$_{BH}$-$\sigma_e$ relation for Low Surface Brightness Galaxies}
\author[S.Subramanian, S.Ramya, M.Das,K.George, T.Sivarani, \& T.P.Prabhu]{S.Subramanian$^{1}$\thanks{E-mail:
smithaharisharma@gmail.com (SS)};S.Ramya$^{2}$\thanks{E-mail:
sramyaraman@gmail.com (SR)}; M.Das$^{1}$\thanks{E-mail:
mousumi@iiap.res.in (MD)};K.George$^{1}$\thanks{E-mail:
koshy@iiap.res.in (KG)};
T.Sivarani$^{1}$\thanks{E-mail:
sivarani@iiap.res.in (TS)};T.P.Prabhu$^{1}$\thanks{E-mail:
tpp@iiap.res.in (TPP)}\\
$^{1}$Indian Institute of Astrophysics, Bangalore 560034, India\\
$^{2}$Shanghai Astronomical Observatory, Shanghai, China\\
}
\begin{document}
 \date{Received/Accepted}

\pagerange{\pageref{firstpage}--\pageref{lastpage}} \pubyear{2014}

\maketitle

\label{firstpage}
\begin{abstract}

We present an analysis of the optical nuclear spectra from the active galactic nuclei (AGN) in a sample of low surface brightness 
(LSB) galaxies. Using data from the Sloan Digital Sky Survey (SDSS), we derived the virial black hole (BH) masses of 24 galaxies from 
their broad H$\alpha$ parameters. We find that our estimates of nuclear BH masses lie in the range $10^{5}-10^{7}~M_{\odot}$, with a median 
mass of 5.62 x 10$^{6}~M_{\odot}$. The bulge stellar velocity dispersion $\sigma_{e}$ was determined from the underlying stellar spectra. 
We compared our results with the existing BH mass - velocity dispersion ($M_{BH}-\sigma_{e}$) correlations and found that the majority of 
our sample lie in the low BH mass regime and below the $M_{BH}-\sigma_{e}$ correlation. We analysed the effects of any systematic bias in 
the M$_{BH}$ estimates, the effects of galaxy orientation in the measurement of $\sigma_e$ and the increase of $\sigma_e$ due to the 
presence of bars and found that these effects are insufficient to explain the observed offset in M$_{BH}$ - $\sigma_e$ correlation. Thus 
the LSB galaxies tend to have low mass BHs which probably are not in co-evolution with the host galaxy bulges. A detailed study of the nature of the bulges 
and the role of dark matter in the growth of the BHs is needed to further understand the BH-bulge co-evolution in these poorly evolved and 
dark matter dominated systems. 
\end{abstract}


\section{Introduction}
Low surface brightness (LSB) galaxies are late type 
galaxies that have
a central disk surface brightness of $\mu_{B(0)}~\ge~$22 to 23 mag arcsec$^{-2}$
(\citealt{IB97}; \citealt{ibs2001}; \citealt{das13}). Optically they
are distinguished by their diffuse, low luminosity stellar disks. The LSB disks 
have low star formation rates and are low in metal content (\citealt{M94}, \citealt{IB97}).
Morphologically they span over a wide range of galaxies from dwarfs and irregulars to
very large disk galaxies (\citealt{MSB95}). However loosely they can be
divided into two types the (i)$\sim$ LSB dwarf and irregular galaxies and (ii)$\sim$ disk LSB
galaxies. The LSB dwarfs and irregulars form the larger fraction of LSB galaxies and can be
found in different environments such as galaxy groups or even within voids \citep{p11}.
Disk LSB galaxies are more rare and usually found in low density environments \citep{b93} and vary over
a range of sizes \citep{bei99}. But the LSB galaxies with really large disk scale lengths (like Malin 1 which has a scale length 
of 55 kpc, \citealt{Bothun1987}), referred to as giant LSB (GLSB) galaxies, are generally found to lie close to the edges of voids
\citep{R09}. A recent study by \cite{Shao2015} on the environment, morphology and stellar populations of a sample of bulgeless 
LSB galaxies suggest that their evolution may be driven by their dynamics including mergers rather than by their large scale environment. 
Another defining characteristic of all LSB galaxies is their large H{{\sc i}} gas 
content. Large or small, LSB galaxies usually have large H{\sc i} disks that can extend over several 
optical disk scale lengths (\citealt{dmv96}; \citealt{obv04}; \citealt{das07}). The H{\sc i} rotation 
curves clearly show that their disks are dark matter dominated \citep{P97} and even their centers
have significant dark matter content \citep{kmd08}. The dark matter halo 
can suppress the formation of disk
instabilities such as bars and spiral arms, which slows down the rate of star formation in these systems
(\citealt{mw04}, \citealt{GJ2014}).

Nuclear activity is not common in LSB galaxies. This is in marked contrast to high surface 
brightness disk galaxies where the percentage having active galactic nuclei (AGN) can be as high as 
50$\%$, depending on the mean luminosity of the sample \citep{h97}. The most probable explanation for this 
low fraction is that LSB disks generally lack two structural features that facilitate inward gas flows and the 
formation of a central mass concentration in galaxies - bars and strong spiral arms in disks \citep{bim97}. 
These structural features may be are suppressed by the dominant dark matter content in these galaxies \citep{GJ2014}. 
 Most of the GLSB galaxies have significant bulge component along with the large LSB disk \citep{bei99}.  
Whereas a large fraction of normal LSB disk galaxies are relatively bulgeless (\citealt{MB94}). But there is often a bright core 
due to nuclear star formation which appears as a luminous point in a 
featureless low luminosity disk \citep{mat99}. 
Strong H$\alpha$ and [O {\sc i}] emission lines in their spectra indicate ongoing 
nuclear star formation which is often in the form of kiloparsec (kpc) scale rings, as observed in the galaxy 
NGC 5905 (\citealt{com10}; \citealt{raichur13}; \citealt{Raichur2015}). Hubble
space telescope (HST) observations have shown that the nuclear star formation can lead to the
formation of compact nuclear star clusters that may sometimes co-exist with AGN activity \citep{seth08}. This activity 
can contribute to the formation of a central massive object (CMO) and
lead to the build up of a bulge in an otherwise bulgeless galaxy \citep{dmb11}.

However, although we do not generally see AGN in LSB galaxies, a significant fraction of bulge dominated 
GLSB galaxies do show AGN activity (\citealt{spray95}; \citealt{gal11}). Some of them are even radio bright 
and visible in X-rays 
(\citealt{das09}, \citealt{n10}, \citealt{mishra15}). 
 \cite{gal11} compared the stellar mass, radius, absolute magnitude and the 
red shift distributions of LSB galaxies and High Surface Brightness (HSB) galaxies hosting AGN. 
They found that most of the distributions for LSB galaxies and HSB galaxies are similar, except the distribution of 
petrosian radius, r90. 
AGN activity in bulge dominated GLSB galaxies is not surprising as 
studies indicate that the growth of
nuclear black holes (BHs) in galaxies is intimately linked to the growth of their bulges (e.g. \citealt{sr88}; 
\citealt{H04}). The strong correlation of BH mass (M$_{BH}$) with bulge mass or
bulge luminosities in galaxies (M$_{BH}$ - $\sigma_e$) is due to this super massive black hole (SMBH) - bugle
co-evolution (\citealt{G09} and \citealt{MM13}). But in LSB galaxies, their bulge velocity dispersion 
and disc rotation speeds suggest that they lie 
below the M$_{BH}$ - $\sigma_e$ correlation for bright galaxies \citep{p05}. X-ray studies also suggest that 
GLSB galaxies do not lie on the radio X-ray correlation \citep{das09} and their BH masses may be quite low 
\citep{n10}.  

\cite{mei09} estimated the BH masses of three GLSB 
galaxies and found them to lie close 
to the M$_{BH}$ - $\sigma_e$ correlation and the values of M$_{BH}$ are found to be in the order of $\sim$ 10$^7$ M$_{\odot}$. 
But later, the estimates from the study of another three GLSB galaxies 
by \cite{Ram11} showed that their sample galaxies lie offset from the M$_{BH}$ - $\sigma_e$ correlation, with 
M$_{BH}$ values of 3-9 x 10$^5$ M$_{\odot}$.  Thus these studies suggest that LSB galaxies tend to have low mass BHs $\sim$ $\le$ 10$^7$ M$\odot$ and some 
of them lie below the M$_{BH}$ - $\sigma_e$ correlation. A systematic study of a larger sample of LSB galaxies is needed to show 
a convincing trend.  

\cite{Ram11} suggested that the bulges of their sample might be well evolved, but the M$_{BH}$ values   
are lower than those found in bright galaxies. \cite{bei99} could not find any distinction between 
the structural parameters (bulge to disk ratio and size) of the bulges in the LSB galaxies and normal galaxies. 
\cite{Mor12} found that the stellar populations in the bulges of LSB galaxies have similar properties as that of the bulges of normal 
galaxies. Based on this they concluded that the formation and evolution history of bulges in LSB galaxies are similar to that of the 
bulges in normal galaxies. \cite{pizzella2008} found signatures of pseudo bulges in 6 galaxies with LSB disk. 
\cite{KH2013} and references therein suggest that these pseudo bulges do not correlate with BHs in the same way 
as classical bulges correlate.
The two LSB galaxies, UGC 6614 and Malin 2 studied by \cite{Ram11} which lie off from the M$_{BH}$ - $\sigma_e$ relation are more likely 
to host classical bulges (\citealt{das13}, \citealt{sd2014}). If they host classical bulges then the reason of their observed offset is not 
clear. 
The significant contribution of 
dark matter in the inner regions may also affect the growth of the BHs in these systems. The total sample of bulge dominated 
LSB galaxies with AGN activity, specifically studied so far is very small (6 galaxies) to say any statistically significant trend on the 
mass of the BHs, their location in the M$_{BH}$ - $\sigma_e$ correlation and the nature of the bulges they host. 
Though some of the LSB galaxies have been analysed in other studies of a large sample of AGN dominated systems 
(BH mass estimates for the galaxies, 1226+0105 and LSBC F727-V01, are available from \citealt{GreeneHo2007}), there is no specific study 
dedicated to analyse a large sample of LSB galaxies and to understand the BH-bulge co-evolution in these systems.  
 A systematic study of a large sample of LSB galaxies is needed to understand the AGN activity, typical range 
of M$_{BH}$ values, nature of their bulges and the BH-bulge co-evolution in these poorly evolved and dark matter dominated systems.  
Low mass BHs in isolated LSB galaxies are also very interesting candidates for the study of seed BHs in galaxies \citep{vol10}. 

In this paper, we present a detailed analysis of 24 bulge dominated LSB galaxies with signatures of type - I AGN (broad H$\alpha$ component)  
identified from a large sample of 558 LSB galaxies to understand the AGN activity and to estimate the mass of the BHs 
present in their centers. Based on their location in the M$_{BH}$ - $\sigma_e$ plot and from the nature of their bulges we discuss 
the presence/absence of BH-bulge co-evoloution in these systems. In the following sections (Section: 2 and Section: 3) we discuss the sample and analysis. 
The results are described in Section: 4. Discussion and conclusions are presented in Sections: 5 and 6 respectively. 

\section{Data and Sample} 
The Sloan Digital Sky Survey (SDSS) Data Release 10 (DR10) 
include hundreds of thousands of new galaxy and quasar spectra from the Baryon Oscillation Spectroscopic 
Survey (BOSS), in addition to all imaging and spectra from prior SDSS data releases. 
The spectral range of the SDSS multi-object spectrograph and the BOSS spectrograph is 
almost similar and the spectra resolution is $\sim$ 70 km/s for both instruments. 
The SDSS spectrograph fibre aperture is 3 arcsec in diameter (corresponding to 2.9 kpc at a redshift of 0.05) 
and the fibre aperture of BOSS spectrograph is 2 arcsec in diameter (corresponding to 1.9 kpc 
at a redshift of 0.05). Thus the size of the nucleus/bulge sample in the BOSS spectra will be less 
than that of the SDSS spectra. 

The literature (\citealt{Impey96}, \citealt{spray95}, \citealt{Schom88}, \citealt{Schom92}, \citealt{burk01}, 
\citealt{Gal02}) provides a large sample ($\sim$ 1200) of 
LSB galaxies. As an initial sample we selected a sample of 558 LSB galaxies from the literature, mentioned above, 
for which the SDSS DR10 nuclear spectra are available. The emission line 
fluxes of these 558 galaxies are available for public in the SDSS DR10 database. The Galspec product in the SDSS database 
provides the emission line 
fluxes for SDSS-I/II galaxies estimated based on the methods given by \cite{K03}, \cite{B04} and \cite{T04}. 
The emission line fluxes for SDSS III BOSS spectra are provided by the team in 
University of Portsmouth \citep{Thom13}. 

\begin{figure}
\resizebox{\hsize}{!}{\includegraphics{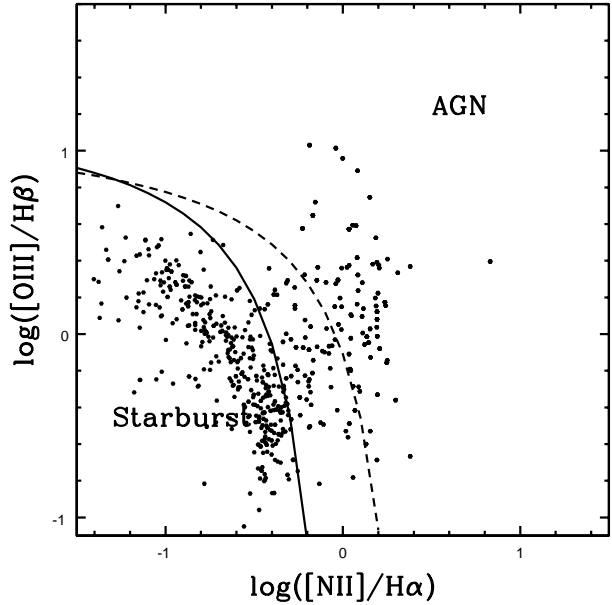}}
\caption{The BPT AGN diagnostic diagram,[O{\sc iii}]/H$\beta$ versus [N{\sc ii}]/H$\alpha$ for the sample of 558 
galaxies. The emission line fluxes for these 
sample galaxies are taken from the SDSS database. The solid line is the starburst-Seyfert demarcation line taken from \citep{Kauff03} and the dotted line represents 
the extreme starburst line taken from \citealt{kew01}. The sample to the left of the solid line are considered as purely star forming 
galaxies and the sample to the right of the dotted line are considered as pure AGN candidates. The sample between 
the two lines are classified as composite. For these galaxies the nuclear activity can be due to both star formation and 
AGN activity.}
\end{figure}

The H${\alpha}$, H${\beta}$, [N{\sc ii}]$_{6583}$ and [O{\sc iii}]$_{5007}$ emission line fluxes of the 558 galaxies, 
were extracted from SDSS database using SQL query and the [N{\sc ii}]$_{6583}$/H${\alpha}$ vs [O{\sc iii}]$_{5007}$/H${\beta}$ 
Baldwin-Phillips-Terlevich (BPT) [first given by \citealt{BPT81} and improved further by \citealt{vo87}, also see \citealt{kew01} and \citealt{kew06} and references therein]
diagram is constructed. This is shown in Figure: 1. The solid line is the starburst-Seyfert demarcation line taken from 
\cite{Kauff03} and the dotted line represents the extreme starburst line taken from \cite{kew01}. 
This is used as our primary diagnostic to select the AGN candidates for our further 
investigation. The sample to the left of solid line are considered as purely star forming 
galaxies and the sample to the right of the dotted line are considered as pure AGN candidates. The sample between 
the two lines are classified as composite and the nuclear activity of these galaxies are expected to have significant 
contribution from both star formation and 
AGN activity \citep{kew06}. As the spectra of composite sample in general indicate some contribution from an AGN 
(\citealt{Trou11}; \citealt{jia11}), they are included in our 
sample to investigate the AGN activity and to estimate the mass of the BH in their center. 
Out of 558 galaxies shown in diagnostic diagram, 160 galaxies ($\sim$29\%, with almost equal contribution from the 
composite and purely AGN candidates) constitute our AGN sample and the remaining 398 ($\sim$71\%) galaxies 
are purely star forming systems. Previous studies suggested a range of values for the fraction of LSB galaxies which host AGN, from 
5\% (\citealt{ibs2001}, \citealt{gal11}) to 50\% (\citealt{spray95}, \citealt{Schombert88}). Our estimate of $\sim$ 29\% is 
including the composite and AGN sample. If we consider only the sample which are classified as pure AGN candidates, the 
fraction is $\sim$ 16\%. This value is closer to the estimates of \citealt{mei09}) who found a fraction of 10-20\% LSB galaxies have 
AGN. 

From the sample of 160 galaxies (which constitute the purely AGN candidates and composite candidates) 
we are particularly interested in those which host broad H$\alpha$ emission. 
The broad H${\alpha}$ emission line full width at half maximum (FWHM) and luminosity can be used to 
estimate the masses of active BHs in the centers of these galaxies \citep{R13}. 
To identify the sample with broad H$\alpha$, good quality spectra are required and hence, 
out of 160 galaxies we selected only those galaxies which have median S/N $>$ 15. 
Thus, finally we used an initial sample of 115 SDSS DR10 spectra of LSB galaxies to identify sources 
with broad balmer lines, specifically the H${\alpha}$ line which is the 
strongest balmer emission line in the optical wavelength region. The steps involved in the analysis of these 115 galaxies are 
described in the next section. 


\begin{figure*}
\resizebox{\hsize}{!}{\includegraphics{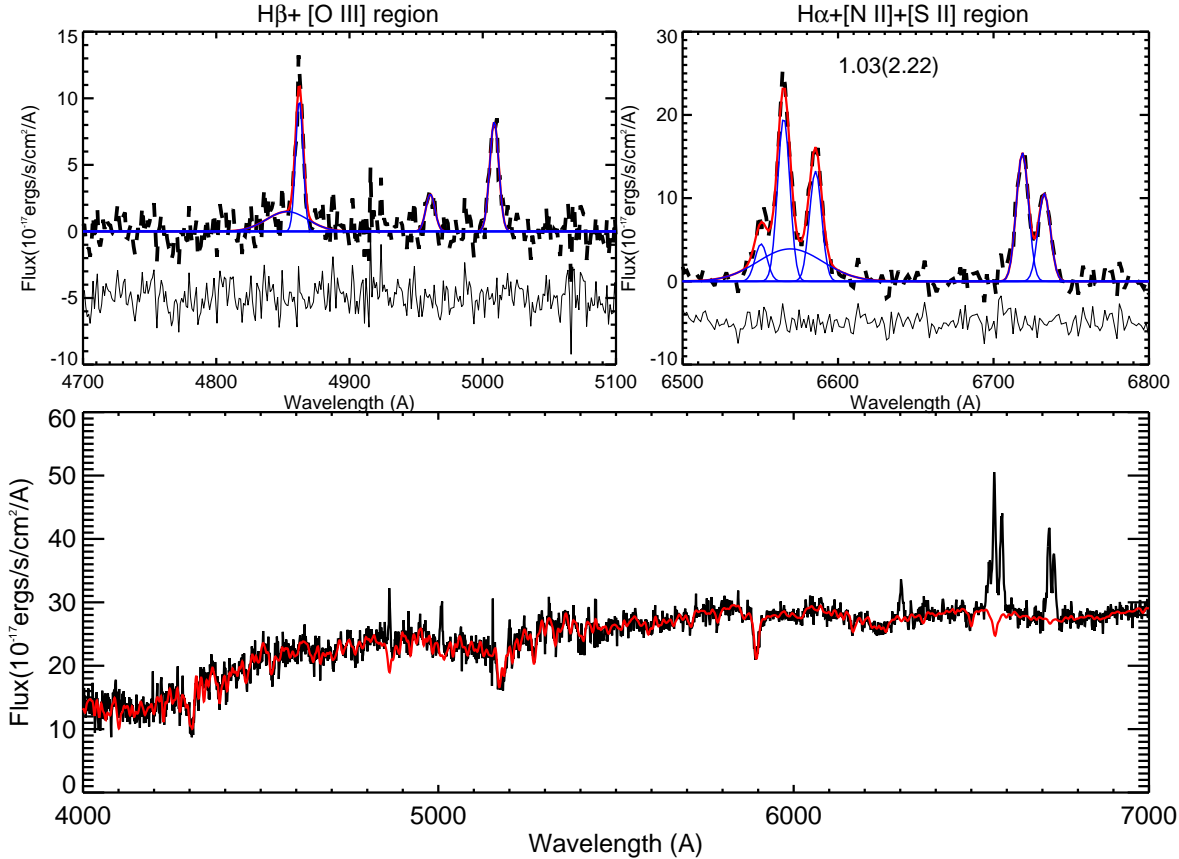}}
\caption{The bottom panel shows the pPXF fit for the galaxy, Malin 1 (identification number 13). The black line is the observed 
 spectrum and the red line is the best fit optimal template for the underlying stellar population. 
The Gaussian fits to the different emission lines are shown in upper left and upper right panels. The black line is  
the reddening corrected pure emission line spectra and the red line is the fit to the spectra. The blue lines 
show the separate components of the spectral region. The residue of the fits are also shown in the upper panels. 
In the upper right panel, the reduced $\chi^2$ values of the fits with and without (inside the parenthesis) 
broad H$\alpha$ component are 
shown.}
\end{figure*}

\section{Analysis}
\subsection{Spectral Decomposition}
In order to isolate the AGN emission lines and identify 
the sample with broad H$\alpha$ feature, we first removed the spectral contribution from 
underlying stellar population. The observed 
spectra of 115 galaxies were corrected for galactic reddening using the E(B-V) values provided by \cite{SF11}.
Then we used the pPXF (Penalized Pixel-Fitting stellar kinematics 
extraction) code by \cite{cap04} to obtain the best fit model for the underlying stellar 
population. The emission lines in the observed spectra were masked and the several 
masked spectra were modelled as a 
combination of single stellar population templates of different ages. The MILES single stellar population models
are used as templates of the underlying stellar population. The templates are available for a large 
range of metallicity ([M/H] $\sim$ -2.32 to +0.22) and age (63 Myr to 17 Gyr). 
pPxF uses Gauss-Hermite parametrization (\citealt{ger93}; \citealt{vf93}) and works in the pixel space, 
which makes it easy to mask gas emission lines or bad pixels from the fit and the 
continuum matching can be done directly \citep{cap04}. pPxF creates an 
algorithm with initial guesses for systemic velocity of the system, V (redshift, z) 
and stellar velocity dispersion, $\sigma$. The model spectra are 
convolved with a broadening function using initial $\sigma$ values. 
The residuals of each of the data points are fed into a non-linear least squares optimization 
routine to obtain V, $\sigma$ and 
Gauss-Hermite polynomials. 
Though the major 
contribution to the observed spectra is the underlying stellar population, there are other contributors like power 
law continuum from AGN, Fe lines etc. While estimating the best fit to the underlying population, ppxf  
fits a polynomial, along with the optimal template, to account for the contributions from power law continuum.  
And the Fe lines in these galaxies are too weak to affect our further analysis. 
In all our sample galaxies the stellar population contribution to the best fit ranges from 80 - 95 \% 
and the remaining factor is the contribution from AGN continuum and Fe lines. The output of this step of analysis is 
shown in the bottom panel of Figure: 2 (for the galaxy Malin 1 with identification number, 13). The reddening corrected spectra is shown 
in black and the best fit model is overplotted in red colour. The best fit model is subtracted from the reddening corrected spectrum to 
get the pure emission line spectrum (the residue of the pPXF fit).

\begin{table*}
\caption{Basic parameters of the sample which are flagged as galaxies with broad H$\alpha$ component. 
The distances shown here are the luminosity distances taken from NED. The weight of de Vaucouleurs 
component to the total, (de Vaucouleurs + Exponential component), fit which is a proxy for the B/T ratio is 
obtained from the SDSS DR10 database. Bulge Sersic index, n$_b$ is taken from \citep{Simard2011}.}
 \begin{tabular}{lllllll}
{\bf No} & {\bf Galaxy} & {\bf Morphology} & {\bf Galaxy} & {\bf Distance} & {\bf fracDev} & {\bf n$_b$}\\
 & & &{\bf Coordinates} & {\bf (Mpc)} & {\bf B/T} & \\
\hline

1&1226+0105 or 2MASX J12291286+0049042& Sc & 12$^h$29$^m$12.8$^s$ +00$^{\circ}$49'04''& 349& 1.0 & 7.95\\
2&VIII ZW 318&Sc (Merging pair?)&13$^h$30$^m$32.0$^s$ -00$^d$36$^m$14$^s$&237& 0.87 &  1.43 \\
3&2MASX J14384627+0106576&Sc&14$^h$38$^m$46.3$^s$ +01$^d$06$^m$58$^s$&365& 1.0 &  7.47\\
4&Leda 135884&Sb&23$^h$15$^m$01.7$^s$ +00$^d$04$^m$24$^s$&214&  0.61 &  7.97\\
5&NGC 7589&SAB(rs)a&23$^h$18$^m$15.7$^s$ +00$^d$15$^m$40$^s$&120& 0.98 &  5.91  \\
6&UGC 00568&Sd&00$^h$55$^m$08.9$^s$ -01$^d$02$^m$47$^s$&189&  1.0 &  4.88\\
7&2MASX J00534265-0105066&SBc&00$^h$53$^m$42.6$^s$ -01$^d$05$^m$07$^s$&193& 1.0&  6.36\\
8&UGC 00514&Sb&00$^h$50$^m$42.7$^s$ +00$^d$25$^m$58$^s$&291& 0.34& 8.0\\
9&LSBC F611-03&Sc&01$^h$13$^m$20.1$^s$ +14$^d$43$^m$40$^s$&180& 1.0 & 5.35 \\
10&ZW 091&Interacting&10$^h$37$^m$23.8$^s$ +02$^d$18$^m$43$^s$&173& 0.93 & 3.41 \\
11&ZW 437&Sc(r)&14$^h$36$^m$02.0$^s$ +02$^d$51$^m$06$^s$&124& 1.0& 8.0\\
12&LSBC F727-V01&Sc&15$^h$51$^m$40.5$^s$ +08$^d$52$^m$26$^s$&307& 1.0& 6.16\\
13&Malin 1& S &12$^h$36$^m$59.3$^s$ +14$^d$19$^m$49$^s$&366& 1.0& 5.62\\
14&UGC 10050&Spiral&15$^h$49$^m$44.9$^s$ +18$^d$31$^m$37$^s$&58.7& 1.0& --\\
15&LSBC F500-V01&Spiral&10$^h$23$^m$33.1$^s$ +22$^d$47$^m$46$^s$&320&  0.98& 5.06\\
16&NGC 0926&SB(rs)bc&02$^h$26$^m$06.7$^s$ -00$^d$19$^m$55$^s$&86& 0.56& --\\
17&UGC 09087&S0&14$^h$12$^m$16.8$^s$ +18$^d$17$^m$58$^s$&74.5&0.91& --\\
18&UGC 08828&Spiral&13$^h$54$^m$26.8$^s$ +21$^d$49$^m$48$^s$&120& 0.38& --\\
19&2MASX J10255577+0200154 &SBb&10$^h$25$^m$55.7$^s$ +02$^d$00$^m$16$^s$&304& 0.93& 6.24\\
20&IC 2423 &SAB(s)b&08$^h$54$^m$47.1$^s$ +20$^d$13$^m$13$^s$&132& 1.0& --\\
21&LSBC F564-01 &Galaxy Pair&09$^h$10$^m$29.4$^s$ +20$^d$33$^m$53$^s$&123& 1.0& 5.16\\
22&SDSS J103704.86+202627.3&Spiral&10$^h$37$^m$04.9$^s$ +20$^d$26$^m$27$^s$&184& 1.0& 3.87\\
23&2MASX J14565122+2152295&Spiral&14$^h$56$^m$51.2$^s$ +21$^d$52$^m$30$^s$&195& 0.79& 8.0\\
24&UGC 06614&SA(r)a&11$^h$39$^m$14.9$^s$ +17$^d$08$^m$37$^s$&93.2& 1.0& -- \\
25&LSBC F508-02&Sb&13$^h$06$^m$59.0$^s$ +22$^d$44$^m$15$^s$&150& 1.0& 4.21\\
26&KUG 0012-000&Sb& 00$^h$14$^m$55.1$^s$ +00$^d$15$^m$08$^s$&160& 0.57& 3.68\\
27&CGCG 006-023&Spiral&09$^h$16$^m$13.7$^s$ +00$^d$42$^m$02$^s$&166& 0.98& 8.0 \\
28&UGC 06284 &Galaxy pair&11$^h$15$^m$49.1$^s$ +00$^d$51$^m$36$^s$&200& 0.89&3.53\\
29&LSBC F573-04&Sb&12$^h$18$^m$15.2$^s$ +20$^d$00$^m$40$^s$&127&0.42& 5.77\\
30&UGC 05035&SBa&09$^h$27$^m$10.2$^s$ +21$^d$35$^m$38$^s$&161& 1.0& 5.4\\
\hline
 \end{tabular}
\end{table*}

\subsection{Emission line analysis}
The pure emission line spectra are carefully 
analysed by modelling the narrow lines and simultaneously looking for the presence of broad balmer line profiles. 
Broad balmer lines indicate the presence of dense gas orbiting the central BH within the broad line 
region. For the sample with broad balmer lines, M$_{BH}$ can be estimated using scaling relations.  
The H$\alpha$ line is the strongest balmer line in the optical wavelength region and hence to identify the final 
sample with broad balmer lines, we analysed the spectra of the initial sample in the wavelength range, 6400-6880 {\AA}.  
As the H$\alpha$ line is blended with the [N{\sc ii}] doublet, 
we need to carefully fit this region to extract the broad line component of H$\alpha$. The method adopted 
here is similar to that described in \cite{R13}. 
The method is based on the assumption that the [S{\sc ii}] doublet lines close to the H$\alpha$ region and well separated, 
are a good representation of the shape 
of the [N{\sc ii}], H$\alpha$ and H$\beta$ narrow lines. The [S {\sc ii}] doublet is fitted with a single Gaussian 
model with the width of the two lines assumed to be equal and the relative separation between the two lines 
held fixed by their laboratory wavelengths. The [S{\sc ii}] model is used as a template to fit the narrow H$\alpha$ 
and [N{\sc ii}] doublet lines. The H$\alpha$ + [N{\sc ii}] region of our emission line galaxy spectra are first modelled 
with three Gaussian components, one for H$\alpha$ narrow and two for [N{\sc ii}] doublet. The widths of [N{\sc ii}] 
lines are assumed to be the same as that of [S{\sc ii}] lines. The separation between the centroids of the [N{\sc ii}] 
narrow components are held fixed using laboratory wavelengths and the flux of [N{\sc ii}]$\lambda$6583 to 
[N{\sc ii}]$\lambda$6548 is fixed at the theoretical value of 2.96 \citep{Lud12}. The reduced $\chi^2$ value for the fit is obtained. 
Then the region is again fitted including a fourth component representing the broad H$\alpha$ line. The final 
sample with broad H$\alpha$ component is selected based on the three criteria, that the the inclusion of the broad 
component should improve the reduced $\chi^2$ value by at least 10\%, the broad H$\alpha$ FWHM should be greater 
than 
600 kms$^{-1}$ (after correcting for the instrumental resolution) and also the broad H$\alpha$ peak flux should 
be at least two times larger than the residue of the fit. Thus based on the above criteria, 30 LSB galaxies 
were flagged as galaxies with broad H$\alpha$ component. The basic parameters of these  
30 galaxies are given in Table 1. More careful examination of the model fits 
of each of these 30 galaxies are performed before making the final sample of broad line AGN candidates for 
which the BH masses are estimated. This is described in Section: 4. 3. 

The other emission lines present in the spectra of these 30 galaxies 
were also analysed. The narrow line 
profile derived from the [S {\sc ii}] doublet is also used as a template for fitting H$\beta$, using the same 
approach as that for H$\alpha$. The H$\beta$ line is fit twice, with and without a broad component. 
Here the width of broad H$\beta$ line is assumed to be the same as that of the broad H$\alpha$ line. 
Based on the same criteria used to identify the broad component in H$\alpha$, the broad component in 
H$\beta$ is also identified. 
Since the [O {\sc iii}] profile commonly exhibits a broad,
blue shoulder (\citealt{h81}; \citealt{w85}), the [O {\sc iii}] doublet are fitted independently 
with one or two Gaussian components (the two component model is included if the reduced $\chi^2$ improves 
by 10 \%. The [O {\sc i}] doublet lines are also independently fitted with single Gaussian component.  
We have used the standard 'MPFITFUN' of IDL to fit the emission lines employing the 'GAUSS1' program where the parameters need to be 
fit for each Gaussian are centroid, peak value and sigma(fwhm/2.35). The input parameters are wavelength, flux and 
sqrt(1/inverse variance of flux) as error.
The fitting routine provides the best fitting parameters along with errors associated with these parameters. 
The fluxes of all the emission lines are estimated using the peak and sigma values provided by the fit. Uncertainties in the fitted 
parameters of peak and sigma are added in quadrature to obtain the error on fluxes. The forbidden narrow emission line fluxes of 
30 galaxies are tabulated in Table: A1. 
The different emission line profile fits of the galaxy, Malin 1 (with identification number 13) are shown in the upper right and upper 
left panels of Figure: 2.

\section{Results}
\subsection{BPT Classification}
  As described in Section: 2, the initial sample of AGN candidates (which includes the broad line AGN candidates) 
  were selected based on their location in the [O{\sc iii}]/H$\beta$ versus [N{\sc ii}]/H$\alpha$ AGN diagnostic diagram (Figure: 1). 
After further analysis we found that of these only 30 galaxies have an extra-broad component in their H$\alpha$ emission region 
along with their narrow lines. 
The narrow emission line fluxes of these 30 galaxies are estimated after carefully deconvolving them from 
the other components. In this section, we discuss the location of the sample galaxies in the three AGN diagnostic 
diagrams. In Figure: 3, we revisit the location of these 30 galaxies in the primary 
diagnostic diagram, [O{\sc iii}]/H$\beta$ versus [N{\sc ii}]/H$\alpha$. {\bf From the figure, 
we can see that 7 galaxies are in the composite region and the remaining 23 galaxies are in the AGN region.} 
Further classification of the sample into Seyfert/LINER are made using the other two AGN diagnostic 
diagrams, [O{\sc iii}]/H$\beta$ 
versus [S{\sc ii}]/H$\alpha$ and [O{\sc iii}]/H$\beta$ versus [O I]/H$\alpha$. These diagrams are shown in 
Figures: 4 and 5. 
The demarcation lines between 
starbursts, Seyferts and LINERs were obtained from \cite{kew01}, \cite{kew06} and \cite{Kauff03} and are shown in these diagrams.  
The [O{\sc iii}]/H$\beta$ versus [O I]/H$\alpha$ plot shown in Figure: 5 has only 25 points as the remaining 5 galaxies do not 
have [O {\sc i}] in emission. The galaxies, 1, 2, 5, 8, 12, 21 and 23 ($\sim$ 20\% of the sample) 
are classified as Seyfert type based on Figures 4 and 5.  The galaxies in the starburst region in Figure 4 (26, 27 and 29) 
and Figure 5 (26, 27) are those which are classified as composite in Figure: 3. The galaxy with identification number 29 is not present in 
Figure: 5 
as it does not have [O {\sc i}] in emission. The galaxy, 17 which is classified as LINER in Figure: 4 lies closer to the starburst/Seyfert 
demarcation line in Figure: 5. This discrepancy may be due to the very low amount of flux (with relatively large error) 
associated with the [O {\sc i}] emission line for this galaxy (see Table A1).  

%
%
\begin{figure}
\resizebox{\hsize}{!}{\includegraphics{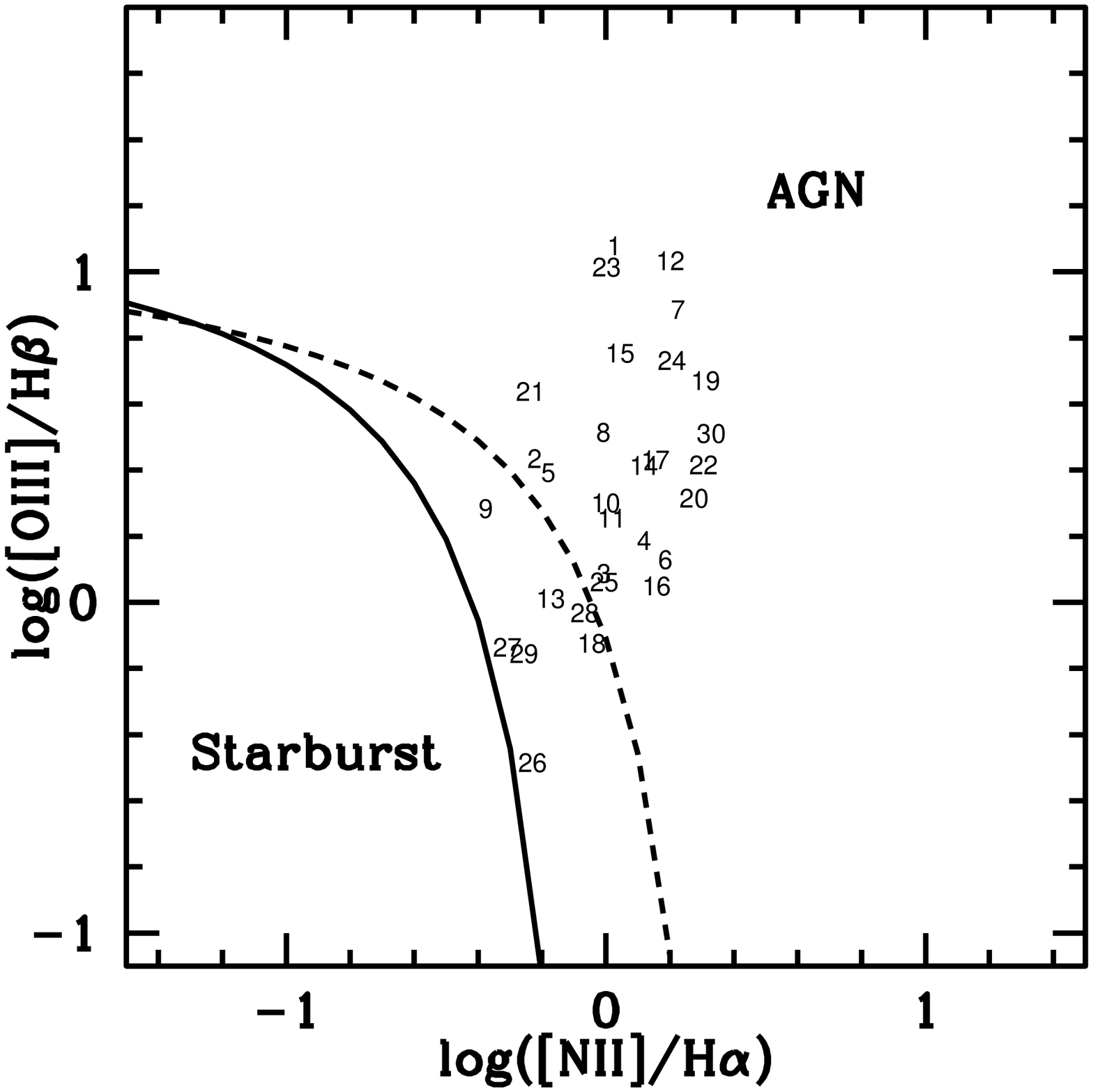}}
\caption{The AGN diagnostic diagram plotted for O III]/H$\beta$ versus [N{\sc ii}]/H$\alpha$. }
 \end{figure}
 \begin{figure}
 \resizebox{\hsize}{!}{\includegraphics{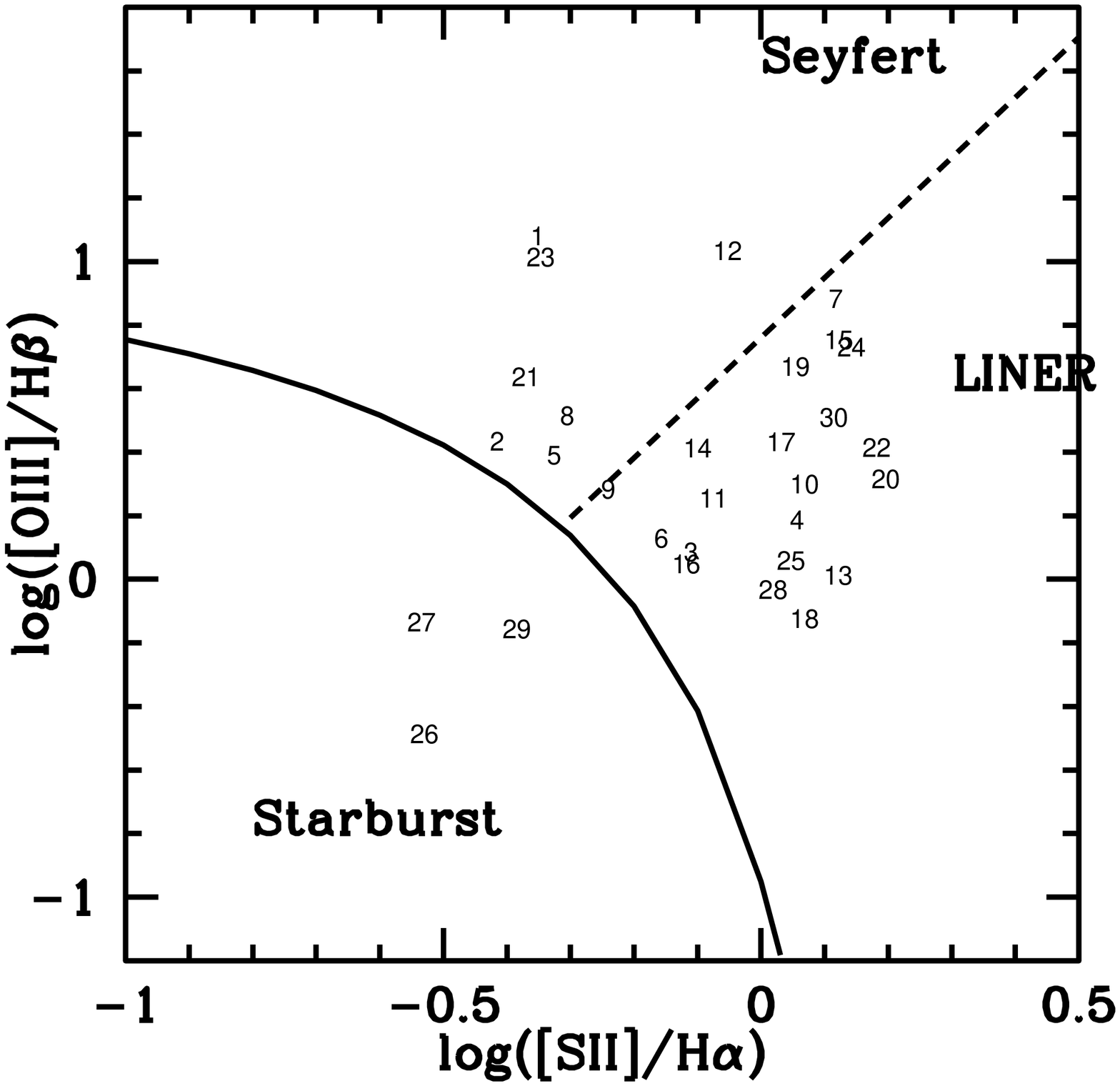}}
 \caption{The AGN diagnostic diagram plotted for O III]/H$\beta$ versus [S{\sc ii}]/H$\alpha$. }
 \end{figure}
 \begin{figure}
 \resizebox{\hsize}{!}{\includegraphics{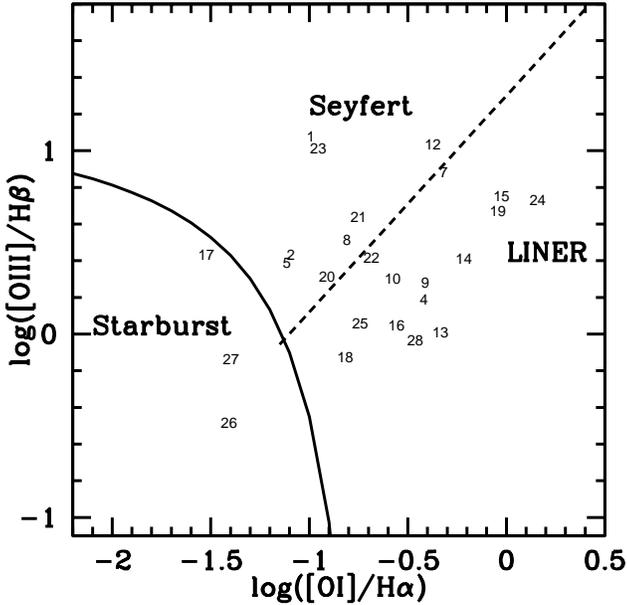}}
 \caption{The AGN diagnostic diagram plotted for O III]/H$\beta$ versus [O I]/H$\alpha$.}
 \end{figure}
\subsection{Metallicity of the sample}
The [N{\sc ii}]$_{6583}$, [O{\sc iii}]$_{5007}$, H$\alpha$ and H$\beta$ narrow emission line fluxes are used to 
estimate 
the oxygen abundance of the nuclear/bulge regions of the final sample using the empirical relation 
derived by \cite{PP04}. The relation is given as, \\
 12 + log (O/H) = 8.73 - 0.32 x O3N2, where O3N2 is defined as 
log [([O {\sc iii}$_{5007}$]/H$\beta$)/ ([N {\sc ii}$_{6583}$]/H$\alpha$))]$^2$.\\
Here we would like to mention that the estimated metallicities are 
gas phase metallicity. The metallicity distribution of the sample galaxies is shown 
in Figure: 6 and the metalicity value has a range, from log[O/H] = 7.94 to 8.8. The mean metallicity is $\sim$ 8.6, which is only slightly lower than the solar value of 8.69 
\citep{Asplund09}. A few of the galaxies {with identification numbers, 26, 18, 16 and 6} in the sample are 
slightly super solar. This initially appears surprising as LSB galaxies 
are generally low in metallicity. However, these results are derived from nuclear spectra. 
Nuclear star formation may result in higher metallicities in the central regions compared to the 
disks of LSB galaxies. 
Hence due to recent star formation in the central regions, the metallicity 
is probably higher than the disks of LSB galaxies.  The underlying stellar population obtained from pPXF fits, of these four super solar 
metallicity galaxies, except the galaxy with identification number 18, have contributions from young (500 Myr-2 Gyr) and metal rich 
(M/H = 0.00 to 0.22) population. 
\begin{figure}
\resizebox{\hsize}{!}{\includegraphics{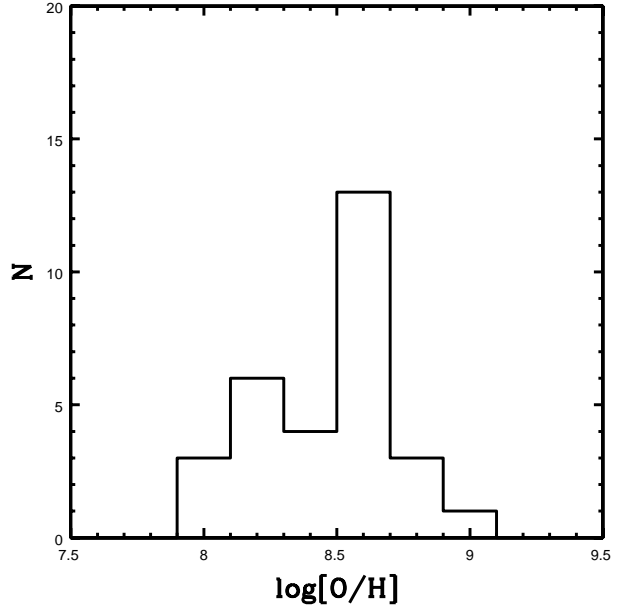}}
\caption{The gas phase metallicity distribution of 30 LSB galaxies.}
\end{figure}
\subsection{Broad line AGN candidates}
 From Section: 3.2 we found that 30 galaxies were identified to host AGN with broad H$\alpha$ component in their 
emission line spectrum.  
Each of these 
30 galaxies were examined carefully to define a sample of broad line AGN candidates.
Out of 30, 3 galaxies (with identification numbers, 26, 27 
and 29) are classified as star forming galaxies (Figures: 3, 4 and 5). 
This may be due to the reason that the star formation dominated the narrow emission lines 
within the SDSS fibre and the underlying broad component is real. But the broad component in H$\alpha$ may as well be 
due to the contribution from stellar phenomenon. In these three galaxies, the broad component detection was 
marginal and we removed them from our final secure sample of broad line AGN candidates. 

 The broad H$\alpha$ component of 
other two galaxies (identification numbers, 28 and 30, Figures: 17 and 18) are significantly blue shifted (15 - 17 {\AA} which is 
$\sim$ 10 - 11 times the error associated with the central wavelength) which 
can be more likely to be associated with an outflow event rather than with a BH accretion. One of 
these galaxies (identification number 28) has 
a similar blueward component in [O {\sc iii}] along with the blue shifted H$\alpha$ balmer line, 
which also supports the presence of outflow. These two galaxies are also removed from our final sample.  
 For the remaining 25 sample galaxies, except the galaxy with 
identification number 2, the central wavelengths of the broad and narrow H$\alpha$ components are separated by less than 3 times the error 
associated with the line centers. For the galaxy with identification number 2, the broad and narrow components are separated by $\sim$ 6 
times the error associated with the line centers and the broad component is blue shifted. This galaxy also shows a blue shifted outflow 
signature in [O {\sc iii}]. It can be assumed that for 
the galaxy, 2 the broad H$\alpha$ component may have contribution  from both the broad line region due to BH accretion and from 
an outflow. We could not deconvolve the components due to the outflow and due to BH accretion using numerical profile fitting. 
Another galaxy, UGC 6614 (identification number 24) has a clear signature of outflow, blue shifted component 
in [O {\sc iii}] doublet lines as well as in H$\alpha$ and H$\beta$ emission lines, along with a broad H$\alpha$ 
component. The candidates (identification number 2, 24, 28 and 30) with outflow signatures are discussed in Section: 5.3. 

In order to check whether the broad H$\alpha$ feature is due to the over 
subtraction of the H$\alpha$ absorption line, we modelled the H$\alpha$ absorption line of each sample (obtained 
from the best fit model, Section: 3.1) with a single Gaussian and obtained the FWHM. The FWHM of H$\alpha$ absorption 
line in our sample galaxies is less than 600 kms$^{-1}$, which is one of the criteria to identify the broad 
H$\alpha$ component. For all our sample galaxies except 
the galaxy, LSBC F500-V01, the FWHM of 
broad H$\alpha$ component is above $\sim$ 2-3 times the FWHM of the H$\alpha$ absorption line. 
For LSBC F500-V01 (with identification number, 15), the FWHM of the H$\alpha$ absorption 
line is 550 kms$^{-1}$ and the FWHM of broad H$\alpha$ is 995$\pm$286 kms$^{-1}$. Also, the absorption line fit 
has large residue in the wings. Thus the broad H$\alpha$ component of this galaxy may not be real. 
As the two values are comparable and the absorption line fit is poor, we removed this galaxy 
from our final sample of broad line AGN candidates. The broad H$\alpha$ detection in other sample galaxies appears to be 
real and not due to the over subtraction of H$\alpha$ absorption line.  
Thus we have 24 confirmed  
broad line AGN candidates for which the mass of the BH can be estimated using scaling relations. Out of 
24 broad line AGN candidates, 8 of them ($\sim$ 30\%) showed the presence of broad H$\beta$ component as well. The 
broad line parameters are given in Table A2.  The emission line fits, similar to Figure: 2, of some of the remaining galaxies 
are shown in Section: 5.4 and the rest in appendix, A1.   

 \begin{figure}
\resizebox{\hsize}{!}{\includegraphics{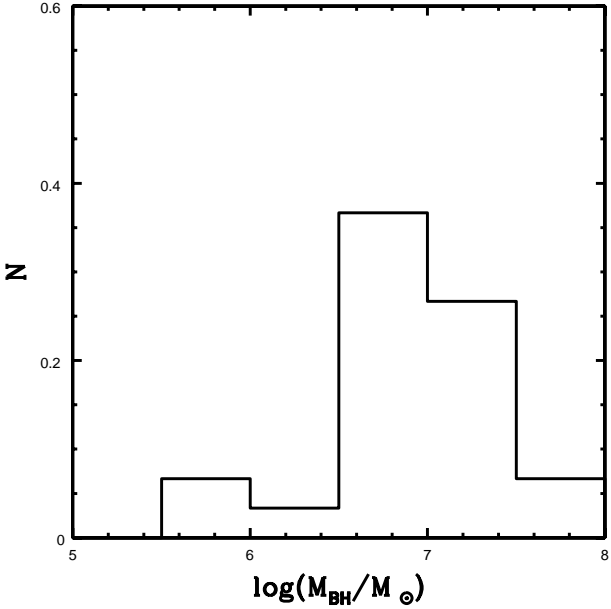}}
\caption{The M$_{BH}$ distribution of 24 broad line AGN candidate LSB galaxies.}
\end{figure}

\subsection{Mass of the Central Black Hole}
The virial M$_{BH}$ of 24 broad line AGN candidates was calculated using the equation given in \cite{R13}, 
using the luminosity and FWHM of broad H$\alpha$. The equation is given below. \\\\
$log {M_{BH} \over M_{\odot}} =  6.57 + 0.47 log {L_{H\alpha} \over 10^{42} ergs s^{-1}} + 2.06 
log {FWHM_{H\alpha} \over 10^{3} km s^{-1}}$\\

In the original equation given in \cite{R13} the term log($\epsilon$) in the right hand side  
is the scale factor that depends on the broad line region geometry. There is a range of values 
available in the literature for $\epsilon$ ($\sim$ 0.75 - 1.4). We have used $\epsilon$ = 1. The variation in the estimated 
M$_{BH}$ values with the adopted values of $\epsilon$ and its effect on the observed M$_{BH}-\sigma_e$ correlation 
is discussed in Section: 5.1.\\
 
The masses estimated are in the range 4.8 x 10$^5$ M$_{\odot}$ - 3.6 x 10$^7$ M$_{\odot}$. The median mass 
was found to be 5.62 x 10$^6$ M$_{\odot}$. 
The median mass is in the low end 
of the SMBH mass regime. These low mass BHs in the poorly evolved and isolated LSB galaxies 
are good candidates to study the evolution of heavy seed BHs. The mass distribution is shown in Figure: 7. 
The figure shows that the LSB galaxies have an asymmetric distribution 
in their BH masses, with a sharp decrease in the galaxies with BH masses less than 
$10^{6}M_{\odot}$. This can be due to the selection effect on the sample. 
The two galaxies with BH masses less than 
10$^{6}M_{\odot}$ are good candidates to study the nature of seed BHs in 
galaxy nuclei.  
The Eddington ratio of these accreting BHs are estimated using the bolometric luminosity estimated 
from [O{\sc iii}]$_{5007}$ flux. The value associated  with each galaxy is given in Table A2. 

\begin{figure}
\resizebox{\hsize}{!}{\includegraphics{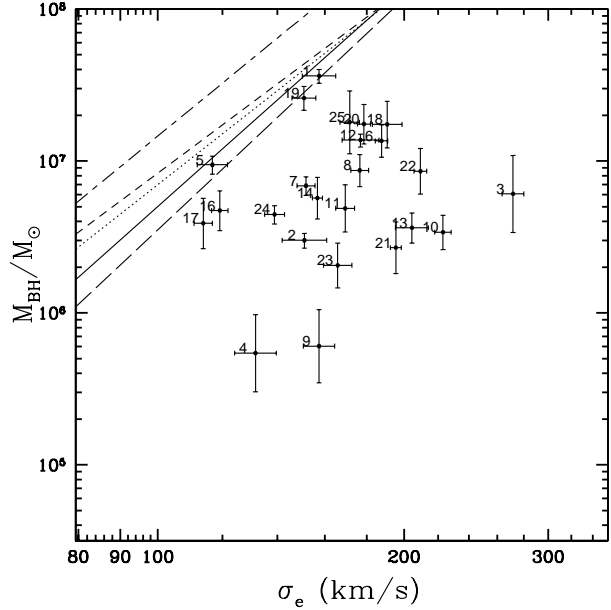}}
\caption{The M-$\sigma_e$ plot with broad line AGN candidates. The linear regression lines given by \citep{T02}, 
\citep{FM00}, \citep{G09}, \citep{KH2013} relation for classical bulges/elliptical galaxies and \citep{MM13} relation for late type 
galaxies (dashed, solid, dotted short-long dashed and long-dashed lines, respectively) for M$_{BH}$ against $\sigma_e$ are also 
shown.}
\end{figure}

\subsection{M$_{BH}$ vs $\sigma_e$ plot}
The stellar velocity dispersion values obtained using pPXf were corrected for aperture effects 
due to the finite size of SDSS fibre.  The equivalent velocity
dispersion $\sigma_e$
at a radius of $r_e$/8 in the galaxy bulge, where $r_e$ is the effective radius of the bulge, is calculated 
using the
transformation equation given by \cite{Jorg95}. The effective radius of the bulge corresponding to each galaxy in our sample is 
obtained from the SDSS database where a de Vaucouleurs profile is fitted to the surface brightenss profile and the 
effective radius of the bulge is obtained. The BH mass estimates for the 24 galaxies in our sample
plotted against the equivalent velocity dispersion values, $\sigma_e$ are shown in 
the M$_{BH}$ -$\sigma_e$ plot in Figure: 8.  Also plotted
in the figure are the linear regression lines given by \cite{T02}, \cite{FM00}, \cite{G09}, \cite{KH2013} for only classical 
bulges/elliptical galaxies and \cite{MM13} for late type galaxies (dashed, solid, dotted, short-long dashed and long-dashed lines, 
respectively) for M$_{BH}$ against $\sigma_e$, based on a quiescent galaxy sample. 
The M$_{BH}$ - $\sigma_e$ relations obtained for AGN sample appears to be shallower/flatter than that obtained 
for inactive/quiescent galaxies (\citealt{Woo2010}, \citealt{GH2006}). The slopes of active and inactive sample are consistent 
only with in 2$\sigma$ level. \cite{Woo2013} found that this discrepancy is due to 
inherent selection effects in the observed sample and intrinsically the M$_{BH}$ - $\sigma_e$ relations for both the active and 
inactive galaxies are similar. \cite{Xiao2011} investigated the low mass end of the M$_{BH}$ - $\sigma_e$ relation of active galaxies 
and found that they tend to follow the extrapolation of the M$_{BH}$ - $\sigma_e$ relation of inactive galaxies. Hence we use only 
the scaling relations obtained from quiescent galaxy sample as reference frame for the comparison of our sample. 

From the plot we can see that most of the galaxies are offset from the 
M-$\sigma_e$ correlation.  Here we would like to point out that the errors given with the $\sigma$ values are obtained from the pPXF fit. 
The default errors given by pPXF are known to be an underestimate of the true uncertainty associated with the $\sigma$. But even an  
increase in the errors obtained from the pPXF fit by three times is not sufficient to explain 
the observed offset of our sample galaxies in the M$_{BH}$ - $\sigma_e$ plot.  

The possible reasons for the observed off-set are discussed in Section: 5. 

\section{Discussion}

 \subsection{Possible reasons for the observed offset in the M-$\sigma_{e}$ plot}
It is now well established that BHs and bulges co-evolve
(e.g. \citealt{Heck04}; \citealt{V11}), though there are exceptions found in bulgeless galaxies 
(Seacrest et al. 2013; Simmons et al. 2013). One of the strongest observational indicators for this co-evolution is the 
$M-\sigma_e$ correlation (\citealt{G09}; \citealt{MM13} and references therein). 
Using the bulge velocity dispersions, we 
plotted our galaxies on the BH mass - velocity dispersion ($M-\sigma_e$) correlation (Figure: 8). 
Most of our sample galaxies ($\sim$ $>$ 80\%) lie below the correlation.

The observed offset of our sample in the M-$\sigma_e$ plot 
could be explained due to an 
over-estimation of $\sigma$ and/or an under-estimation of M$_{BH}$. 
In this section we describe these possibilities 
in detail and then explain why we think that the observed offset of our sample galaxies 
in the M-$\sigma_e$ correlation is real. 
\subsubsection{Methods adopted for the estimation}

{\bf Offset in M$_{BH}$}: The M$_{BH}$ of three LSB galaxies in our sample, with identification numbers, 1 (1226+0105), 12 
(LSBC F727-V01) and 24 (UGC 6614) have been previously estimated by various groups. \cite{GreeneHo2007} estimated the BH masses of 
galaxies, 1226+0105 and LSBC F727-V01 as 2.51 x 10$^7$ $M_\odot$ and 0.63 x 10$^7$ $M_\odot$ respectively. The error bars associated with the 
M$_{BH}$ estimates are not provided. They used the formalism provided by \cite{GH05} for the M$_{BH}$ estimation. Our 
M$_{BH}$ estimates for 1226+0105 and LSBC F727-V01 are 3.63$^{+0.40}_{-0.35}$ x 10$^7$ $M_\odot$ and 
1.37$^{+0.14}_{-0.12}$ x 10$^7$ $M_\odot$ respectively and both the values are slightly higher than the estimates of \cite{GreeneHo2007}. 
Our estimates were obtained using the relation given by \cite{R13} who have also followed the approach outlined in \cite{GH05} to estimate the M$_{BH}$, but with the modified 
radius-luminosity relationship of \cite{bentz2013}. This modification causes a difference in M$_{BH}$ estimates in these galaxies, 
of the order of $\sim$ 1.7 x 10$^7$ M$_\odot$. The differences between our estimates and that from \cite{GreeneHo2007} are within this 
range. For UGC 6614 with identification number 24, the M$_{BH}$ estimates are provided by \cite{n10} and \cite{Ram11}. \cite{Ram11} 
estimated the M$_{BH}$ using the same technique that we used (from the single epoch measurements of 
H$\alpha$ emission line from the spectra obtained from Himalayan Chandra Telescope and used the Starburst 99 model 
to remove the stellar continuum of the host galaxy) and obtained a BH mass of 
3.89 $^{+ 1.21}_{-1.04}$ x 10$^6$ M$_\odot$. From Table 3, we can see that our M$_{BH}$ estimate for 
UGC 6614 is 4.44 $^{+ 0.63}_{-0.58}$ x 10$^6$ M$_\odot$ and these two values match well within errors. 
For all these three galaxies, the M$_{BH}$ estimates from previous studies based on single epoch measurements 
are in the same order of magnitude with our estimates. Out of these three galaxies the galaxy number 1 lie on the observed M-$\sigma_e$ correlation, whereas the galaxies, 12 and 
24 lie below the correlation. The small differences in the estimates are not sufficient to explain the observed offset 
of these two galaxies. \cite{n10} used the technique of excess variance in X-ray emission and determined a mass of of 0.12 x 10$^6$ M$_\odot$.
The estimate of M$_{BH}$ from \cite{n10} is lower than our estimate. This can be attributed to the uncertainties in the 
method adopted. But this order of difference in the mass estimates are also not enough to explain the observed offset in the 
M -$\sigma_e$ correlation. 

In the present work, the virial BH masses of broad line AGN candidates are 
estimated using the single epoch H$\alpha$ emission line measurements. The formalism (\citealt{R13}, 
\citealt{GH05}) is based on the strong empirical correlation between the 
radius of the broad line region and the continuum luminosity and also the scaling relations using the 
reverberation mapping M$_{BH}$ estimates. The virial BH masses are subject to a number of 
uncertainties, even the different procedures to measure luminosities and line widths used in the virial 
mass estimators can cause discrepancies \citep{shen08}. The broad line region geometry varies considerably 
from object to object \citep{barth11} and assuming a single value for the geometric scaling factor, 
$\epsilon$ (here we assume a value, $\epsilon$ = 1) can cause uncertainties in the M$_{BH}$ estimations. 
 The value of scale factor which is physically associated with the geometry of the 
broad line region has a range ($\epsilon$ $\sim$ 0.75-1.4,\citealt{grier13}) and the geometry varies for each galaxy.  
If we adopt the values 0.75 and 1.4 then we obtain the lower and upper 
limits for the M$_{BH}$ estimates. 
M-$\sigma_e$ plot with the upper mass limits is shown in the upper left panel of 
Figure: 9. From the plot we can see that even the upper limit of 
the BH masses are insufficient or falls short to explain the observed offset in the 
M-$\sigma_e$ relation.  
\begin{table}
\caption{Comparison of $\sigma$ estimates based on our method for the sample with 
independent $\sigma$ measurements. The previous estimates of $\sigma$ values of all the galaxies in the table, 
except NGC 5273 are taken from \citep{grier13}. 
For NGC 5273, the $\sigma$ estimates are from \citep{bentz14}. For NGC 5273 the $\sigma_e$ 
(74.1 $\pm$ 3.7) was provided in \citep{bentz14} and we converted that to $\sigma$ using the effective radius (r$_e$ = 18.2) 
values taken from SDSS database.}
 \begin{tabular}{lcc}
{\bf Galaxy} & {\bf Previous estimate of }  & 
Our estimate \\
              & {\bf $\sigma$ (km/s)}  & $\sigma$ (km/s) \\   
\hline
NGC 5273& 72.8$\pm$3.7 & 73.3$\pm$7.3\\\\
Mrk 202 &78.0$\pm$3.0&83.6$\pm$14.6\\\\
Arp 151 &118.0$\pm$4.0&129.8$\pm$9.0\\\\
SBS1116+583A&92.0$\pm$4.0&86.0$\pm$16.8\\\\
Mrk 590&189.0$\pm$6.0&196.4$\pm$5.8 \\\\
\hline
 \end{tabular}
\end{table}

\begin{figure*}
\resizebox{\hsize}{!}{\includegraphics{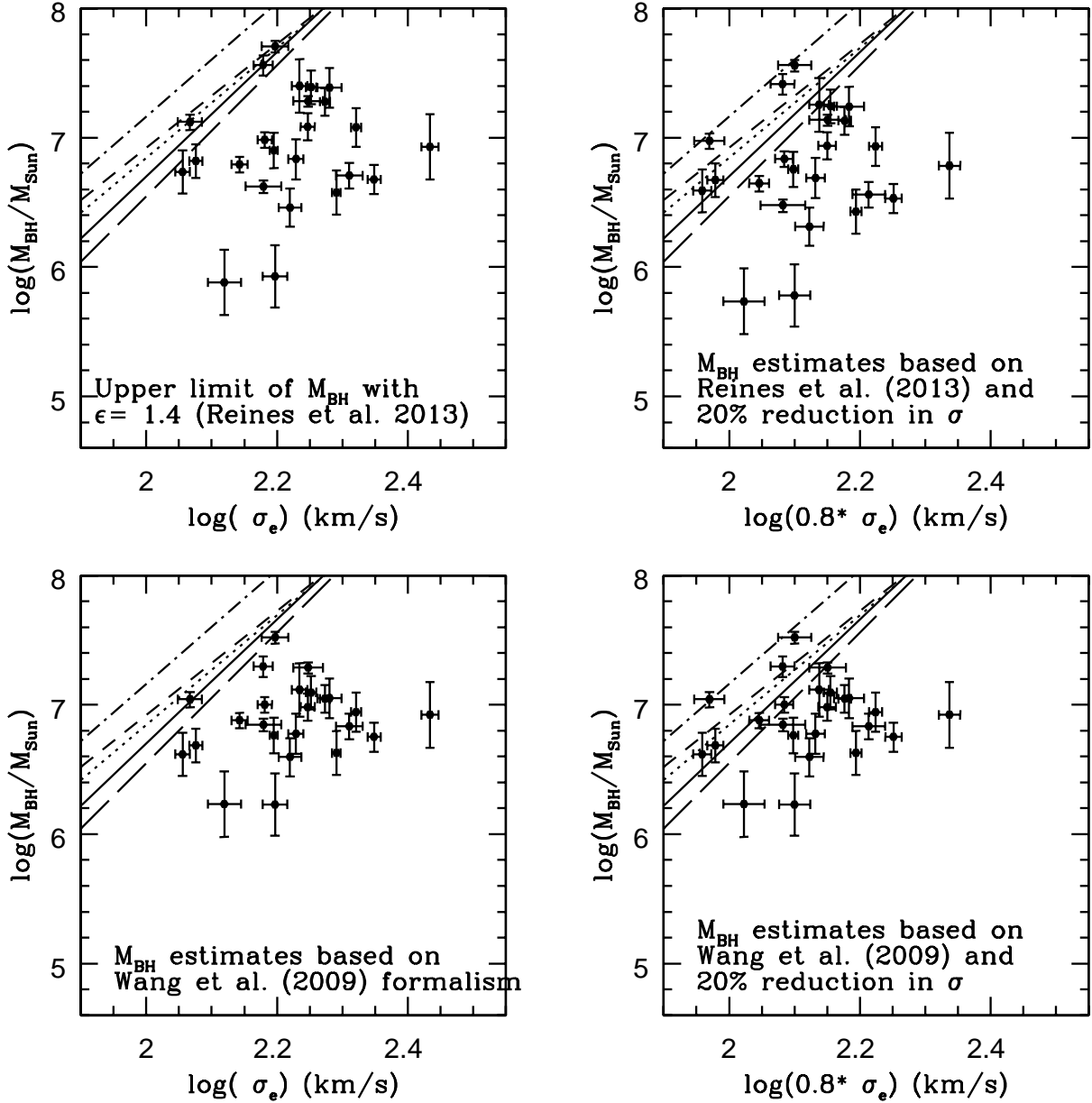}}
\caption{The lower left panel shows the M$_{BH}$ - $\sigma_e$ correlation with M$_{BH}$ estimates based on 
the formalism by \citep{Wang2009}. The lower right panel 
shows the M-$\sigma_e$ plot with the BH masses estimated using the formalism by \citep{Wang2009} and with a 20\%
reduction in the observed $\sigma$ values.  
The upper left panel shows the M-$\sigma_e$ plot for the 
BH masses estimated using $\epsilon$ value of 1.4. The upper right panel shows the M$_{BH}$ - $\sigma_e$ plot with 20\%  
reduction in the observed $\sigma$ values.
 The linear regression lines given by \citep{T02}, \citealt{FM00} \citep{G09}, \citep{KH2013} relation for the classical bulges/elliptical 
galaxies and \citep{MM13} relation for late type 
 galaxies (dashed, solid, dotted, short-long dashed and long-dashed lines, respectively) for M$_{BH}$ against $\sigma_e$ are 
 also shown in all the panels.}
\end{figure*}


 \cite{Wang2009} derived a new formalism to estimate the M$_{BH}$ using broad H$\beta$ line. They 
found that the M$_{BH}$ $\propto$ (FWHM of H$\beta$)$^{1.09\pm0.22}$. This departs significantly from the relation, 
M$_{BH}$ $\propto$ (FWHM of H$\beta$)$^{2}$, which was widely used in the previous studies. They claimed that their 
formalism reduced the internal scatter between the mass estimations based on reverberation mapping and the single 
epoch estimators. They found that the previous single epoch M$_{BH}$ estimators based on M$_{BH}$ $\propto$ (FWHM of H$\beta$)$^{2}$ 
underestimated the M$_{BH}$ at lower end and overestimated the M$_{BH}$ at higher mass end. 
Recently \citep{wd14} reported that the M$_{BH}$ estimated using \cite{Wang2009} formalism are 
more consistent with those from the M$_{BH}$ - $\sigma_e$ relation than those from previous single epoch 
mass estimators. They suggested that the the BLR of AGN are multi-componential with at least two components, an intermediate 
line region producing the emission line core and a very broad line region which produces the wings of the emission line. 
They argued that the contribution of very broad line region affects the M$_{BH}$ - FWHM (of the emission line which is 
the tracer of virial velocity) relation and induces a systematic bias in the BH masses of AGN estimated from 
single epoch estimators. They found that the formalism based on \cite{Wang2009} is less affected by this bias. 
Our M$_{BH}$ estimates are based on the relation given by \cite{R13} who have used the assumption, 
M$_{BH}$ $\propto$ (FWHM of H$\beta$)$^{2}$ in their formalism. 
In order to check whether the systematic bias, suggested by \cite{Wang2009} and \cite{wd14}, 
can explain the observed offset of our sample in the M$_{BH}$ - $\sigma_e$ 
correlation we re-estimated the M$_{BH}$ of our sample using the formalism by \cite{Wang2009}. 
We used the formalism derived by \cite{Wang2009}, \\\\
$log {M_{BH} \over M_{\odot}} = 7.39 + 0.5 log {L_{5100} \over 10^{44} ergs s^{-1}} + 1.09 log {FWHM_{H\beta} \over 10^{3} km s^{-1}}$ \\\\
and the empirical relations provided by \cite{R13}, between broad H$\alpha$ luminosity and continuum luminosity, L$_{5100}$, 
and the FWHM of H$\alpha$ and H$\beta$,\\
L$_{H\alpha}$ = 5.25 x 10$^{42}$ ($L_{5100}\over10^{44} ergs s^{-1}$)$^{1.157}$ ergs s$^{-1}$ and \\
FWHM$_{H\beta}$ = 1.07 x 10$^3$ ($FWHM_{H\alpha}\over1000 km s^{-1}$)$^{1.03}$ km s$^{-1}$ \\
to obtain a relation between M$_{BH}$ and the broad H$\alpha$ parameters. The relation hence obtained was given by, \\\\
$log {M_{BH} \over M_{\odot}} =  7.11 + 0.43 log {L_{H\alpha} \over 10^{42} ergs s^{-1}} + 1.1227
log {FWHM_{H\alpha} \over 10^{3} km s^{-1}}$\\\\
and was used to re-estimate the M$_{BH}$ of our sample. The re-estimated M$_{BH}$ is plotted against $\sigma_e$ and is shown in the lower left 
panel of Figure: 9. From the plot we can see that the M$_{BH}$ of the sample galaxies have slightly increased and the points have moved 
closer to the M$_{BH}$ - $\sigma_e$ correlation. But the majority of our sample galaxies are still significantly off from the 
correlation and hence the systematic bias in the estimation of M$_{BH}$ due to the assumption, M$_{BH}$ $\propto$ (FWHM of H$\beta$)$^{2}$ is 
insufficient to explain the observed offset of our sample from the M$_{BH}$ - $\sigma_e$ correlation.

{\bf Offset in $\sigma$}:  \cite{bentz15} have provided a 
database of AGN BH masses, for a sample of 60 galaxies, estimated 
using reverberation mapping. From the sample of 60 galaxies  
we selected the bulge dominated galaxies which have SDSS DR10 spectra and also have 
BH masses in the range of our estimates ($\sim$ 10$^6$ - 10$^7$ M$_{Sun}$). We got a sample of 
5 galaxies (NGC 5273, Mrk 202, Mrk 590, Arp 151 and SBS1116+583A) which satisfy our selection criteria.
For this sample of five galaxies (in Table: 2), independent stellar velocity dispersion measurements 
(Mrk 202 and SBS1116+583A have estimates from Keck/ESI observations, Arp 151 has estimates from double 
spectrograph in Palomar Hale 5 m telescope, Mrk 590 has measurements from 4 m telescope at Kitt Peak National 
Observatory and NGC 5273 has $\sigma_e$ measurements from the IFU 4.2 William Herschel Telescope) are 
available. The previous estimates of $\sigma$ values of all these 5 galaxies, 
except NGC 5273 are given in \citep{grier13}. 
For NGC 5273, the $\sigma$ estimate is from \citep{bentz14}. For NGC 5273 the $\sigma_e$ 
(74.1 $\pm$ 3.7) was provided in \cite{bentz14} and we converted that to $\sigma$ using the effective radius (r$_e$ = 18.2) 
values taken from SDSS database. The $\sigma$ values are given in Table. 2. 
We applied our procedure to estimate the $\sigma$ values to these 5 galaxies and the values are given in Table: 2. 
The values estimated by us are consistent 
with the above $\sigma$ values and this suggests that the estimation of $\sigma$ from SDSS matches with the estimates from other 
observations. 
It should be noted that all the five galaxies in Table 2 
are QSO's and are dominated by the AGN emission which makes the stellar velocity measurements difficult. 
The LSB galaxies in our sample are low luminosity AGNs and the SDSS spectra have sufficient S/N to estimate 
the stellar velocity dispersion.  
Thus the methodology adopted to compute the $\sigma$ value is appropriate 
and the uncertainties in the method are not likely to be the reason for the observed offset of our 
sample in the M-$\sigma_e$ correlation. 



Recent studies by \cite{bel14} suggested that line of sight effects due to galaxy orientation can 
affect the measured $\sigma$. They found that the edge-on orientations give higher velocity 
dispersion values, due to contamination by non-bulge/disk stars when measuring 
line of sight quantities. Studies by \cite{hart14} and \cite{deb13} also suggest that the $\sigma$ measurements 
can be affected by the contamination of line of sight disk  
stars. According to all these studies the disk contamination in highly inclined systems 
can increase the bulge velocity dispersion upto 25\%.    
The ellipticity (b/a) based on SDSS images of all of our sample galaxies are available in the SDSS database and ranges from 0.54 - 0.95. 
Ellipticity can be used as a proxy of galaxy inclination and it can be used to correct the observed $\sigma$ for 
orientation effects. Though \cite{bel14} find a correlation between the $\sigma$ and galaxy inclination they 
did not find any trend of $\sigma$ with ellipticity. \cite{bel14} 
suggested that this may be due to the reason that the ellipticity changes with radius and can be affected by the non-axisymmetric 
shapes as well as inclination. They recommend to use the kinematic inclination estimates for the correction of orientation effects 
in the observed $\sigma$. \cite{hart14} found that in barred galaxies, the effect of redistribution of angular momentum due to bars can 
also cause an increase of $\sigma$ in the central region (average of 12\% increase with a maximum of 20\%). 

Here we consider the extreme case, of 25\% increase of bulge velocity dispersion suggested due to the effect of orientations effect of host 
galaxy. This in turn suggests that a reduction up to 20\% 
in the observed $\sigma$ measurements is required to correct the effect of galaxy orientation. We considered this extreme case of 
reduction of 
20\% in the $\sigma$ values of all the galaxies in our sample and re-plotted the M-$\sigma_e$ plot. 
 This is shown in the upper right panel of Figure: 9.  
The figure shows that though the sample galaxies move close to the M-$\sigma_e$ correlation, majority of our sample 
are still off from the correlation. The identification number of our 
galaxies are not shown to avoid crowding. They are similar to what is shown in Figure: 8. 
The plot suggests that the orientation effects of galaxy cannot solely explain the observed offset of all the galaxies 
in our sample. 


Thus the effects which can increase the observed $\sigma$, explained above, individually cannot explain 
the observed offset 
in $\sigma$ of all the galaxies in our sample.  A detailed study involving observations and simulations of the nature of 
the different components of these galaxies are necessary to understand the evolution of 
black holes in LSB galaxies. \\

{\bf Combined offset in M$_{BH}$ and $\sigma$}: From the above two sub-sections we found that an under-estimation 
and over-estimation of M$_{BH}$ and $\sigma$ respectively based on the uncertainties in the methods 
adopted for their estimation, cannot explain the observed offset of our sample LSB galaxies from the M-$\sigma_e$ 
correlation (upper left, upper right and lower right panels of Figure: 9). At the same time we need to check the combined effect 
on these galaxies.  In the lower right panel of Figure: 9 we plotted 
the M$_{BH}$ estimates, based on the formalism by \cite{Wang2009}, against the $\sigma_e$ values 
which are reduced by 20\% (which is an upper limit of the effect on $\sigma$ due to the orientation of host 
galaxies). The plot shows that when a combined effect is considered, most of the sample galaxies lie close to the 
M-$\sigma_e$ correlation. But still the offset is present in many of the galaxies. 
Here we would like to point out that, 
the reduction in $\sigma$ considered here is the upper limit of the 
effect of orientation of the host galaxy to the observed $\sigma$. Thus the offset of all our sample from the 
M-$\sigma_e$ correlation may not be explained only by these effects. The intrinsic properties of the LSB galaxies may also 
contribute to the observed offset. This possibility is explained in the Section: 5.2.


\subsection{Is the lack of BH - bulge co-evolution an intrinsic property of LSB galaxies?}
 A significant fraction of bulge dominated GLSB galaxies have large bulges (\citealt{Galaz2006}; \citealt{Mor12})
 and so it is not surprising that GLSB galaxies show AGN activity, despite having low luminosity disks.
 However, what is surprising is that most of GLSB bulges appear to host relatively low mass BHs and many  
not even fall into the SMBH regime ($>~10^{6}~M_{\odot}$) \citep{Ram11}. The galaxies in our sample 
have BH mass in the range $10^{5} - 10^{7}~M_{\odot}$ and majority of them lie below the M$_{BH}$ - $\sigma_e$ correlation of the 
elliptical galaxies.  The precise M$_{BH}$ estimates for a group of spiral galaxies with maser 
 disks (from megamaser measurements, \citealt{kuo2011}) are also $\sim$ 10$^7$ M$\odot$. These galaxies also lie below the 
 M$_{BH}$ - $\sigma_e$ correlation of massive elliptical galaxies \citep{Greene2010}.
 
Theoretical models 
predict a M$_{BH}$ - $\sigma_e$ correlation of the form $M\propto{\sigma}^{\alpha}$ where $\alpha~=~4$ to $5$; the value of the 
constant $\alpha$ depends on 
whether the early BH mass accretion models included momentum driven winds \citep{SilkRees1998},  or thermally
driven winds \citep{Fabian1999}. The winds result in shocked shells of matter that are driven outwards and ultimately 
fragment to form stars that contribute to the growth of the bulge \citep{King2003}. In both models mergers 
play an essential role in driving gas into the nucleus and fueling BH accretion leading to coupled 
BH-bulge growth. Since mergers are important in the theoretical framework for explaining the $M-\sigma_e$ correlation, a possible 
explanation for some of the galaxies lying below the $M-\sigma_e$ (as well as having low mass BHs) could be the lack of major mergers 
in the history of their evolution. The two reasons supporting this hypothesis are,~(i)~Most GLSB disks appear 
relatively undisturbed in appearance; they do not show signs of recent merger event. Although some 
GLSB galaxies such as UGC 6614 have prominent rings that could be the result of galaxy collisions (e.g. \citealt{mapelli2008}), 
secular evolution can also result in the formation of such rings \citep{KK2004}. (ii)~They are usually 
isolated galaxies and often lie close to the edges of voids. 

 \cite{C2011} found that triggers for major BH growth since z$\sim$ 1 are not due to major galaxy mergers, but due to 
alternate mechanisms like internal secular processes and minor interactions. From the study of quasar host galaxies at z$\sim$ 2, 
\cite{SCH2012} found that the predominant driver of massive BH growth in these galaxies are secular processes rather than 
major mergers. Semi-empirical modelling of AGN fueling by \cite{Hop2014}, suggested that the secular processes dominate the black 
hole growth in the low mass end. \cite{Menci2014} investigated the effect of disk instabilities in triggering AGN in isolated galaxies and 
found that it can efficiently provide the BH accretion in the low luminosity end. \cite{KH2013} and references therein suggest 
that the local AGNs in late type spirals accrete at sub-Eddington rates and they have low BH masses (10$^5$ - 10$^7$ M$\odot$). These 
BHs grow mainly by secular processes and they do not co-evolve with their host galaxies. These studies suggest that secular processes 
dominate the growth of low mass BHs in our sample of LSB galaxies. 
Secular evolution leads to formation of pseudo bulges in disk galaxies 
\citep{KK2004}. \cite{Kormendynature2011} did a detailed classifications of the bulges in the galaxies for which dynamical 
BH mass estimations were available and found that the BHs do not co-evolve with pseudo bulges. 
The majority (7 out of 9) of the maser disk galaxies with low mass BHs are found to have pseudo bulges \citep{Greene2010}. 
\cite{Jiang2011} did a systematic study of the properties of the 147 active host galaxies with low mass BHs (10$^5$ - 10$^6$ M$\odot$) 
using HST observations. They found that majority of the sample galaxies with disks host pseudo bulges and only a small fraction 
host classical bulges. But interestingly, based on the structural classification by 
\cite{Simard2011}, the majority of our sample LSB galaxies have bulges which are of classical nature (bulge Sersic index $\ge$ 2.5, 
Table 1). 

\cite{deb13} suggested that when a disk forms around a pre-existing classical bulge, due to gravitational compression 
$\sigma_e$ increases. Then an increase of M$_{BH}$ by an average of 50-65\% is required to satisfy the M-$\sigma_e$ relation. 
If there is no corresponding growth in M$_{BH}$, then sample with classical bulges are expected to have an offset.  
They did not find any such offset in the M-$\sigma_e$ relation of classical bulges in the sample of \cite{G09}. 
Based on this they concluded that the BH in the classical bulges was growing along with the disk 
evolution. They also suggested that growth of  BHs might be regulated by the AGN feedback. If the galaxies in our sample 
which show significant offset have classical bulges in their center, that would suggest that the bulge 
has not evolved with the disk and lack BH - bulge co-evolution. The structure and kinematics of 
the BLR, at least as crudely encoded in the f factor, seems to be related to the large-scale properties or formation history of 
the bulge \citep{Hoandkim2014}. A detailed study on the classification of the bulges is required to better understand their relation 
between the observed offset of the LSB galaxies in the M$_{BH}$ - $\sigma_e$ correlation and we plan to do this as a future work.

The secular processes and formation of pseudo bulges are mainly facilitated by bars and by other instabilities like spiral arms in 
the disk \citep{KK2004}. In the case of LSB galaxies, which are dark matter dominated even significantly in the inner 
regions, have featureless disks due to the suppression of disk instabilities by dark matter halos. 
Thus the formation of bulges 
and their properties in these systems are hence not clearly understood. The effect of dark matter may increase the potential 
experienced by stellar population in the bulge and hence increase the $\sigma_e$ values which can lead to the offset of these galaxies in the M$_{BH}$ - $\sigma_e$ correlation. 
AGN feedback is one of the natural mechanisms for BHs to regulate their growth and 
to couple with the host galaxy properties. \cite{BoothSchaye2010} explored the correlation between SMBH mass and the mass of the hosting dark matter halo.  From their simulations, they suggested that the mass of SMBH is regulated primarily by the dark matter halo mass and not by the stellar mass of the galaxy.  
\cite{Krumpe2015} also suggested a positive correlation of dark matter halo mass on the mass of BHs which is prominent in the BHs with M$_{BH}$ $\sim$ 10$^{8-9}$ \msun.  In the case of LSB galaxies, the effect of dark matter  in the self regulation of black hole growth through AGN feedback and its coupling with the host galaxy properties may be important and it needs to be analysed further. It is necessary to obtain more direct measures of halo properties, such as halo mass to understand BH growth in LSBs. Another mode of BH growth, which is through disk instabilities can also have an impact in the dark matter dominant LSB galaxies.  The low surface gas density in these systems can also affect the growth of BHs and bulges.

\subsection{Do LSB nuclei harbor {\it Pristine} BHs ?}
Seed BHs that are formed in the early universe grow by mass accretion
to become the massive BHs in galaxies that we observe in our local universe; the accretion is driven by galaxy mergers and interactions. 
The most massive BH seeds formed from gravitational collapse of large gas clouds whereas the lighter BH seeds formed from the 
evolution of Population III stars (\citealt{vol10} and references therein). Models predict higher fraction of low mass galaxies to 
contain nuclear black holes if seeds are created from Population III stars \citep{Volonteri2008}. 
From the study of low mass BHS in small galaxies, \cite{Greene2012} found a tentative evidence that the progenitors of SMBHs were 
formed via direct collapse rather than from deaths of Population III stars using a simple argument of the occupation fractions, 
matching with the models of \cite{Volonteri2008}. 
The observational evidence for the presence of SMBH (10$^9$ - 10$^{10}$ \msun) at early epochs (z$\sim$7-10) poses time constraints on the growth efficiency of their seeds. Giving a jump start to the growth process through more massive seeds (10$^3$-10$^6$ \msun) and/or occurrence of super Eddington accretion episodes may overcome this limitation. These have been investigated thoroughly by \cite{PVF2015} and references therein. They found that the seeds in the mass range (10$^3$ - 10$^4$ \msun)  evolve less efficiently and those seeds in the mass range (10$^5$ - 10$^6$ \msun) grow very rapidly . The end product of these massive seeds  depends on the initial seed mass, the halo mass and density profile and the accretion scenario (standard Eddington-limited accretion or slim accretion disk appropriate for super Eddington accretion). The spectrum of the emerging radiation, for a massive BH observed at z=9, will be dominated in the infrared-submm (1-1000 micron) and X-ray (0.1-100 keV) bands and the future instruments like JWST and ATHENA could detect them \citep{PFVD2015}. \cite{Natarajan2011}, using models of direct collapse, predicted that LSB galaxies and bulgeless galaxies with large disks are less likely to form massive black hole seeds at high redshifts. According to \cite{Jarrett2012} extremely massive 
stars ($\sim$ 10$4$ \msun) at z $\sim$ 10 could collapse into BHs which could be the promising seed  candidates  for the origin of SMBHs (of the order of 10$^9$) at z$\sim$ 7.  The recombination emission from the surrounding H II regions of the super massive stars in the early universe may be bright enough to be detected by future missions such as JWST.
Alternatively, BH seeds for IMBHs or SMBHs could be formed from fluctuations in the early Universe during or soon after inflation \citep{kho10}. Another process for seed BH formation is through the evolution of dark stars. This is very relevant for LSB galaxies because they have a high dark 
matter content. Unlike regular stars that are supported by fusion, dark stars are supported by the energy released from dark matter particle or WIMP annihilations \citep{spo09}. Such stars could later evolve into massive BHs in the nuclei of LSB galaxies. Dark stars which have cooler temperatures than the fusion powered stars can accrete baryons all the way up to $\sim$ 10$^5$\msun. The observational signatures and capabilities of future missions like JWST to detect of these dark stars are 
discussed in literature (\citealt{Cosmin2012}, \citealt{Tanja2015}). \\\\

Simulations suggest that both gas accretion and galaxy mergers are important for the growth of seed BHs in the centers of galaxies as both processes lead to gas accretion by the nuclear BH \citep{AA2015}. But at high redshifts the growth is probably driven by gas accretion from the surroundings, resulting in the formation of SMBHs in the centers of massive host galaxies. 
 These SMBHs lie close to the  $M-\sigma_e$ relation. As the gas supply reduces, galaxy mergers and accretion events may play a more important role in BH growth \citep{Dubois2014}.  However, if a galaxy is isolated, the rate of gas accretion maybe too slow and the rate of galaxy mergers too low for the seed BHs to grow beyond $\sim10^{6}~M_{\odot}$ at redshift z=0. These relatively 
 lighter nuclear BHs in isolated galaxies, can grow through slow accretion leading to relatively low mass BHs ($<10^{6}~M_{\odot}$) that lie below the $M-\sigma_e$ relation \citep{VN2009}.  Alternative models have also been proposed for the formation of 
 intermediate mass BHs that have masses less than $<10^{6}~M_{\odot}$ \citep{PF2015}. 
 
Bulge dominated LSB galaxies, such as 
 those presented in this study, may be an example of such systems. These low mass BHs are relatively {\it Pristine} as they have not 
 undergone as much gas processing as SMBHs in massive galaxies. 
Thus they can reveal important clues to the initial BH mass function and help us constrain the early evolution of BHs in galaxies.
The BH masses in LSB galaxies lie in the $10^{6}-10^{7}~M_{\odot}$ range and the majority are outliers on the $M-\sigma_e$ relation. 
Also, the Eddington ratios of majority of them are not high (see Table 3).  Hence their BHs are not accreting 
at a high rate and they fall in the low luminosity AGN (LLAGN) class (\citealt{Nagar2005}; \citealt{H04}). 
Thus, the BHs in LSB
galaxies represent one of the best candidates for {\it Pristine} BHs and a good place to study seed black 
holes in our local universe. Other low mass BH candidates for such studies are those found in bulgeless 
galaxies (e.g. \citealt{ss14}; \citealt{salvo12}) and dwarf galaxies (\citealt{tho08}, \citealt{Reines2011}, \citealt{R13}, 
\citealt{Moran2014}.


\subsection{Nuclear Outflow from AGN in GLSB galaxies}

Theoretical models of galaxy formation and evolution often invoke feedback from AGNs and SNe to explain the quenching of 
star formation in galaxies and hence formation of very massive galaxies in the universe \citep{DiMatteo2005}. 
Outflows are believed to be one of the signatures of AGN feedback and hence forms an important evidence to test the BH-bulge co-evolution.
Numerical simulations indicate that feedback is triggered during an intense and rapidly accelerated BH fueling phase 
during a gas-rich galaxy merger (\citealt{narayanan2010} and references therein). 
So one would expect to see extreme outflowing gas moving with velocities greater than 1500 \kms only 
in a high-z quasar or/and ULIRGs associated 
with AGNs accreting at high eddington ratios.

\begin{figure*}
\centering
\begin{tabular}{cc}
\includegraphics[width=0.45\textwidth]{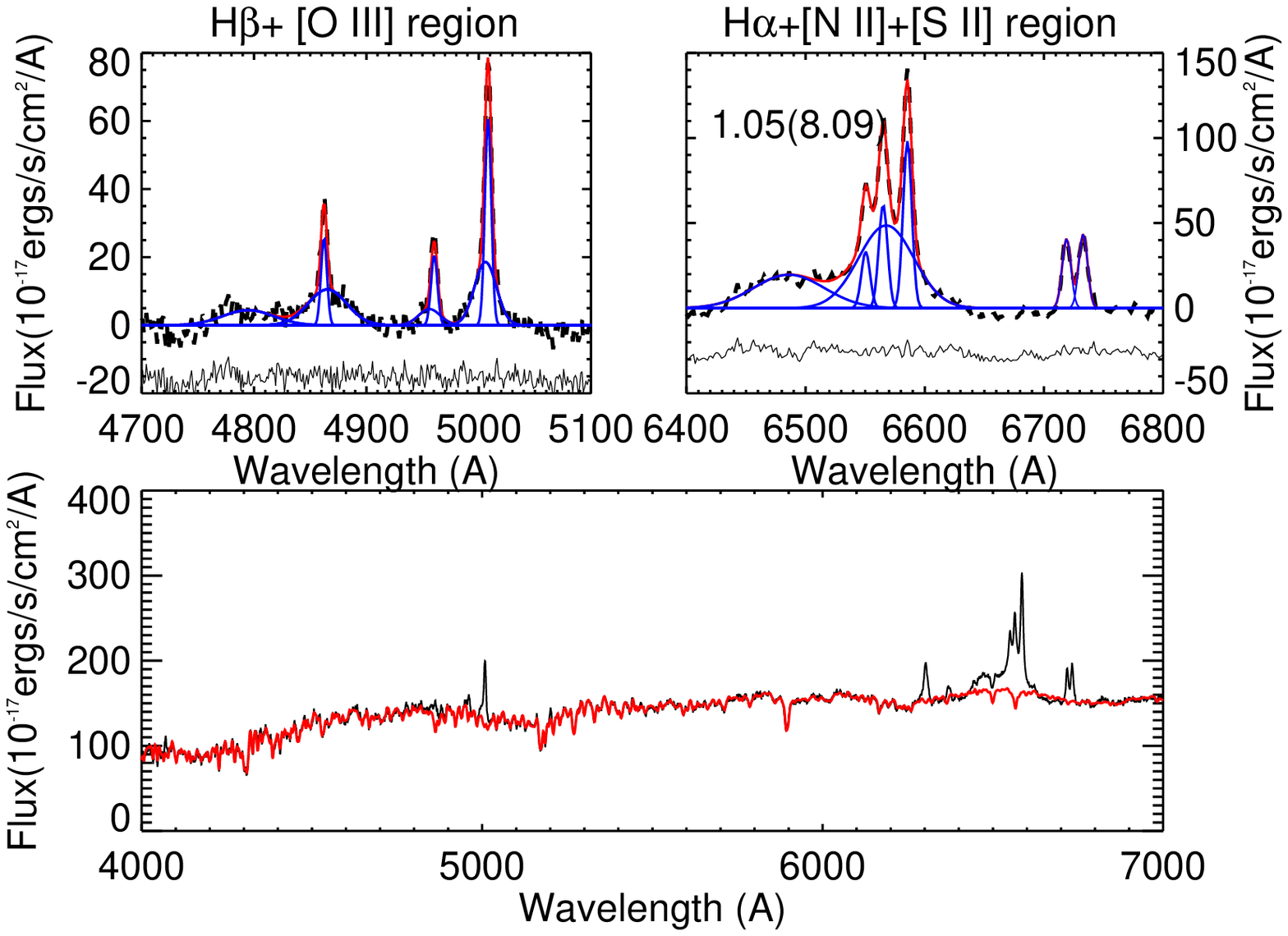}
&
\includegraphics[width=0.45\textwidth]{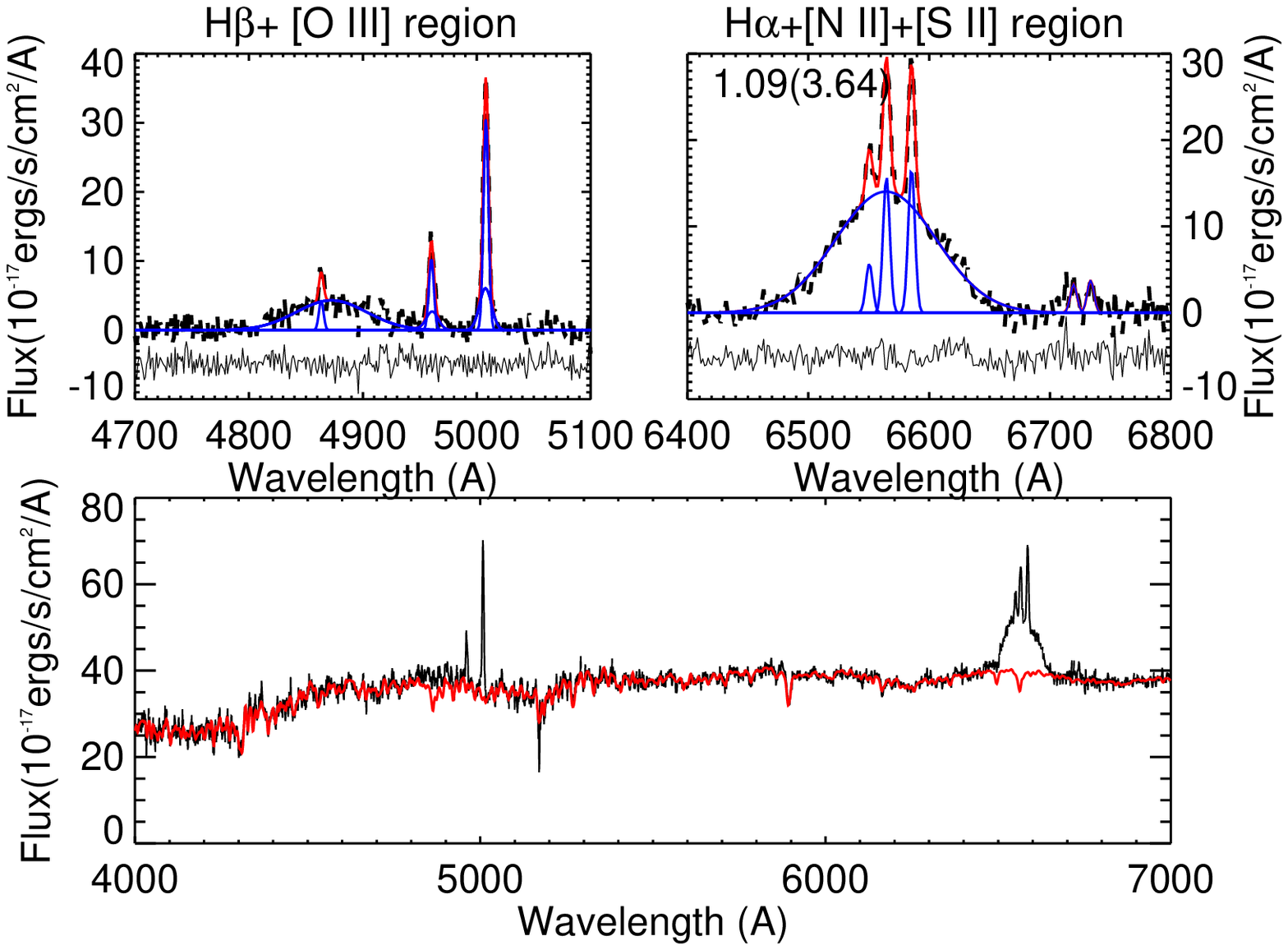} \\
UGC 6614 (No 24) & 1226+0105 (No 1) \\
\includegraphics[width=0.45\textwidth]{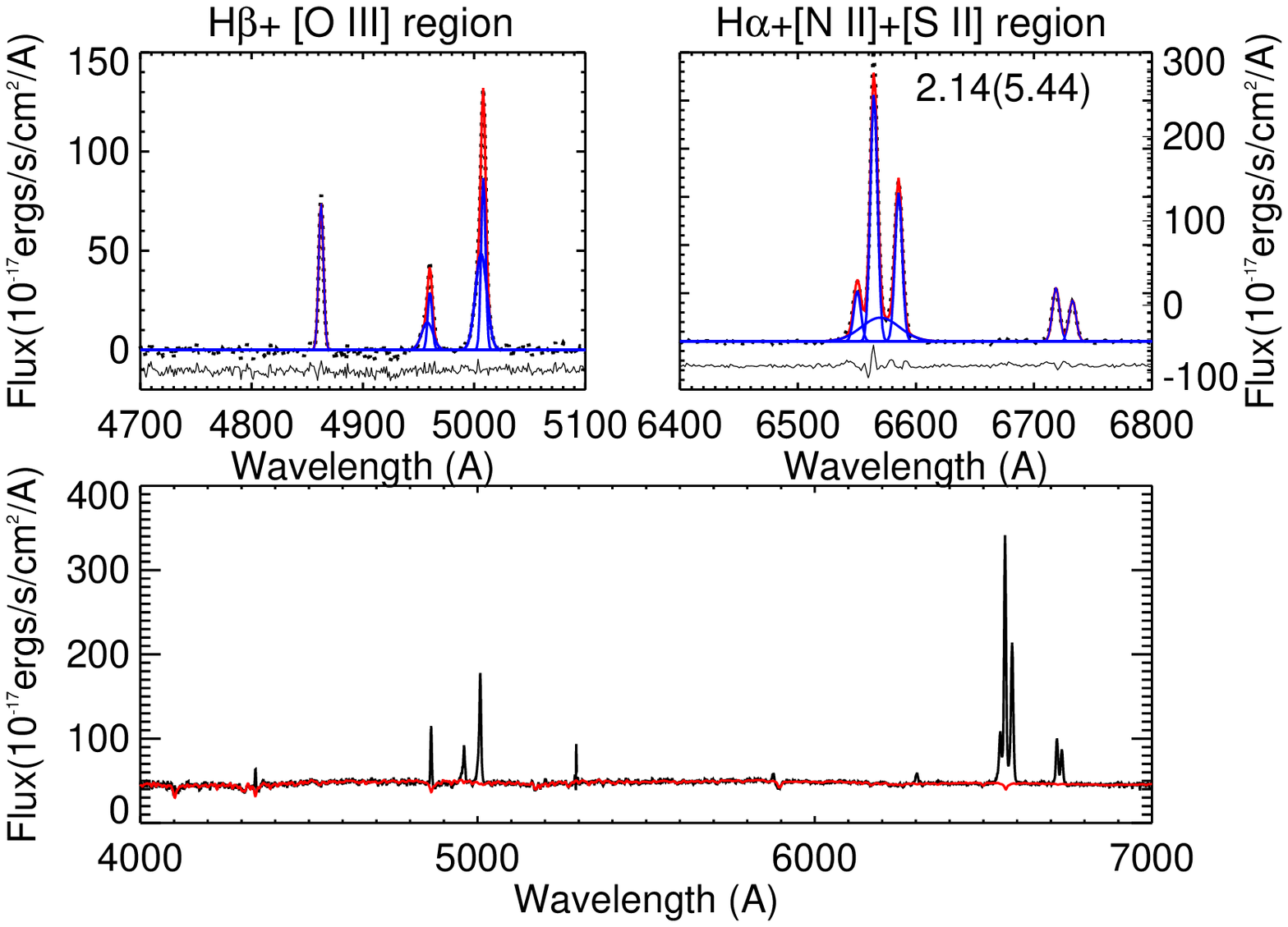}
&
\includegraphics[width=0.45\textwidth]{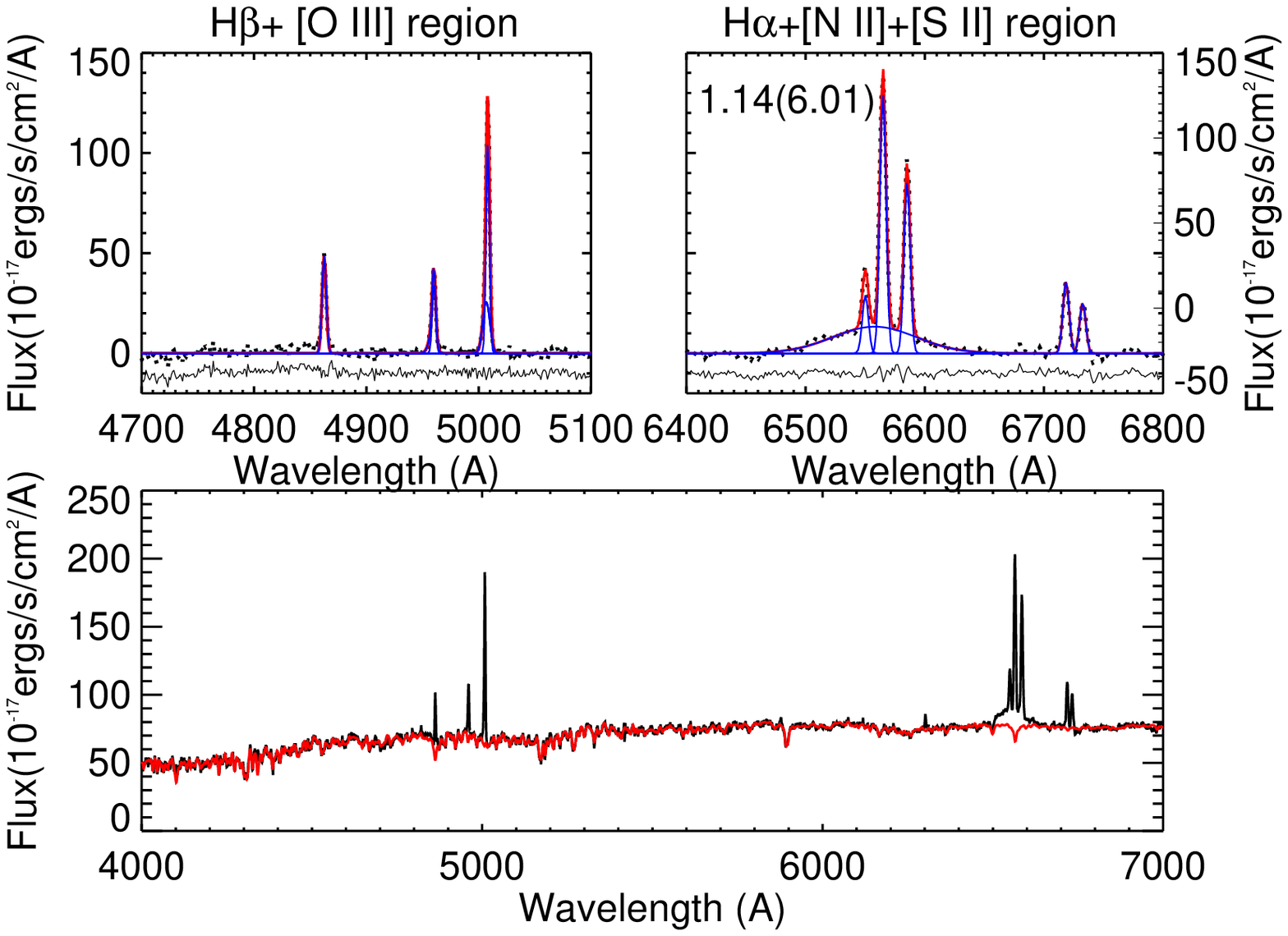} \\
VIII ZW 318 (No 2) & NGC 7589 (No 5) \\
\includegraphics[width=0.45\textwidth]{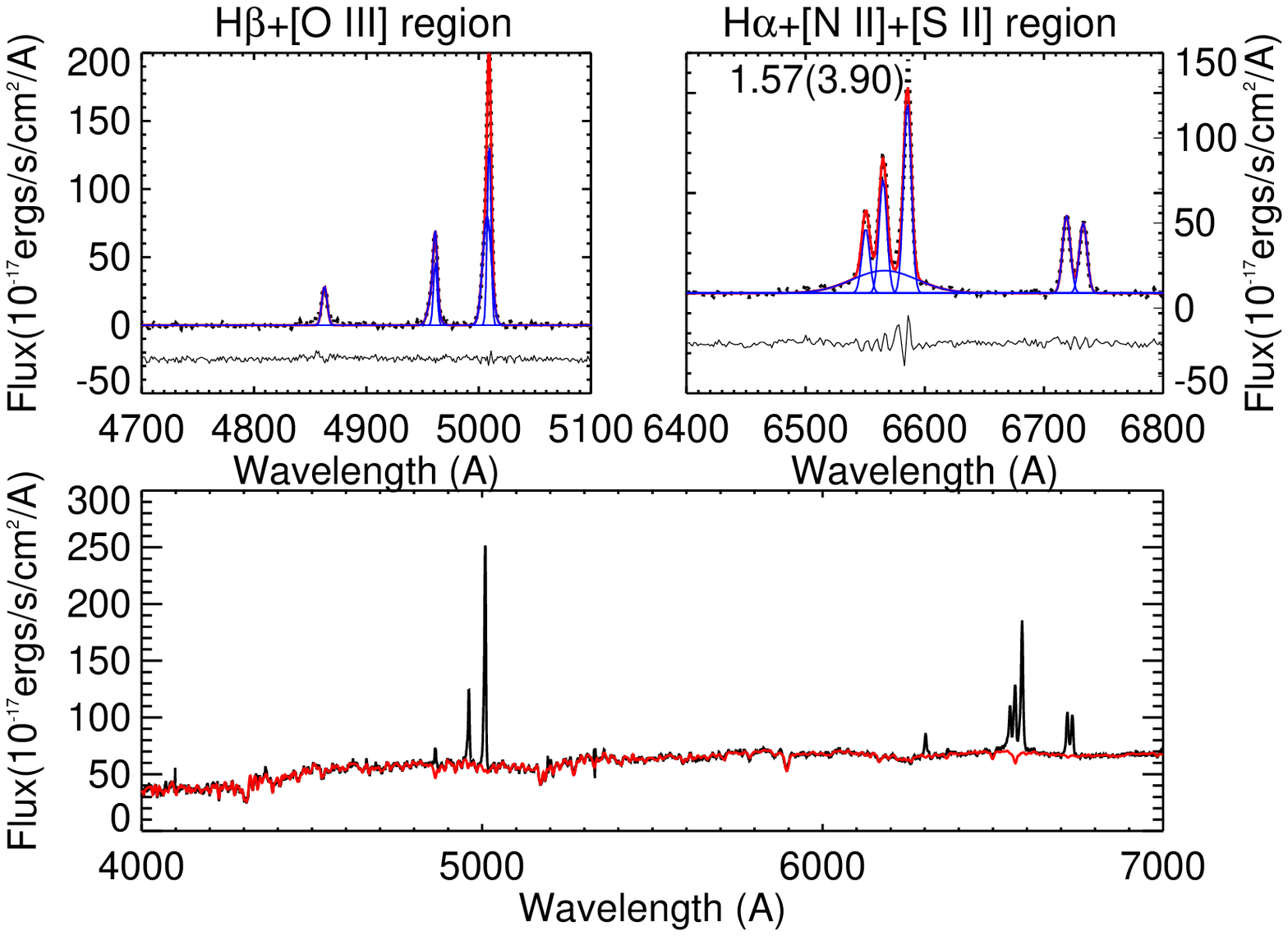}  &  
\includegraphics[width=0.45\textwidth]{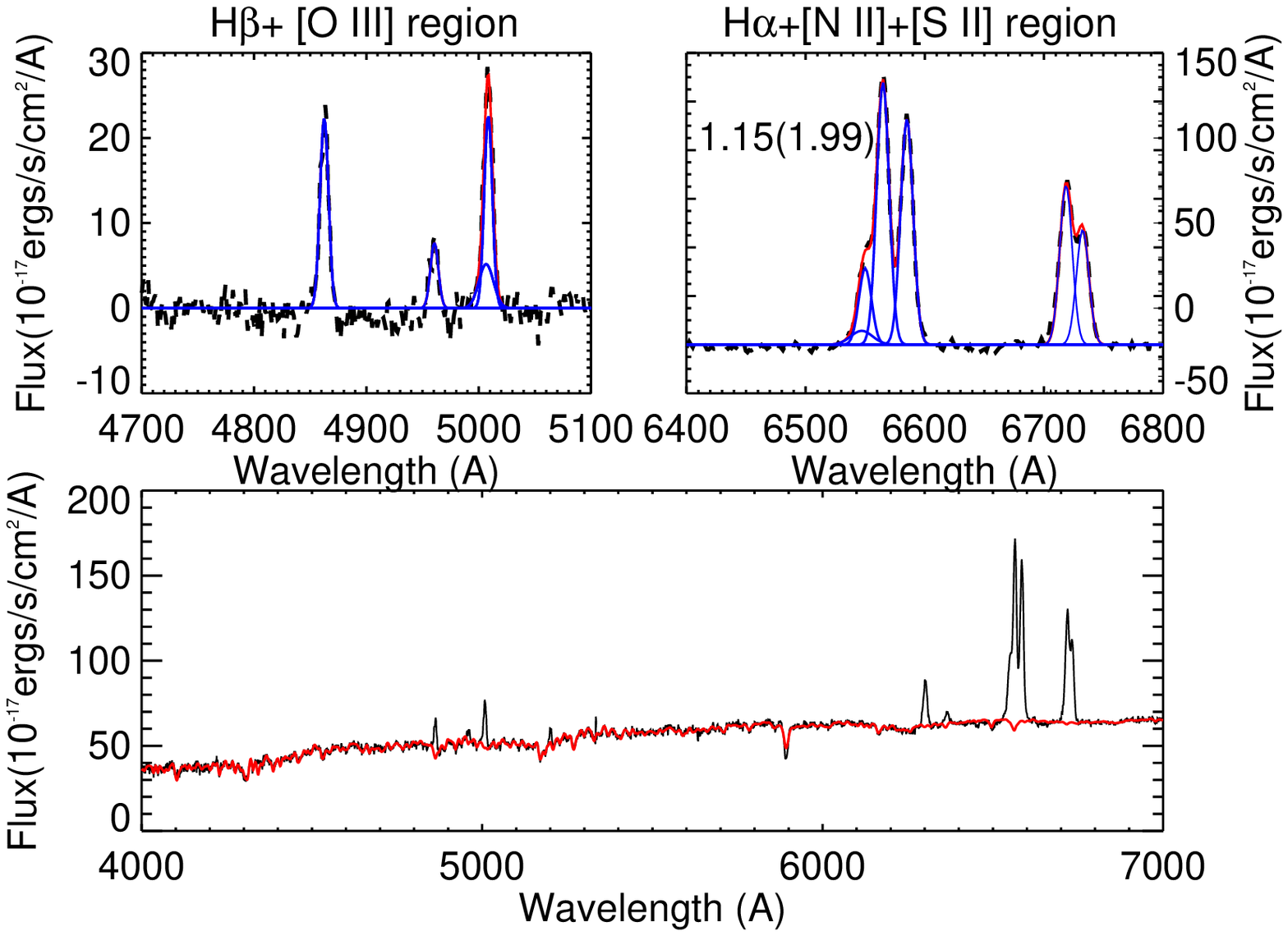}\\
2MASX J00534265-0105066 (No 7) & UGC 6284 (No 28) \\
\end{tabular}
\caption{Same as in Figure: 2, for the galaxies which show signatures of outflow.}
\end{figure*}

Interestingly, a few galaxies in our sample show signatures of outflows. 
The three galaxies in our sample, UGC 6614 (identification no 24), UGC 6284 (identification no 28) and UGC 5035 (identification no 30), 
which are classified to be LINERs show blue-shifted component in H$\alpha$ emission lines.  
For UGC 6614, along with the blue shifted outflow component, there is a 
broad H$\alpha$ component 
which is associated with the BH accretion. The blue shifted H$\alpha$ component in these galaxies suggest the presence of an 
ionised gas outflow. The forbidden [O {\sc iii}] emission lines are ideal tracers of extended ionized outflowing gas which 
cannot be produced by the high density broad line regions \citep{cresci2014}. Thus the asymmetric extended [O {\sc iii}] line profiles 
are considered as powerful signature of the presence of outflowing gas. 
Extended [O {\sc iii}]  emission from quasars 
are ubiquitously found by many authors (\citealt{nesvadba2008},\citealt{greenenadia2014}, Ramya et al. in preparation and 
references therein). 
UGC 6614 and UGC 6284 show the presence of outflow components in [O{\sc iii}] emission line region along with H$\alpha$ region. This 
suggests that outflow signatures in these galaxies are real. Whereas, UGC 5035 does not show 
any outflow signature in the [O{\sc iii}] region. Thus the blue shifted H$\alpha$ component in this galaxy may not be associated with any 
outflow event and may be associated with the broad line region and BH accretion. 
Along with UGC 6614 and UGC 6284, another four galaxies in our 
sample, 2MASX J12291286+0049042 (identification no 1), VIII ZW 318 (identification no 2), NGC 7589 (identification no 5) and 
2MASX J00534265-0105066 (identification no 7) also show blue 
shifted outflow signatures in [O {\sc iii}]. These four galaxies are classified as Seyfert galaxies in the BPT diagrams.  The velocity of 
outflowing gas, $v_{out}$ associated with the [O{\sc iii}] blue shifted components of the galaxies in our sample are given in Table 3. 
 The emission line fits, similar to Figure: 2, of the galaxies which show signatures of outflow are shown in different panels of Figure: 10. 

 \begin{table}
 \caption{Parameters of \othree$_{5007}$ outflow component}
 \centering
 \tiny
 \begin{tabular}{l|l|l|l|l}
 Galaxy& Velocity offset (kms$^{-1}$) & FWHm (kms$^{-1}$) \\
\hline
UGC 6614  & -141.38 $\pm$45.0&  1363$\pm$122.39\\
VIII ZW 318 & -120.04  $\pm$ 10.12& 705.87$\pm$82.12 \\
UGC 6284 & -114.7  $\pm$83.07& 900.07$\pm$ 169.34\\
2MASX J00534265-0105066 & -112.67 $\pm$ 6.34 & 520.32 $\pm$  67.24 \\
NGC 7589 & -79.47 $\pm$  23.23 & 404.26$\pm$55.69\\
2MASX J12291286+0049042 & -27.08 $\pm$  47.71& 727.54$\pm$97.77\\
\hline
\end{tabular}
\end{table}

UGC 6614 shows outflow feature in H$\alpha$, [O{\sc iii}] and H$\beta$ regions. The measured 
 Eddington ratio and BH mass for UGC 6614 are 0.024 and $\sim4.44\times10^6$ \msun \ respectively. 
The outflow component as measured by the blue shifted H$\alpha$ line is moving at 
3604.77 $\pm$ 86.39 kms$^{-1}$ 
with a FWHM of 3213 $\pm$ 200 kms$^{-1}$. Similarly the outflow component as measured by H$\beta$ line is moving 
at 4149 $\pm$ 267 kms$^{-1}$ with a FWHM of 
3304 $\pm$ 667 kms$^{-1}$. The velocity offset and FWHM values of H$\alpha$ and H$\beta$ components are similar. 
\cite{Ram11} had detected outflow components in H$\alpha$ and H$\beta$ regions. Our values are similar to their estimates.  
The H$\beta$ luminosity of the outflowing component can be used to estimate 
the mass of the ionized gas in the outflow.   
For further discussions on outflows and estimation of outflow rate, we  will concentrate on UGC 6614 alone.

It is possible to obtain an order of magnitude estimate for the mass outflow rate assuming that the outflowing gas is moving in a simple 
bi-conical distribution of uniformly filled outflowing clouds. We also assume that the farther side of the bi-conical outflow is obscured and hence we 
detect only the blue shifted emission while the red shifted part is obscured in the observed emission lines \citep{maiolinoet.al.2012}. 
The total ionized gas mass can be estimated using H$\beta$ luminosity and assuming electron number density (n$_e$) 
of 100 cm$^{-3}$. Then, 

$M_{ion} = 2.82\times10^9(\frac{L_{H\beta}}{10^{43}erg~s^{-1}}) ~ (\frac{n_e}{100 cm^{-3}})^{-1}$\msun

The H$\beta$ luminosity of the outflow component in UGC 6614 is $0.24\pm0.06\times10^{40}$ ergs$^{-1}$. Estimated ionized gas mass is  
$6.77\times10^5$ \msun. The $v_{out}$ of \othree$_{5007}$ component is $\sim$ 141.4 kms$^{-1}$. 
Following \cite{cresci2014} and assuming the radius at which the outflow is detected to 
be about $\sim1-3$ kpc (typical size of NLR), we derive the dynamical time $t \approx R_{out}/v_{out}$ to be equal to 
$\sim7.14-21.43$ Myr.
Then, the volume-averaged density is estimated as $\rho_{out} 
\approx \frac{M_{out}}{\Omega/3 . R_{out}^3}$ where $\Omega$ is the solid angle subtended by the outflow. 
The mass outflow rate is then : \\

$\dot M_{out} \approx \rho_{out}.\Omega~R_{out}^2.~v_{out} = 3 v_{out}.\frac{M_{out}}{R_{out}} $

The estimated mass outflow rate is then 0.1-0.28 \msunyr. 
The above outflow rate is estimated from broad [ O {\sc iii}] lines. If we consider the outflow component  
using H$\alpha$ and H$\beta$ lines, then we will have another estimate for the mass 
outflow rate which is $\sim$27 times more than the one estimated using the [O {\sc iii}] emission. 
The mass outflow rate estimated from broad H$\alpha$ (or H$\beta$) component is 2.7-7.695 M$\odot$yr$^{-1}$.
These are order of magnitude estimates and we caution the reader in using them during specific calculations. 
Using the narrow line H$\alpha$ luminosity of UGC 6614, we estimate the SFR in this galaxy to be equal to 0.04 M$\odot$yr$^{-1}$. The mass 
outflow rate is an order of magnitude greater than the SFR in this galaxy and hence clearly indicating that the outflows 
are powered by the AGN. An important question arises regarding the discrepany of outflow velocities as seen from Balmer lines 
($\sim$ 3800 km/s and FWHM $\sim$3000 km/s) and the forbidden [O {\sc iii}] line ($\sim$ 141 km/s and FWHM $\sim$1400 km/s). 
One explanation could be that these outflows are emitted from two different regions, one close to the BLR region or the intermediate 
line regions (ILR) and the other from the NLR. The deceleration in velocities from $\sim$1000km/s (in ILR) to a $\sim$100 km/s (NLR) is 
consistent with that seen in nearby Seyfert galaxy NGC 4151 (\citealt{DasV2005}, \citealt{Crenshaw2015} and references therein).  
The mass outflow rate of $\sim$ 3 M$\odot$yr$^{-1}$ seen in NGC 4151 is similar to the value obtained using the broad balmer lines for 
UGC 6614 and could produce effective feedback on scales where circumnuclear star formation and bulge growth occur. 

While UGC 6614 shows such large outflow velocities, it is not very common in these systems as only 1 out of 30 LSBs with AGN show 
this feature encouraging our search for such systems. This further calls for detailed study of the geometry and kinematics of 
UGC6614 using existing IFUs.

\cite{mapelli2008} suggest that the progenitors for GLSB galaxies such as UGC 6614, Malin1, Malin2 and NGC 7589 are ring galaxies which is 
created after an interaction. The gas and stars are redistributed in such an encounter and hence the formation of large diffuse stellar disks.
This encounter triggers an AGN which could further create such ionized gas outflows. 

The effect of this outflowing gas on the host galaxy is difficult to determine with existing data. 
\cite{cresci2014} using VLT SINFONI observations found that the fast expanding outflowing gas sweeps away expelling most of the gas 
thereby quenching star forming activity 
along its route, while the outflowing gas also triggered star formation by compression of gas clouds by the 
outflow-driven shock at the outflow edges. This was a classical case of both positive and negative feedback by fast moving outflows. 
In UGC 6614 and other LSB galaxies which show signature of outflow, the action of such feedback on the host galaxy can be 
understood using future IFU observations.

\section{Conclusions}

\begin{itemize}
 \item From the analysis of the SDSS optical spectra of 24 LSB galaxies, We derived the virial mass of their central BHs. The median 
value of M$_{BH}$ is 5.62 x 10$^6$ M$\odot$.\\
\item Majority of our sample lie below the M$_{BH}$ - $\sigma_e$ correlation of the normal spiral galaxies. \\
\item The effects of systematic bias in the M$_{BH}$ estimation and the effect of orientation of the host galaxies in the 
measurement of $\sigma$ are insufficient to explain the observed offset of our sample galaxies in the M$_{BH}$ - $\sigma_e$ relation.\\ 
\item  In general, the LSB galaxies tend to have low mass BHs which probably are not in co-evolution with the host 
galaxy bulges. The isolated and poorly evolved LSB galaxies are good candidates to understand the evolution of heavy seed BHs.\\  
\item The lack of bulge - BH co-evolution, indicative of secular evolution in these systems, can be one of the 
probable reasons for the observed offset. But majority of the LSB galaxies in the sample are likely to host classical bulges (Sersic index 
$\ge$ 2.5) which are products of major mergers. A detailed study of the nature of the bulges and the role of dark matter in the growth of 
the BHs is important and planned to be addressed in future.\\ 

\end{itemize}

{\bf Acknowledgements}\\\\
We would like to thank the anonymous referee for the comments and suggestions which improved the quality of the paper immensely.\\\\ 
RS kindly acknowledges the award of NSFC (Grant No. 11450110401) and President's International Fellowship Initiative (PIFI) awarded by the 
Chinese Academy of Sciences. \\\\
This research has made use of the NASA/IPAC Extragalactic Database (NED) which is operated by the Jet Propulsion Laboratory, 
California Institute of Technology, under contract with the National Aeronautics and Space Administration. \\\\
We would like to thank the SDSS team for making the data available in public. Funding for SDSS-III has been provided by the 
Alfred P. Sloan Foundation, the Participating Institutions, the National Science Foundation, 
and the U.S. Department of Energy Office of Science. The SDSS-III web site is http://www.sdss3.org/.
SDSS-III is managed by the Astrophysical Research Consortium for the Participating Institutions of the SDSS-III Collaboration including 
the University of Arizona, the Brazilian Participation Group, Brookhaven National Laboratory, Carnegie Mellon University, University of 
Florida, the French Participation Group, the German Participation Group, Harvard University, the Instituto de Astrofisica de Canarias, 
the Michigan State/Notre Dame/JINA Participation Group, Johns Hopkins University, Lawrence Berkeley National Laboratory, Max Planck 
Institute for Astrophysics, Max Planck Institute for Extraterrestrial Physics, New Mexico State University, New York University, Ohio 
State University, Pennsylvania State University, University of Portsmouth, Princeton University, the Spanish Participation Group, 
University of Tokyo, University of Utah, Vanderbilt University, University of Virginia, University of Washington, and Yale University.
\\\\
{
\bibliography{LSBAGNsub5_23oct}
}
\appendix
\section{}

\begin{figure*}
\centering
\centering
\begin{tabular}{cc}
\includegraphics[width=0.45\textwidth]{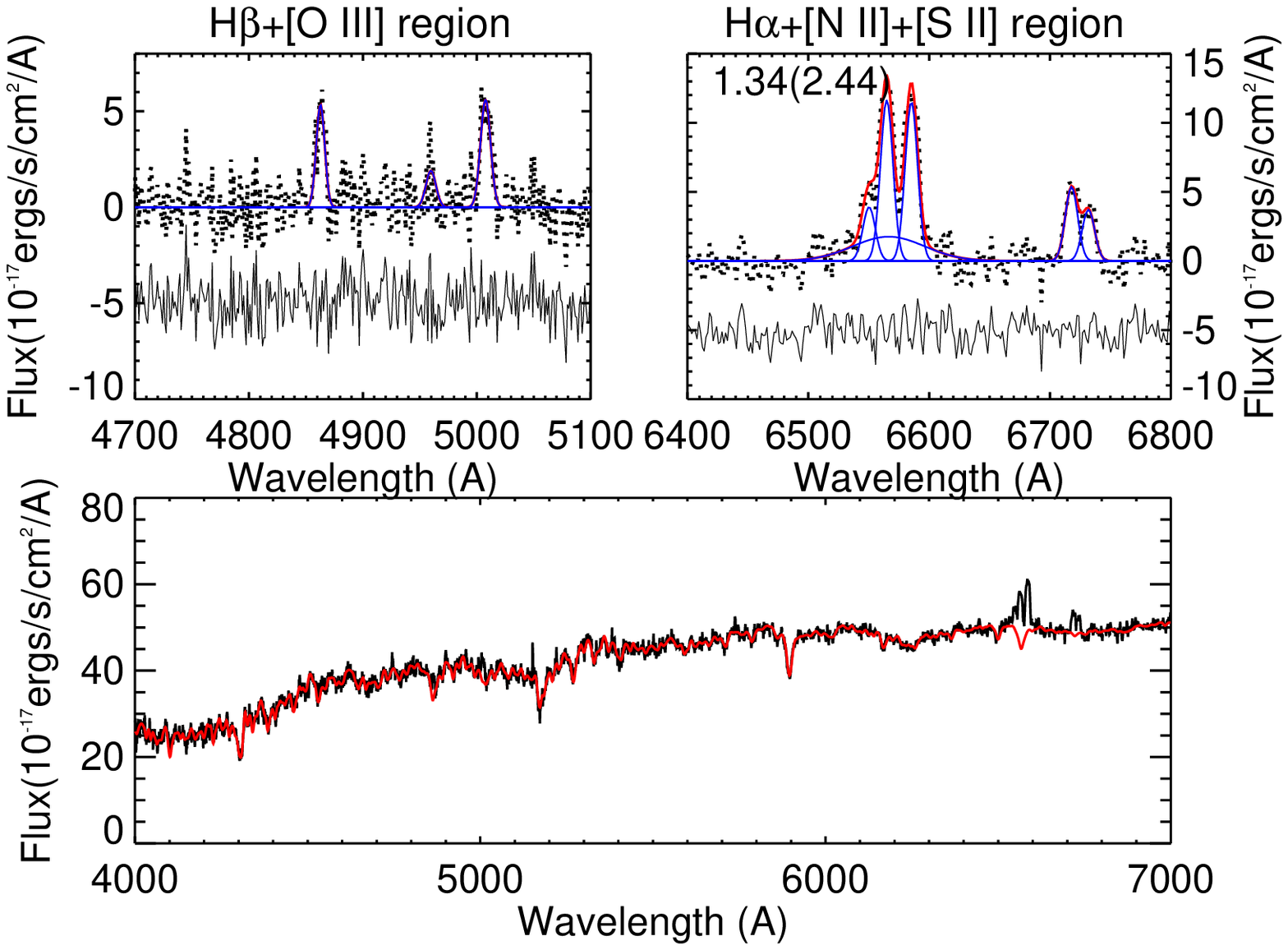}
&
\includegraphics[width=0.45\textwidth]{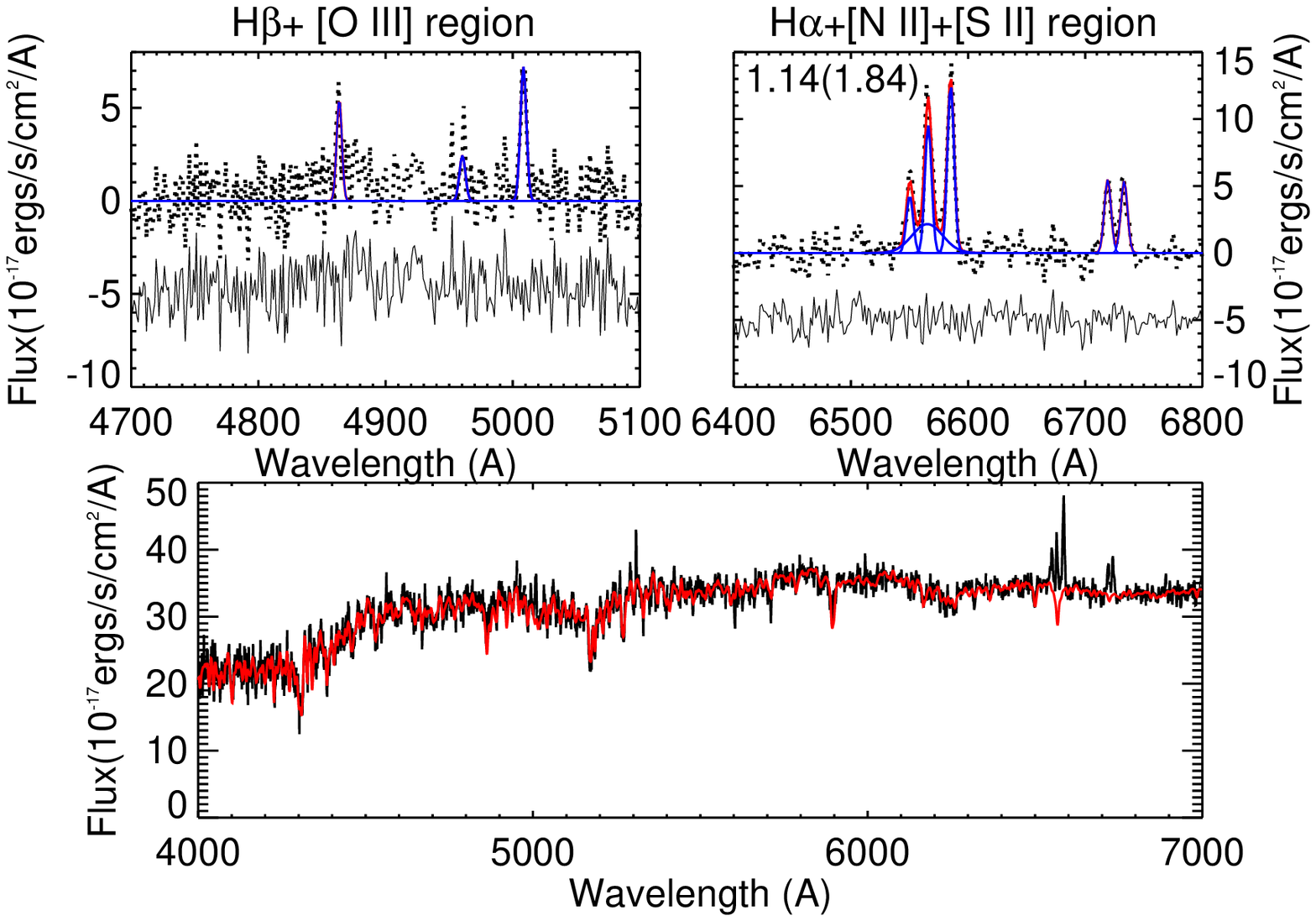} \\
2MASXJ14384627+0106576 (No 3) & Leda 135884 (No 4) \\\\\\
\includegraphics[width=0.45\textwidth]{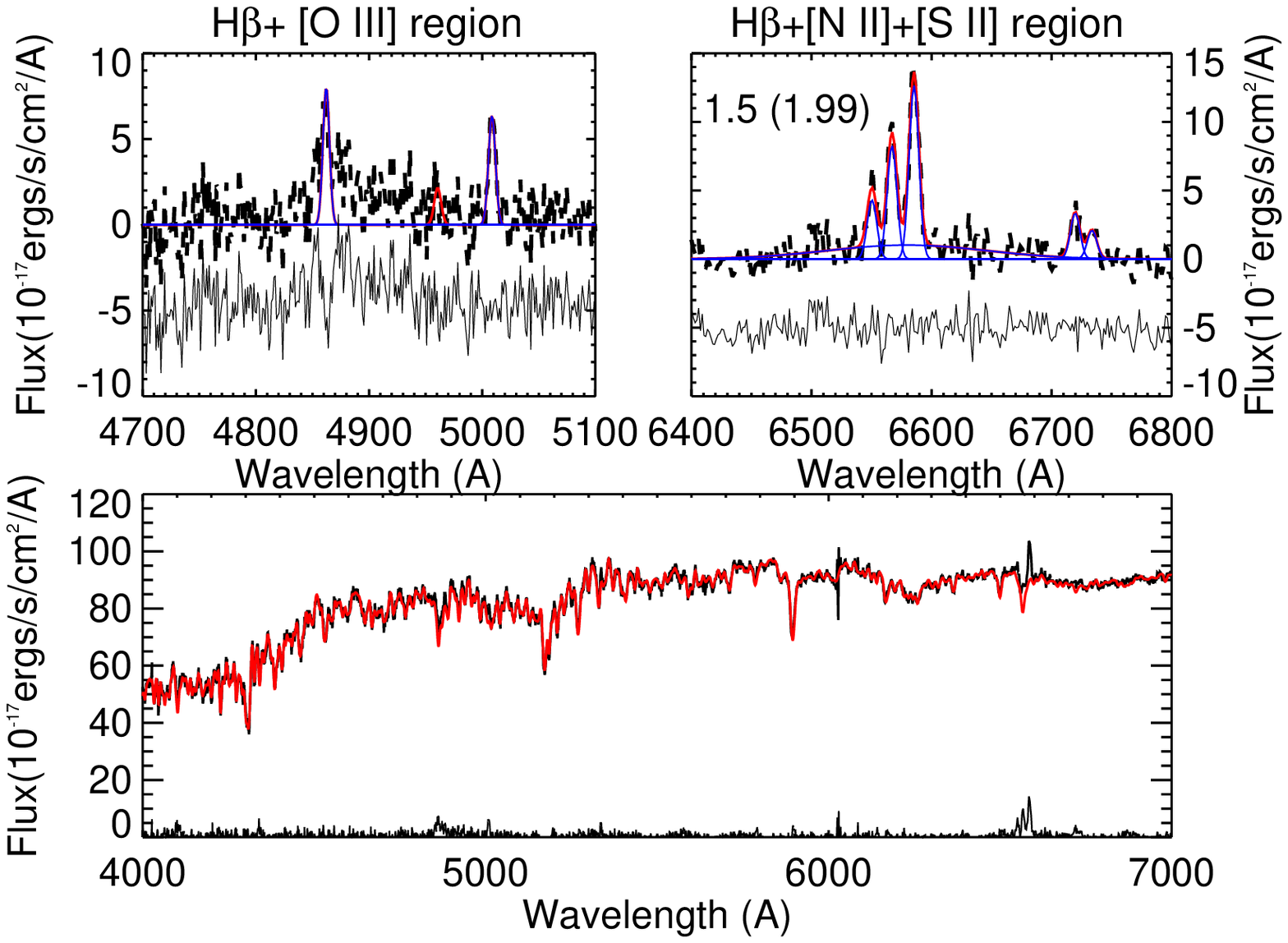}
&
\includegraphics[width=0.45\textwidth]{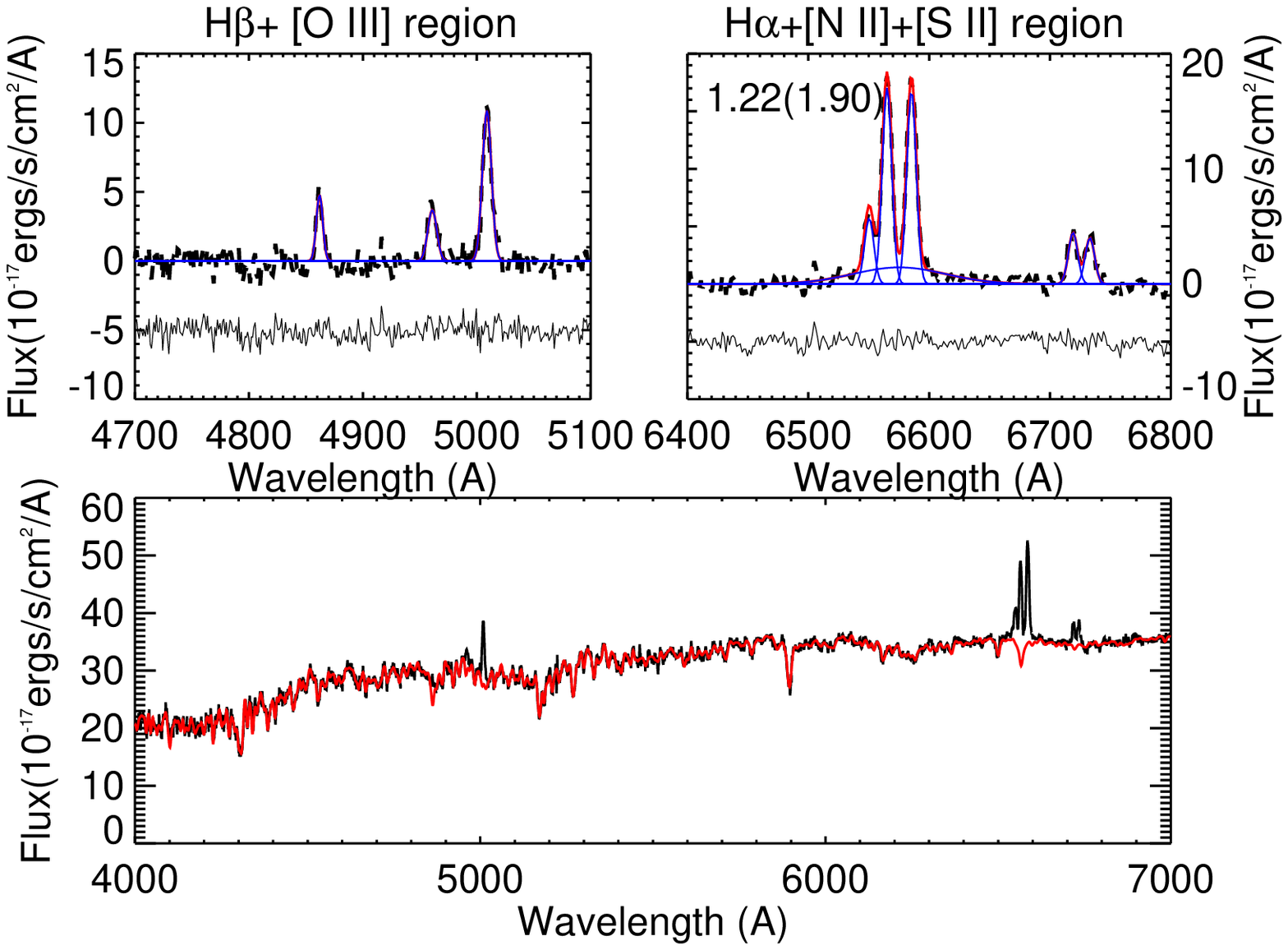} \\
UGC 00568 (No 6) & UGC 00514 (No 8) \\\\\\
\includegraphics[width=0.45\textwidth]{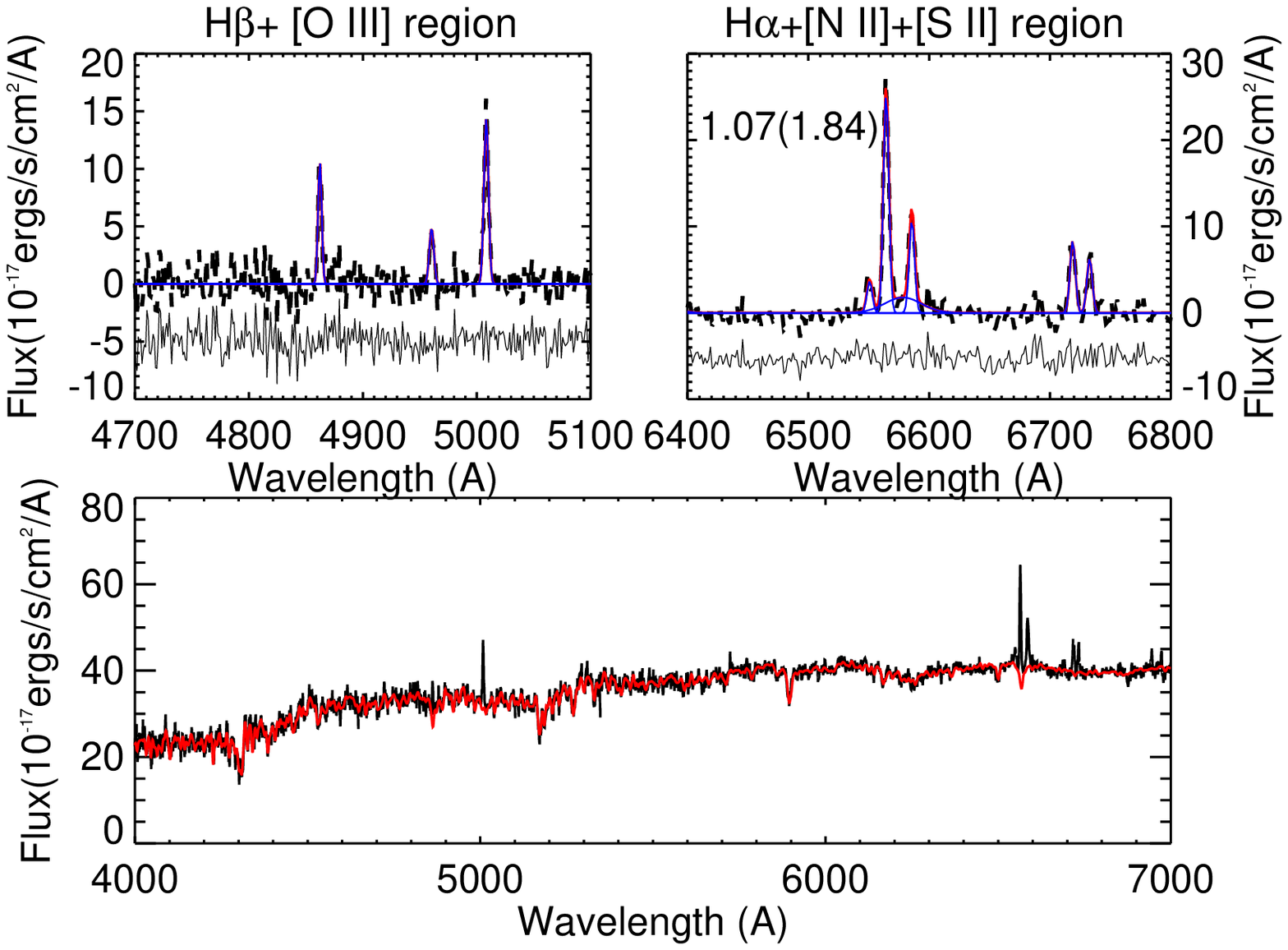}  &  
\includegraphics[width=0.45\textwidth]{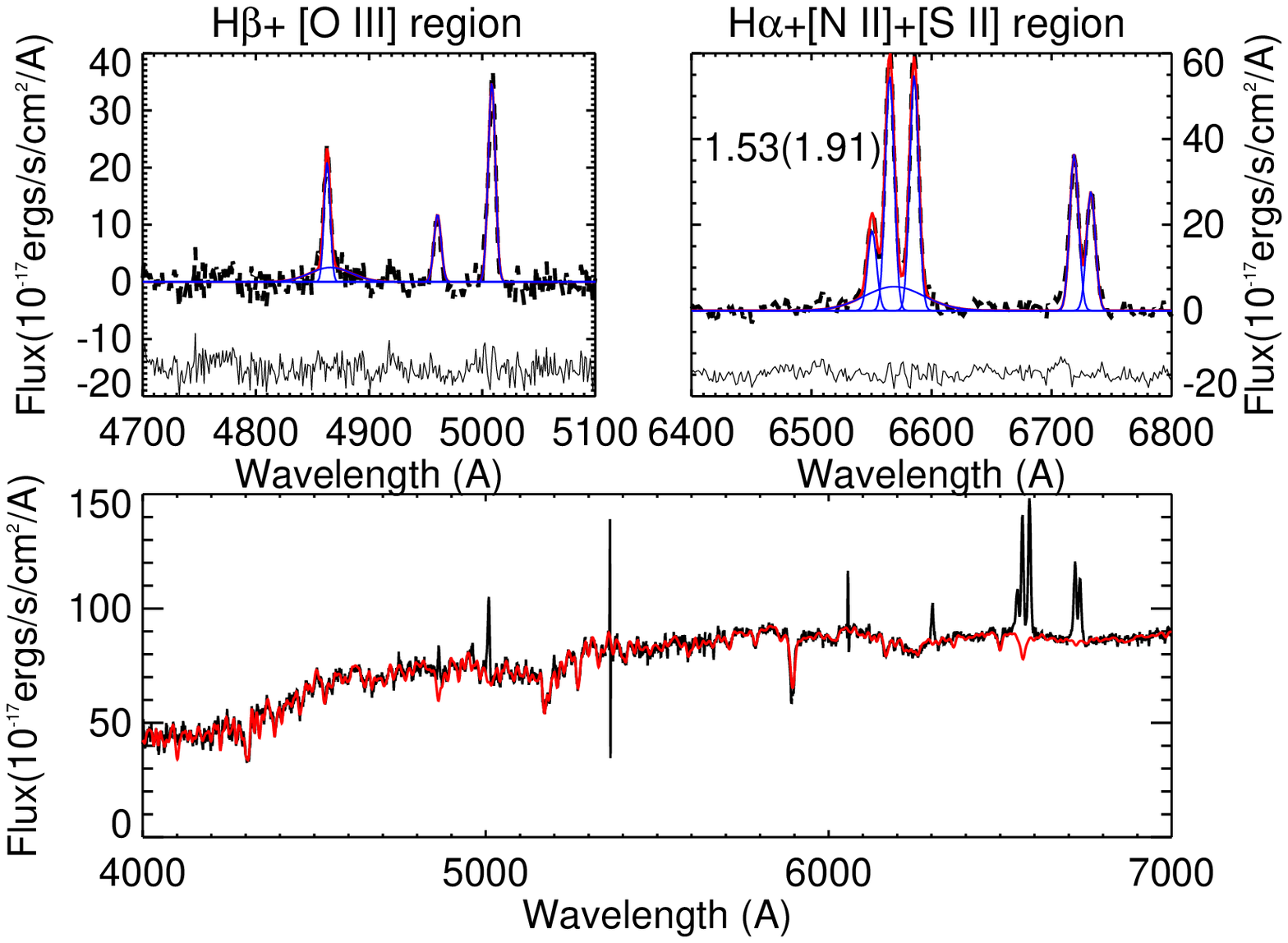}\\
LSBC F611-03 (No 9) & VIII ZW 091 (No 10) \\\\\\
\end{tabular}
\caption{Same as in Figure: 2 for other sample galaxies.}
\end{figure*}
\addtocounter{figure}{-1}
\begin{figure*}
\centering
\centering
\begin{tabular}{cc}
\includegraphics[width=0.45\textwidth]{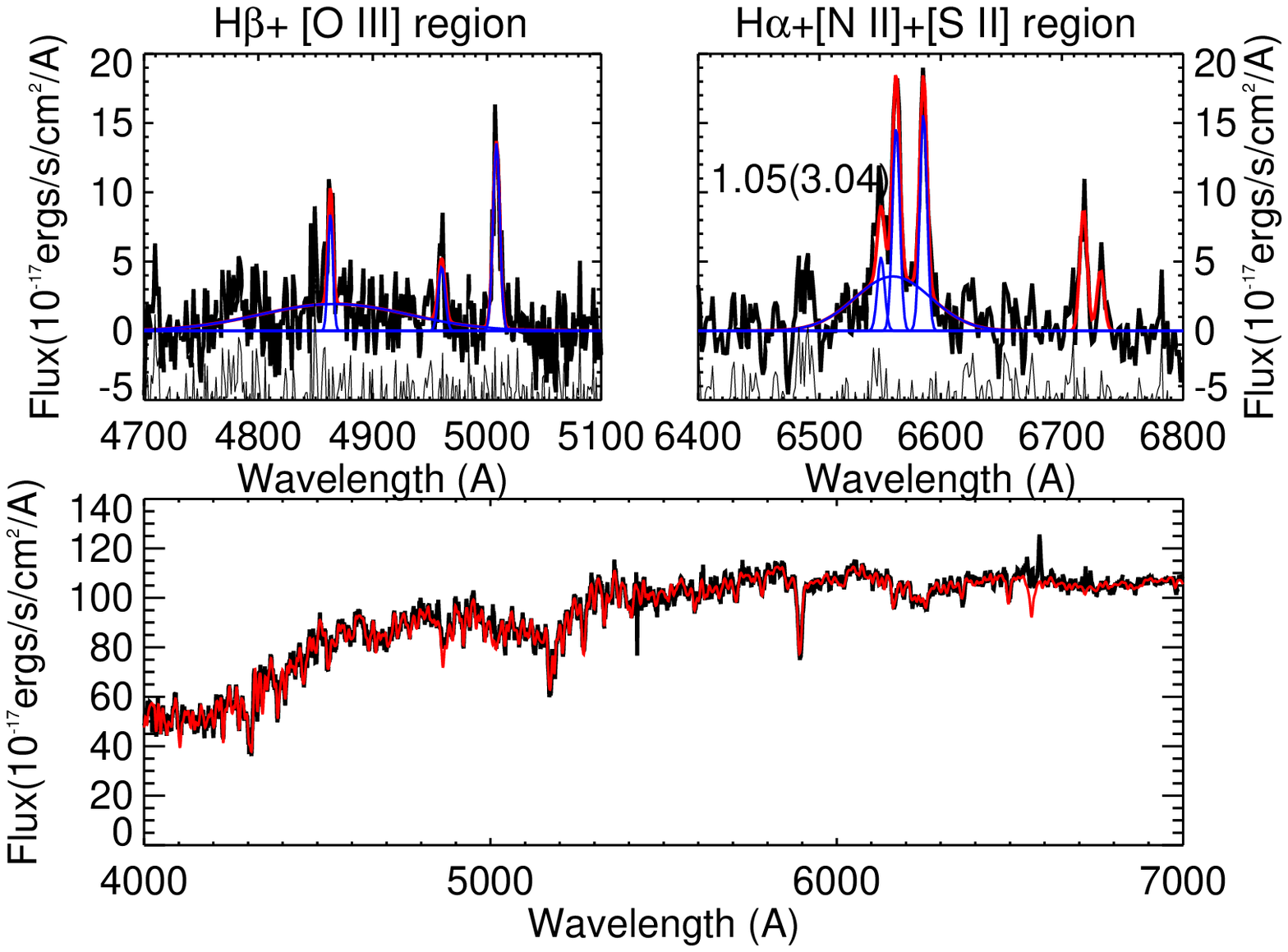}  &  
\includegraphics[width=0.45\textwidth]{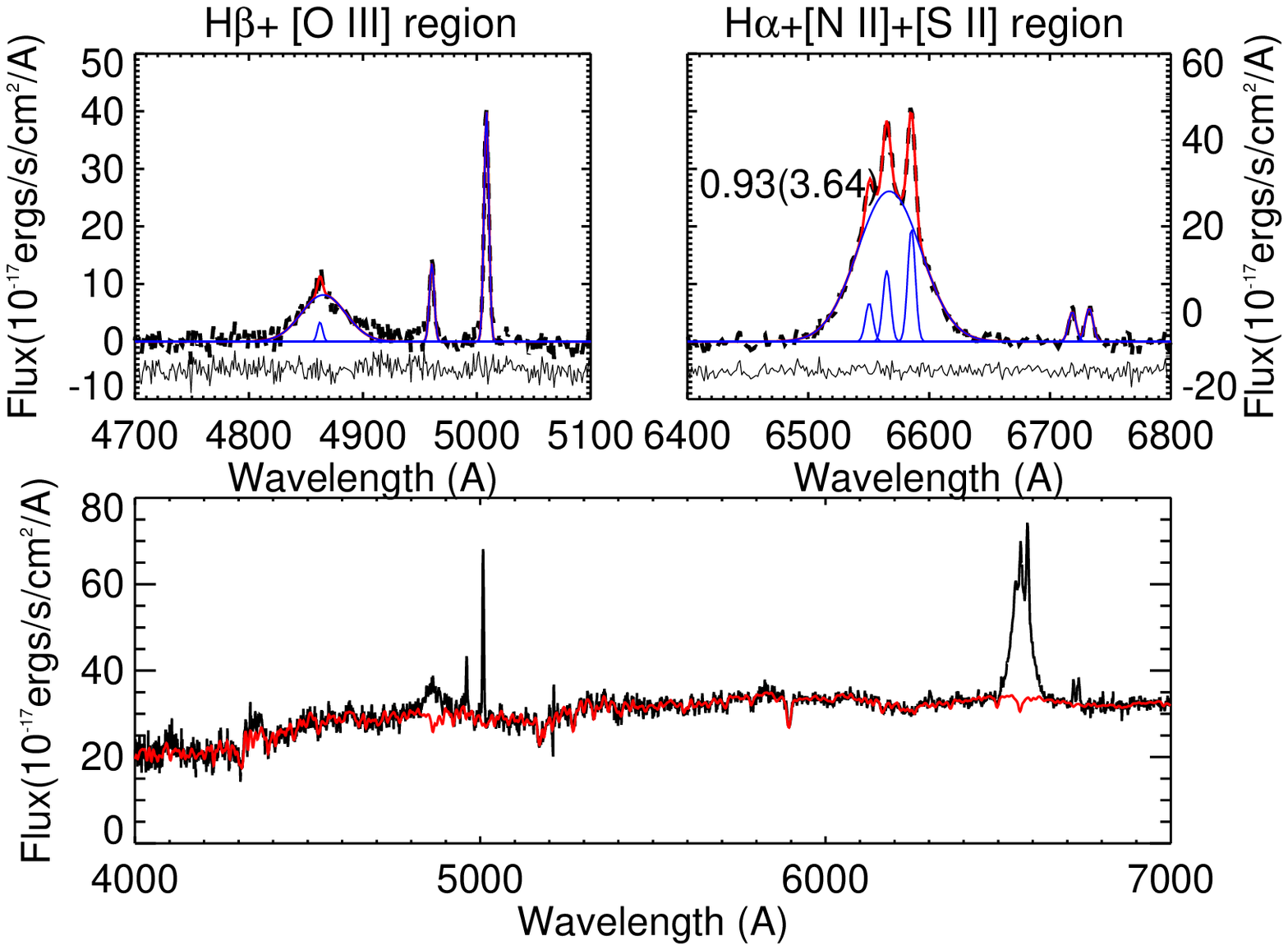}\\
VIII ZW 437 (No 11) & LSBC F727-V01 (No 12) \\\\\\
\includegraphics[width=0.45\textwidth]{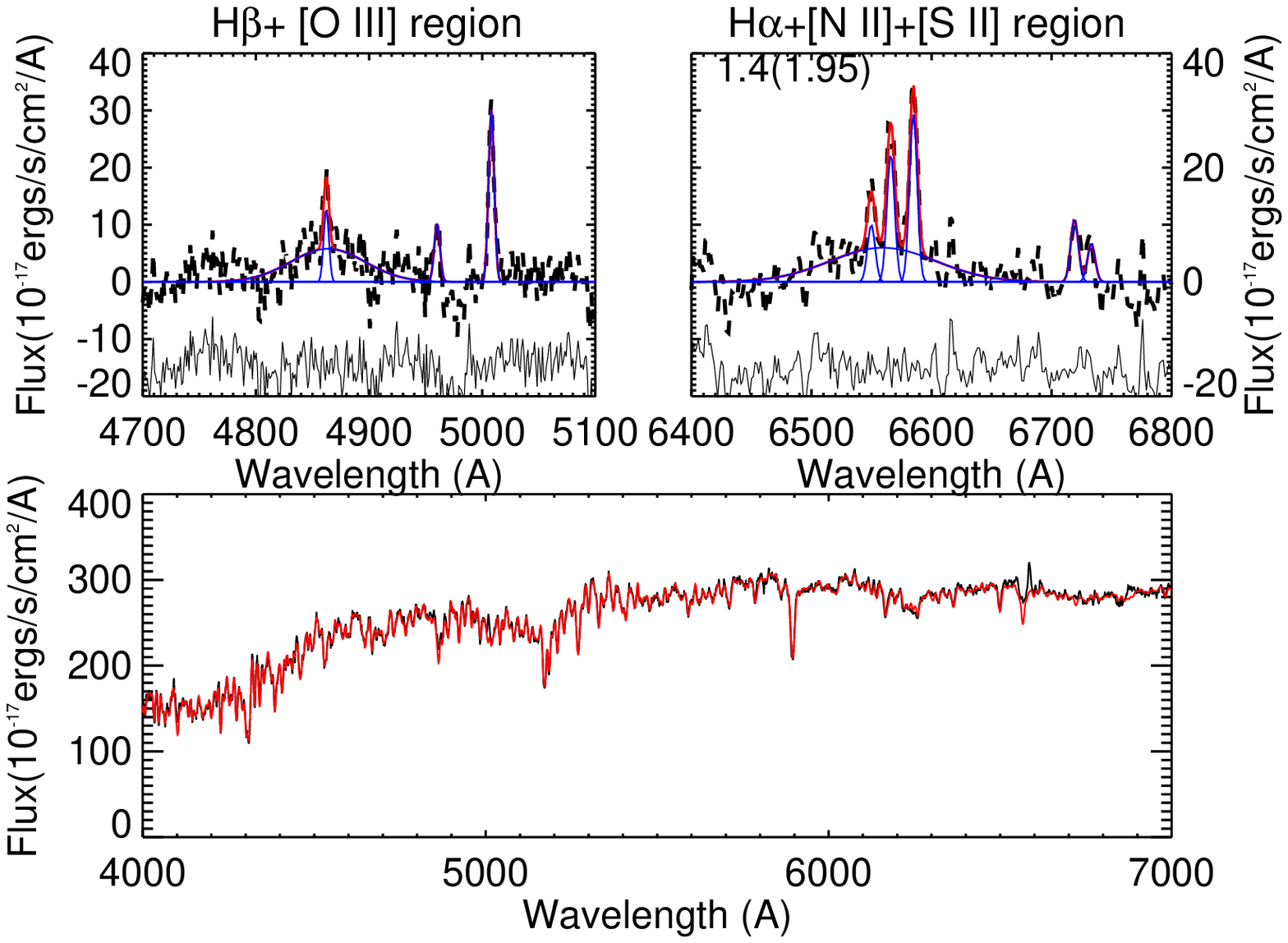}
&
\includegraphics[width=0.45\textwidth]{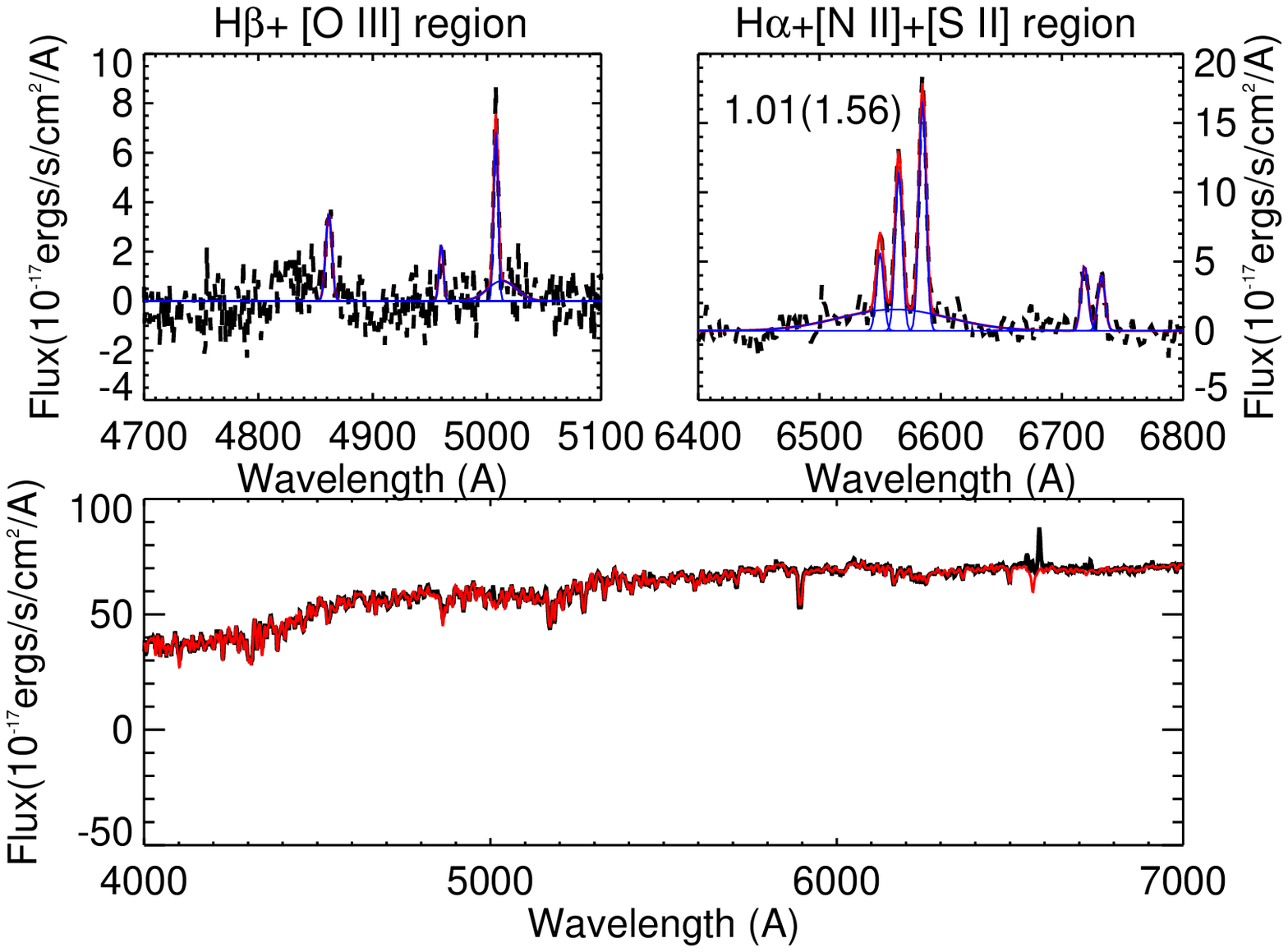} \\
UGC 10050 (No 14) & NGC 0926 (No 16) \\\\\\
\includegraphics[width=0.45\textwidth]{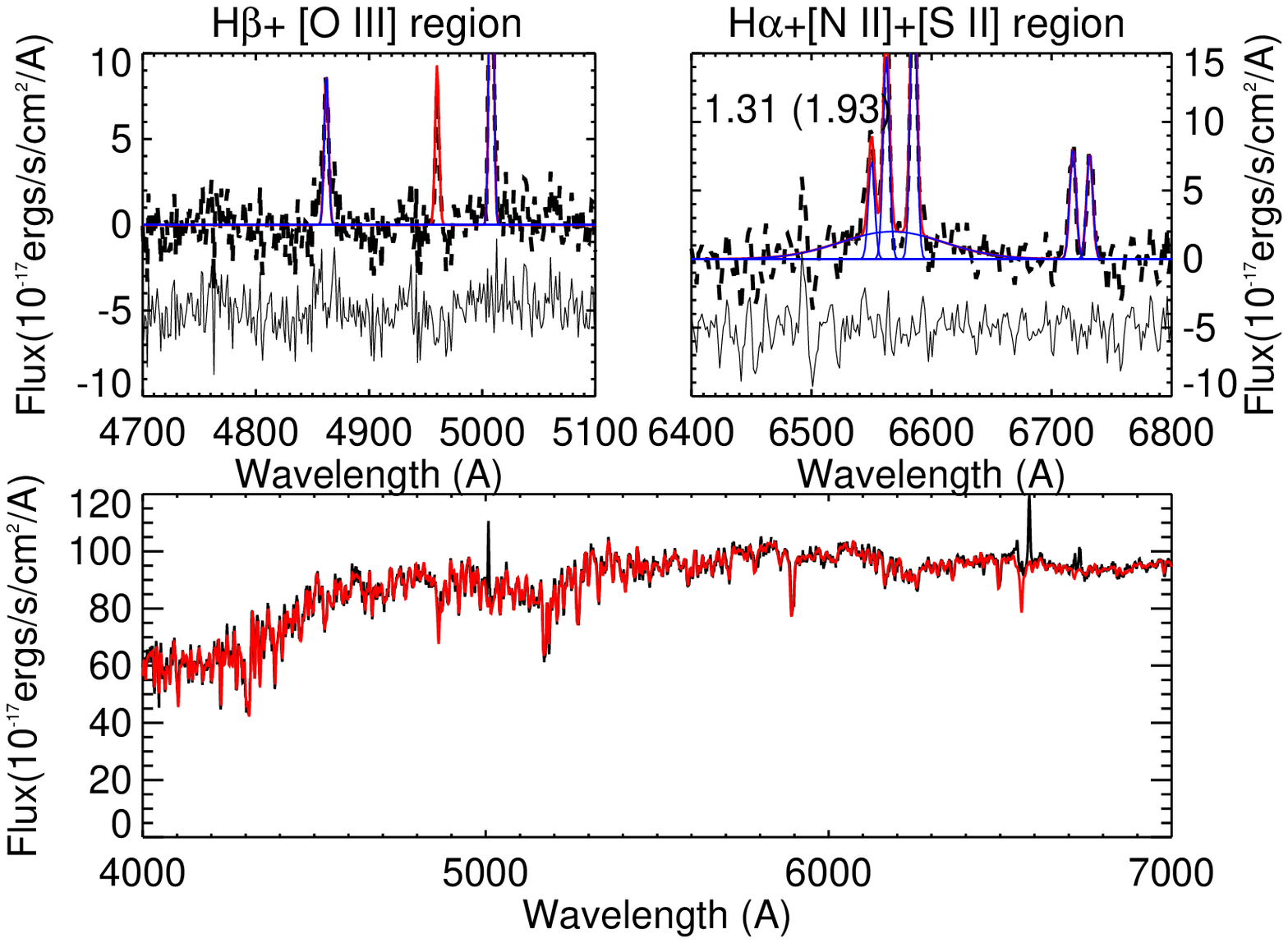}
&
\includegraphics[width=0.45\textwidth]{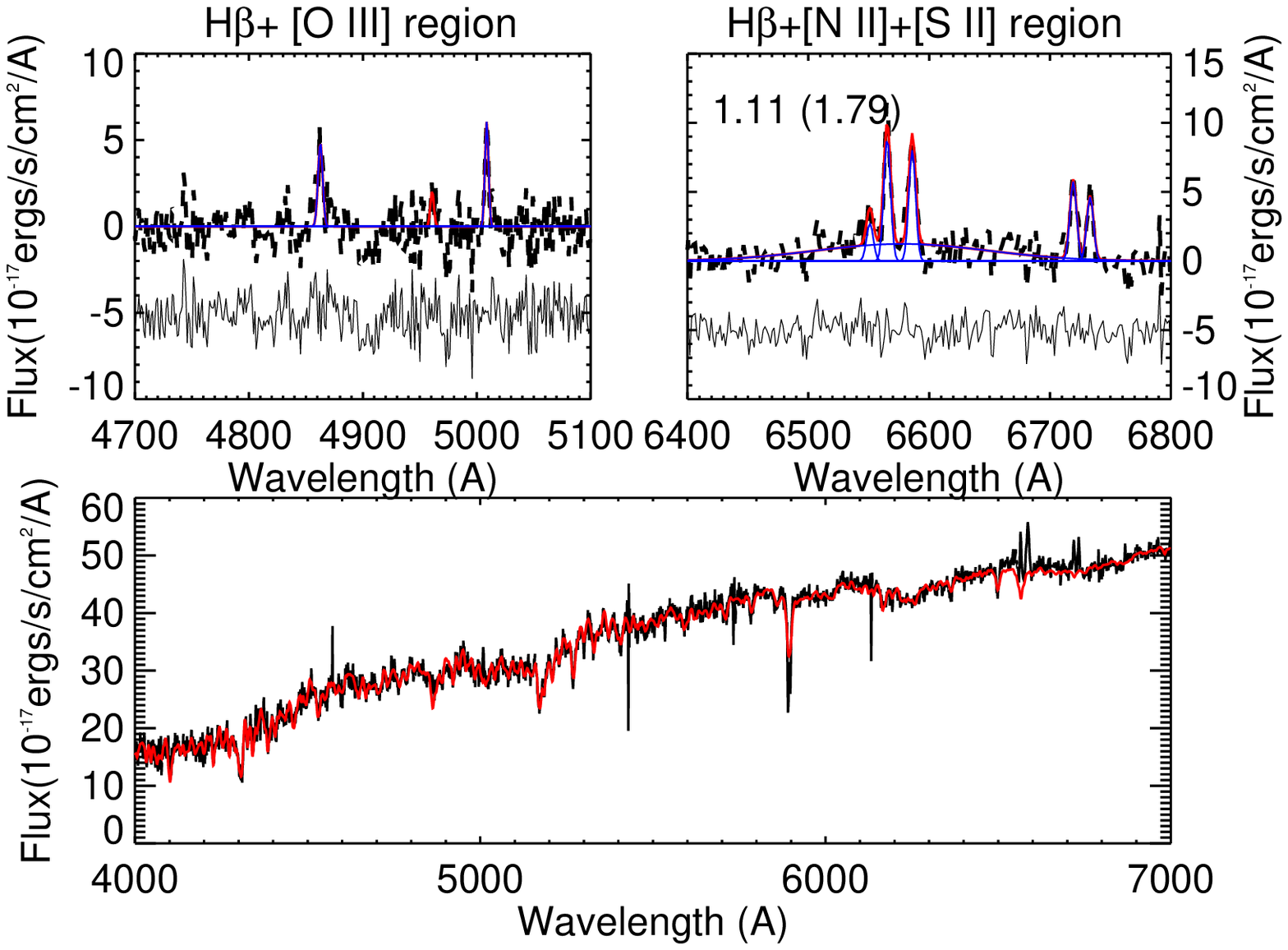} \\
UGC 9087 (No 17) & UGC 8828 (No 18) \\\\\\
\end{tabular}
\caption{Continued}
\end{figure*}

\addtocounter{figure}{-1}
\begin{figure*}
\centering
\centering
\begin{tabular}{cc}
\includegraphics[width=0.45\textwidth]{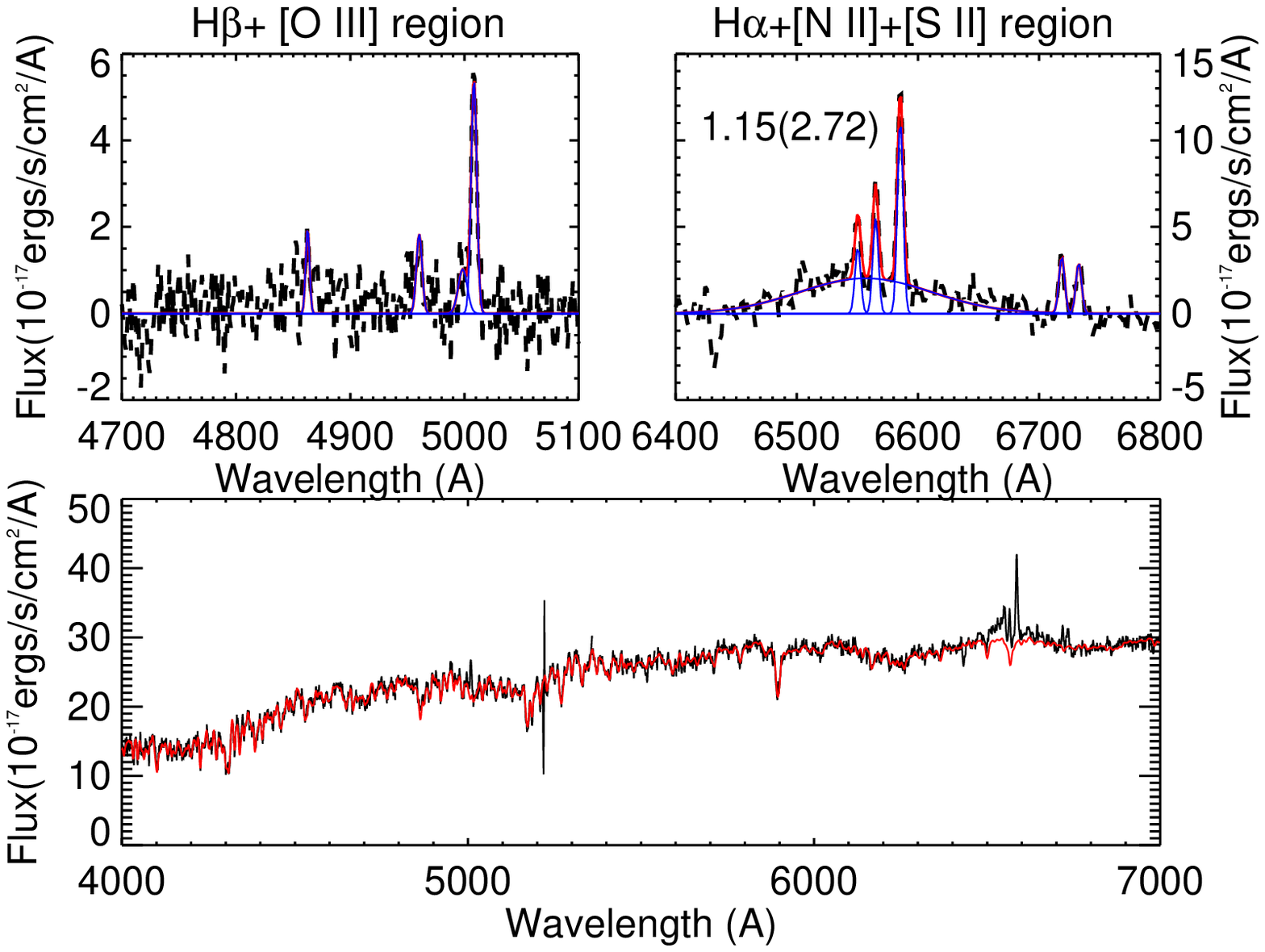}  &  
\includegraphics[width=0.45\textwidth]{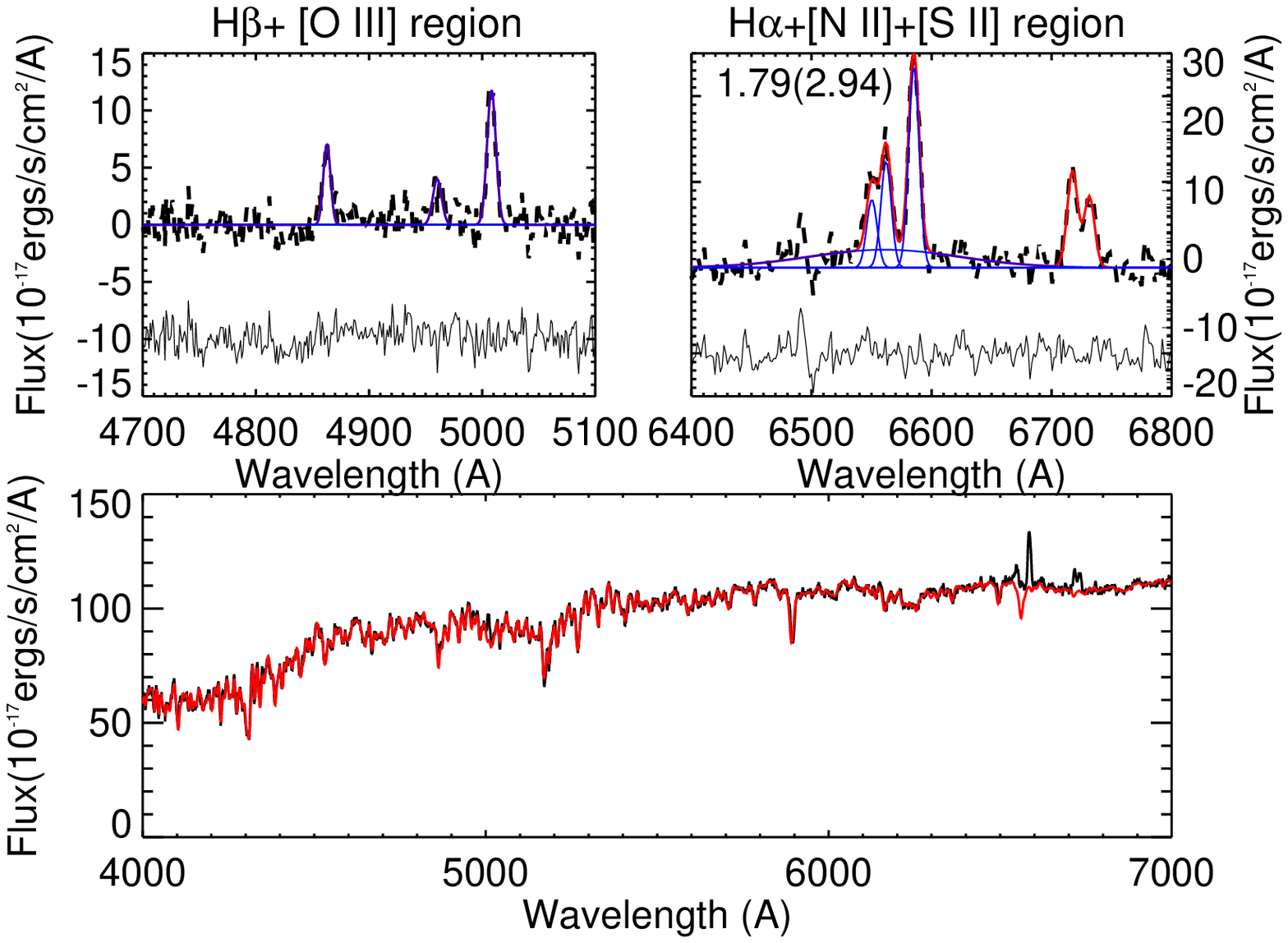}\\
2MASX J10255577+0200154 (No 19) & IC 2423 (No 20) \\\\\\
\includegraphics[width=0.45\textwidth]{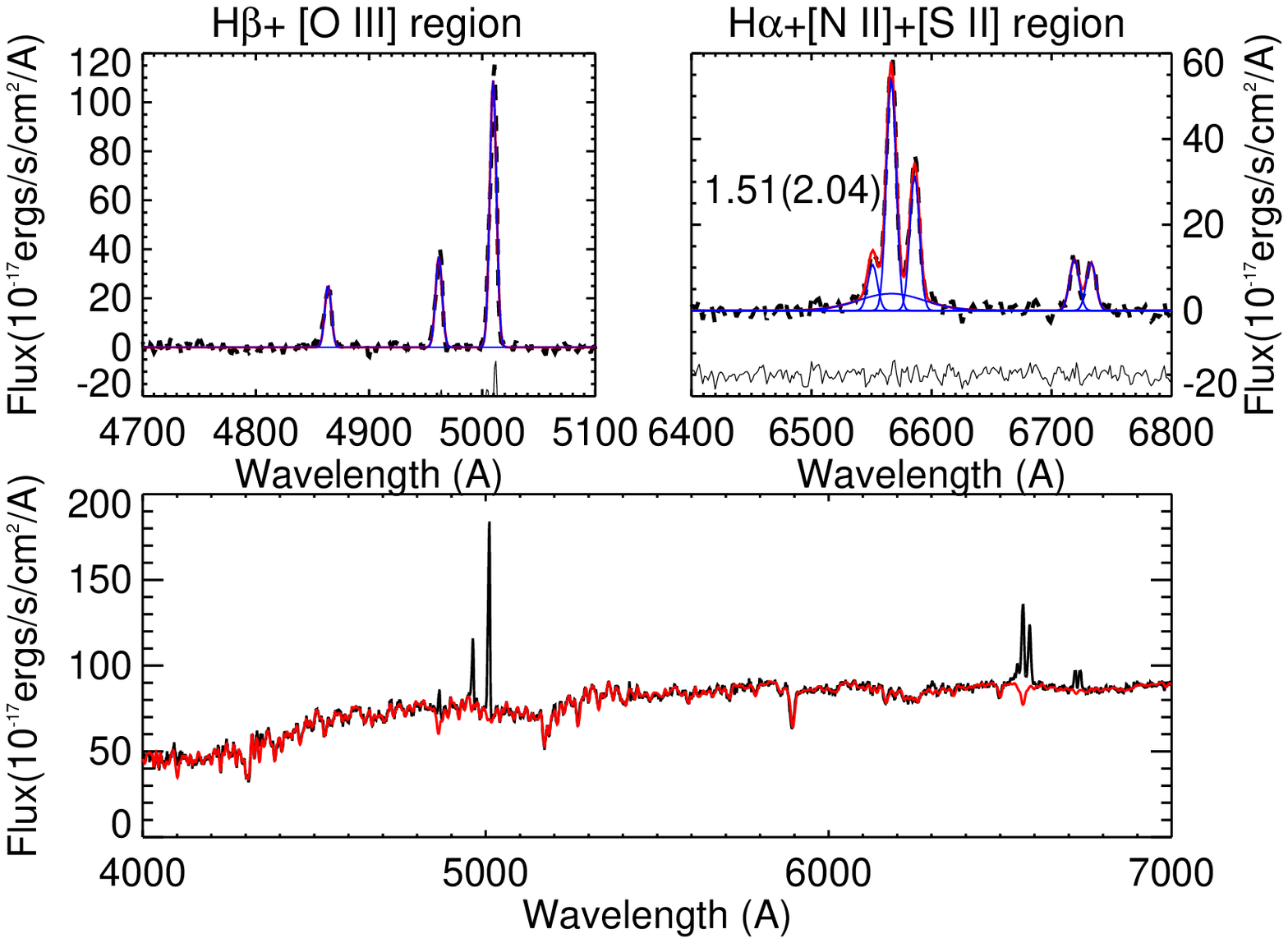}  &  
\includegraphics[width=0.45\textwidth]{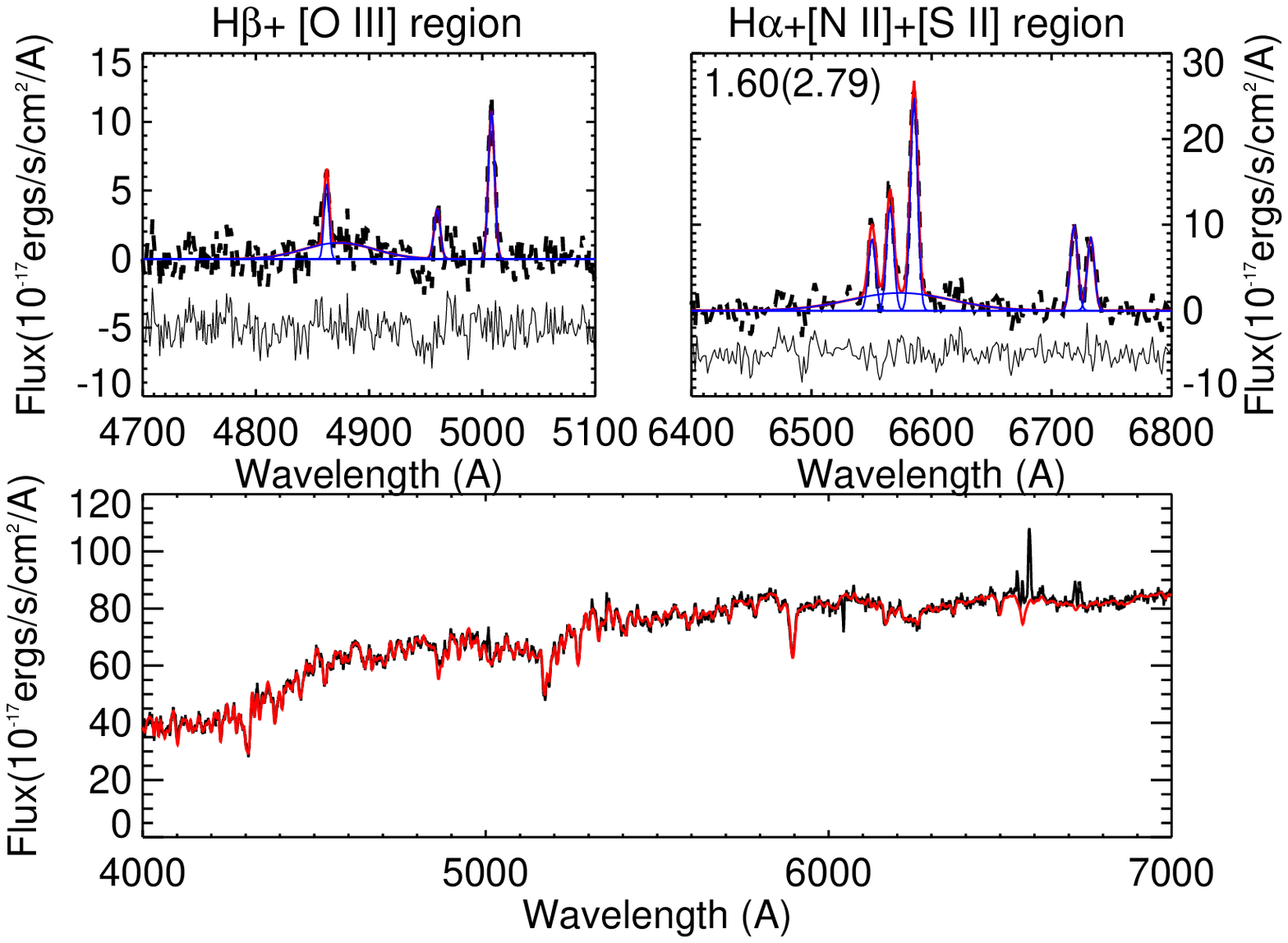}\\
LSBC F564-01 (No 21) & UGC 5770 (No 22) \\\\\\
\includegraphics[width=0.45\textwidth]{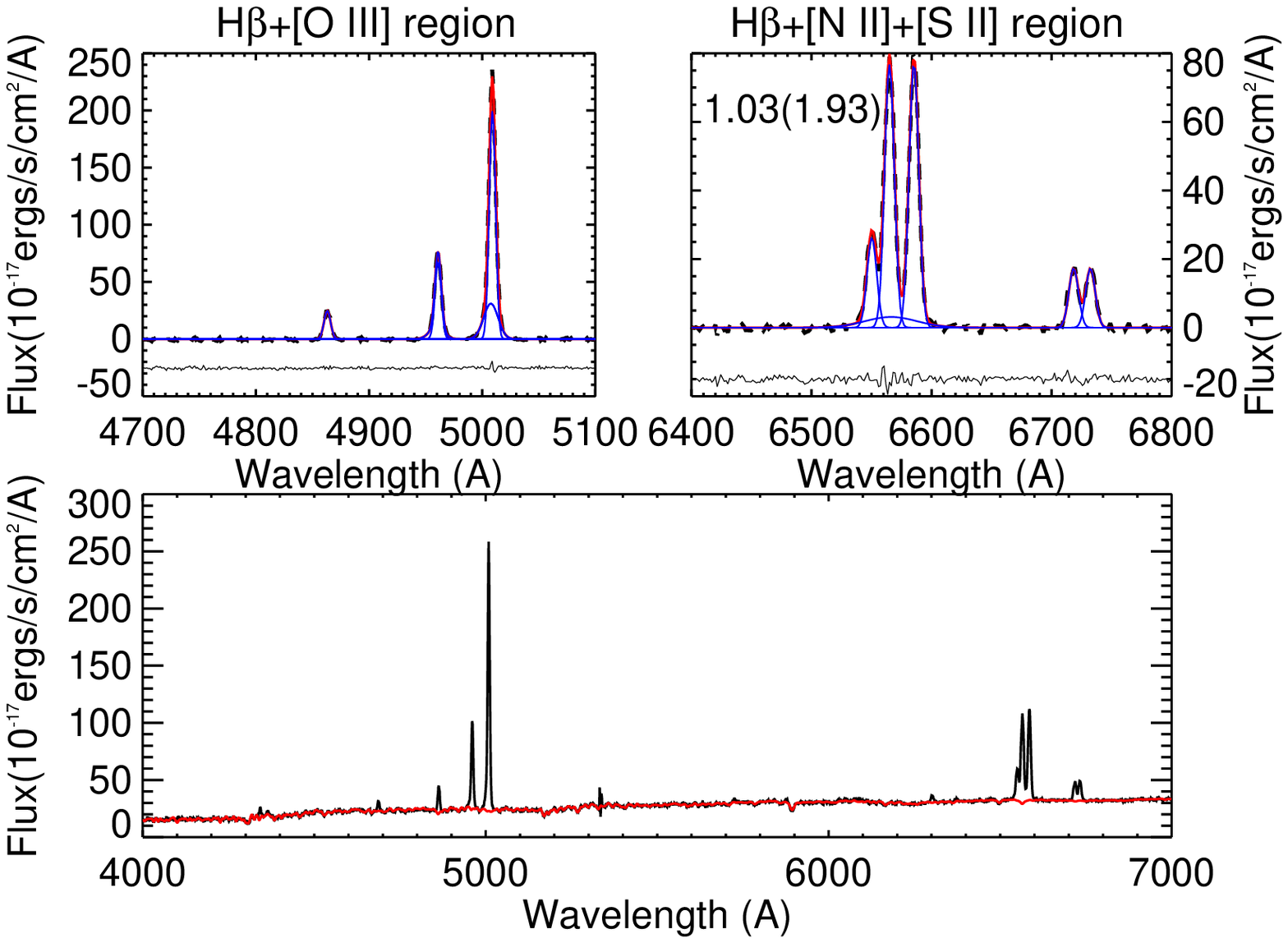}
&
\includegraphics[width=0.45\textwidth]{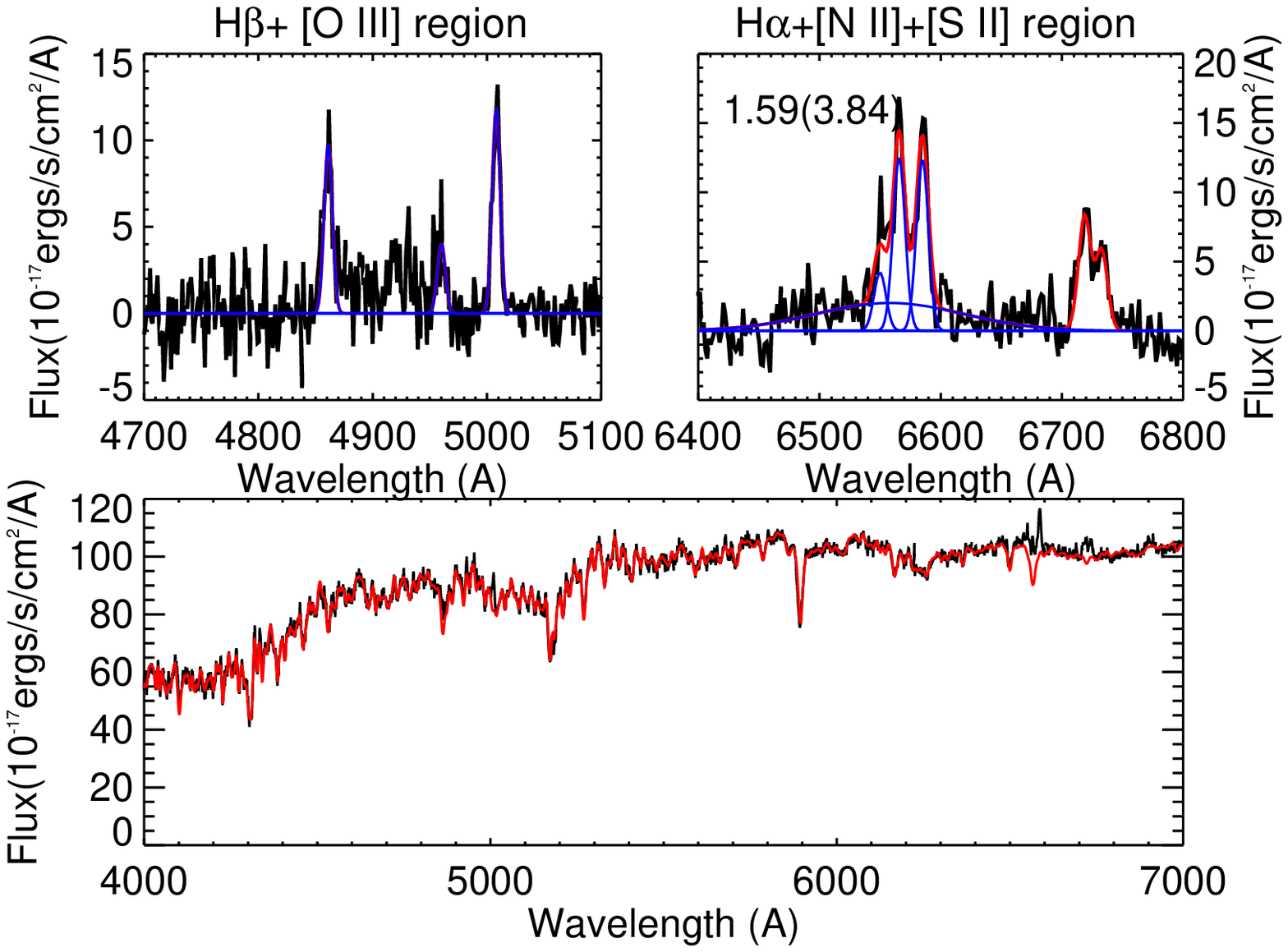} \\
2MASX j14565122+2152295 (No 23) & LSBC F508-02 (No 25) \\\\\\
\end{tabular}
\caption{Continued}
\end{figure*}

\begin{landscape}
 \begin{table}
 \caption{Fluxes of narrow emission lines in the units of 10$^{-17}$ ergs/s/cm$^2$}
 \centering
 \small
 \begin{tabular}{lllllllllll}
 Galaxy&
 H$_{\beta}$&  [O{\sc iii}]& [O {\sc iii}]& [O \sc{i}]& [O {\sc i}]& [N{\sc ii}] &[N{\sc ii}] &H$_{\alpha}$ & 
  [S{\sc ii}] & [S{\sc ii}]\\
 &narrow&&&&&&&narrow&&\\
 &4861&4960&5007&6300&6364&6548&6584&6563&6717&6731\\\\
 \hline
1&     18.13$\pm$      3.72&    82.84$\pm$      19.07&      223.68
$\pm$      46.01 &     9.51$\pm$      6.67&     0.77$\pm$
      3.54&      36.38$\pm$      1.92&      107.31$\pm$ 1.92&
      100.94$\pm$      5.69&      20.85$\pm$      4.52&      24.08
$\pm$      3.91\\

2&      346.28$\pm$      5.98&      293.38$\pm$      17.71&      967.12
$\pm$      43.61 &     108.14$\pm$      9.03&      36.97$\pm$
      4.51&      365.99$\pm$      3.66&      1079.69$\pm$      3.66&
      1800.43$\pm$      16.76&      391.04$\pm$      6.78&      300.47
$\pm$      5.63\\
3&     47.74$\pm$      4.89&      20.31$\pm$      3.35&      59.92
$\pm$      3.35&         --&         --      &49.30$\pm$      3.73&      145.44$\pm$      3.73&
      147.27$\pm$      12.00&      67.62$\pm$      9.25&      46.79
$\pm$      7.70\\
4&      23.96$\pm$      3.39&      12.83$\pm$      2.58&      37.85
$\pm$      2.58&      14.02$\pm$      6.49&      10.05$\pm$
      3.97&      28.24$\pm$      1.97&      83.29$\pm$      1.97&
      63.19$\pm$      6.34&      36.42$\pm$      4.52&      35.64
$\pm$      3.93\\
5&      177.24$\pm$      5.32&      155.14$\pm$      22.32&      449.03
$\pm$      55.47 &      43.17$\pm$      9.68&      14.25$\pm$
      4.19&      166.39$\pm$      3.30&      490.85$\pm$      3.30&
      744.37$\pm$      12.37&      206.37$\pm$      7.55&      144.97
$\pm$      6.39\\
6&     31.61$\pm$      4.61&      14.80$\pm$      2.86&      43.66
$\pm$      2.86&     -- &      ---&      38.40$\pm$      1.95&      113.29$\pm$      1.95&
      73.74$\pm$      5.53&      31.57$\pm$      4.98&      19.78
$\pm$      4.49\\
7&      147.12$\pm$      4.79&      391.31$\pm$      18.04&      1158.35
$\pm$      42.67 &     153.15$\pm$      10.77&      55.95$\pm$
      5.68&      246.66$\pm$      3.73&      727.66$\pm$      3.73&
      433.60$\pm$      10.47&      297.74$\pm$      8.07 &      270.37
$\pm$      7.03\\
8&      30.89$\pm$      1.97&      35.26$\pm$      1.59&      104.03
$\pm$      1.59&      16.59$\pm$      3.63&      8.11$\pm$
      2.09&      52.93$\pm$      1.09&      156.16$\pm$      1.09&
      158.95$\pm$      3.17&      41.50$\pm$      2.34&      37.22
$\pm$      2.21\\
9&      30.15$\pm$      2.93&      20.17$\pm$      2.65&      59.53
$\pm$      2.65&      18.24$\pm$      10.88&      28.80$\pm$
      9.65&      17.35$\pm$      1.68&      51.19$\pm$      1.68&
      121.46$\pm$      5.72&      39.88$\pm$      4.66&      30.07
$\pm$      3.70\\
10&      126.50$\pm$      7.49&      88.02$\pm$      4.78&      259.66
$\pm$      4.78&      128.01$\pm$      14.01&       - &
      163.24$\pm$      4.36&      481.57$\pm$      4.36&      482.11
$\pm$      14.12&      320.42$\pm$      12.06&      244.14$\pm$
      9.21\\
11& 54.42$\pm$      6.34&      34.05$\pm$      5.01&      100.46
$\pm$      5.01&     - &     - &      39.53$\pm$      3.56&      116.61$\pm$      3.56&
      111.71$\pm$      10.31&      63.55$\pm$      10.08&      30.51
$\pm$      8.42\\
12&      16.62$\pm$      4.41&      62.43$\pm$      2.81&      184.18
$\pm$      2.81&      26.61$\pm$      6.29&      12.26$\pm$
      3.50&      49.33$\pm$      2.29&      145.53$\pm$      2.29&
      91.51$\pm$      7.14&      38.21$\pm$      4.30&      42.79
$\pm$      3.87\\
13&     63.76$\pm$      5.69&      22.83$\pm$      3.10&      67.34
$\pm$      3.10&      57.59$\pm$      8.08&      27.99$\pm$
      4.73&      42.12$\pm$      2.38&      124.25$\pm$      2.38&
      184.49$\pm$      8.52&      144.91$\pm$      8.32&      99.08
$\pm$      5.68\\
14&      63.21$\pm$      13.10&      56.99$\pm$      8.95&      168.14
$\pm$      8.95&      56.03$\pm$      29.96&      44.99$\pm$
      19.76&      73.89$\pm$      7.88&      217.99$\pm$      7.88&
      165.69$\pm$      21.06&      81.65$\pm$      20.69&      49.96
$\pm$      18.29\\
15&      46.18$\pm$      3.89&      78.06$\pm$      5.02&      268.83
$\pm$      17.40&      60.73$\pm$      7.04&      40.49$\pm$
      4.76&      40.12$\pm$      3.37&      118.37$\pm$      3.37&
      106.79$\pm$      15.53&      81.94$\pm$      6.71&      59.49
$\pm$      4.41\\
16&       21.70$\pm$      5.33&      8.46$\pm$      1.65&      24.97
$\pm$      1.65&      6.06$\pm$      3.71&      15.12$\pm$
      3.98&      37.24$\pm$      1.34&      109.85$\pm$      1.34&
      76.23$\pm$      3.57&      31.31$\pm$      3.41&      26.87
$\pm$      3.38\\
17&      32.48$\pm$      3.51&      30.55$\pm$      2.76&      90.11
$\pm$      2.76&      5.18$\pm$      7.05&     --&
      41.95$\pm$      2.23&      123.76$\pm$      2.23&      86.39
$\pm$      5.81&      47.45$\pm$      5.83&      45.53$\pm$
      5.22\\
18&     20.28$\pm$      2.64&      5.31$\pm$      1.64&      15.66
$\pm$      1.64&      6.970$\pm$      7.37&      1.64$\pm$       7.37&
      17.36$\pm$      1.52&      51.21$\pm$      1.52&      56.54
$\pm$      4.56&      36.66$\pm$      5.06&      29.57$\pm$
      4.21\\
19&      6.36$\pm$      1.29&      10.35$\pm$      1.73&      30.54
$\pm$      1.73&      13.56$\pm$      4.29&      12.14$\pm$
      2.98&      19.72$\pm$      1.04&      58.18$\pm$      1.04&
      28.37$\pm$      2.54&      16.98$\pm$      2.46&      15.15
$\pm$      2.23\\
20&     51.99$\pm$      4.28&      37.45$\pm$      3.22&      110.47
$\pm$      3.22&      45.98$\pm$      12.39&     ----&      83.31$\pm$      2.79&      245.77$\pm$      2.79&
      129.83$\pm$      7.23&      117.83$\pm$      8.02&      86.29
$\pm$      6.55\\
21&      166.06$\pm$      6.44&      250.92$\pm$       5.52&      740.21$\pm$
      5.52&      67.19$\pm$      14.99&      24.74$\pm$      7.99&
      101.64$\pm$      4.72&      299.83$\pm$      4.72&      515.57
$\pm$      17.16&      113.54$\pm$      9.54&      106.62$\pm$
      8.76\\
22&       25.51$\pm$      3.91&      23.08$\pm$      2.97&      68.08
$\pm$      2.97&      11.27$\pm$      9.03&      6.31$\pm$
      4.85&      58.23$\pm$      2.79&      171.77$\pm$      2.79&
      84.67$\pm$      7.00&      69.17$\pm$      7.17&      59.54
$\pm$      6.17\\
23 &       164.83$\pm$      4.02&      578.50$\pm$      20.15&      1747.51
$\pm$      46.59&      60.29$\pm$      7.95&      21.91$\pm$
      4.21&      252.17$\pm$      3.72&      743.89$\pm$      3.72&
      740.07$\pm$      14.24&      166.52$\pm$      5.75&      166.74
$\pm$      5.22\\
24&      153.36$\pm$      12.82&      248.21$\pm$      33.23&      847.85
$\pm$      66.69&      585.09$\pm$      32.96&      181.70$\pm$
      16.93&      290.32$\pm$      9.21&      856.44$\pm$      9.21&
      530.91$\pm$      30.44&      356.75$\pm$      19.78&      380.67
$\pm$      17.03\\
25&     84.08$\pm$      8.07&      33.65$\pm$      5.14&      99.27
$\pm$      5.14&      7.63$\pm$      9.42&      19.96$\pm$
      10.39&      50.96$\pm$      4.80&      150.33$\pm$      4.80&
      151.84$\pm$      13.54&      99.62$\pm$      14.40&      69.93
$\pm$      12.16\\
26&       392.26$\pm$      5.85&      44.62$\pm$      3.48&      131.63
$\pm$      3.48&      62.11$\pm$      8.70&      19.76$\pm$
      3.99&      417.59$\pm$      4.12&      1231.90$\pm$      4.12&
      2084.34$\pm$      20.56&      316.52$\pm$      6.16&      298.41
$\pm$      5.66\\
27&      306.88$\pm$      5.38&      78.30$\pm$      3.36&      231.00
$\pm$      3.36&      38.64$\pm$      7.82&      21.77$\pm$
      4.32&      252.44$\pm$      2.91&      744.69$\pm$      2.91&
      1512.81$\pm$      14.76&      245.98$\pm$      6.23&      196.46
$\pm$      5.67\\
28&      202.17$\pm$      6.19&      65.17$\pm$      6.19&      192.25
$\pm$      15.25&      368.55$\pm$      15.25&      110.39$\pm$
      7.40&      403.25$\pm$      4.80&      1189.60$\pm$      4.80&
      1383.67$\pm$      21.04&      840.22$\pm$      15.18&      605.96
$\pm$      11.95\\
29&      23.88$\pm$      3.36&      5.80$\pm$      2.24&      17.12
$\pm$      2.24&         ---&         ----&      18.21$\pm$      1.17&      53.73$\pm$      1.17&
      96.84$\pm$      5.09&      23.63$\pm$      3.57&      16.35
$\pm$      3.19\\
30&      27.09$\pm$      3.45&      30.53$\pm$      4.71&      90.07
$\pm$      4.71&     ----&      ---&      69.81$\pm$      4.56&      205.94$\pm$      4.56&
      96.26$\pm$      23.42&      59.06$\pm$      12.11&      66.33
$\pm$      11.17\\

\hline
 \end{tabular}
 
 \end{table}
 \end{landscape}
 
\begin{landscape}
 \begin{table}
 \caption{Estimates of different parameters of galaxies with broad balmer lines.}
 \centering
 \small
 \begin{tabular}{lllllllll}
\hline
Galaxy&  H$_{\beta}$  & H$_{\alpha}$ & H$_{\alpha}$
& M$_{BH}$  &$\sigma$ km/s & Eddington & L$_{H_{\alpha}}$  &L$_{H_{\beta}}$  \\\\
&  x10$^{-17}$ ergs/s/cm$^2$ & x10$^{-17}$ ergs/s/cm$^2$& FWHM km/s 
& x 10$^6$ M$_{\odot}$ &$\sigma$ km/s & ratio & x 10$^{40}$ ergs/s & x10$^{40}$ ergs/s  \\\\ 
\hline
1&     332.46$\pm$      16.82&      1463.00$\pm$      40.39&      4453.21
$\pm$      76.08&       36.30$^{+4.04}_{-3.48}$&
      152.78$\pm$      7.40&     0.014&       18.32$\pm$
      0.51 &     4.16$\pm$      0.21 \\

2& - &      988.09$\pm$      36.89&       1713.59
$\pm$       35.03&       3.00$^{+0.36}_{-0.31}$&
      132.17$\pm$      9.45&      0.20&       5.99$\pm$
      0.22 &      -  \\

3& - &      130.81$\pm$      68.33&      3177.11$\pm$
      792.68&       6.07$^{+3.55}_{-3.53}$&      267.89
$\pm$      8.25&     0.038&       1.79$\pm$      0.94
& - \\

4& - &      70.69$\pm$      26.29&       1398.97$\pm$
       364.71&      0.54$^{+0.32}_{-0.32}$&      131.09
$\pm$      7.70&      0.11&      0.38$\pm$      0.14
       &- \\
5&      - &      1166.11$\pm$      54.55&      3758.59$\pm$
      111.36&       9.44$^{+1.35}_{-1.24}$&      116.74
$\pm$      4.93&     0.015&       2.19$\pm$      0.10
& - \\
6&       ----&      172.08$\pm$      31.59&      5725.13
$\pm$      506.28&       13.56$^{+3.38}_{-3.29}$&
      185.00$\pm$      3.29&    0.004&      0.75$\pm$
      0.14 &    1.09$\pm$     0.09 \\
7&  -&     766.36$\pm$      52.06&      2902.42$\pm$
      108.41&       6.87$^{+0.99}_{-0.92}$&      148.56
$\pm$      3.85&     0.08&       3.46$\pm$      0.24
       & - \\
8& - &      138.01$\pm$      22.08&      4058.77$\pm$
      374.01&       8.67$^{+2.12}_{-2.07}$&      177.86
$\pm$      4.37&     0.032&       1.31$\pm$      0.21
       & - \\
9&   - &      66.81$\pm$      22.90&      1606.80$\pm$
      394.84&      0.60$^{+0.33}_{-0.33}$&    149.96
$\pm$      6.88&      0.107&      0.262$\pm$     0.09
  & - \\
10&      115.74$\pm$      21.77&      344.50$\pm$      69.27&      2642.06
$\pm$      262.85&       3.39$^{+0.89}_{-0.87}$&
      212.05$\pm$      5.13&     0.072&       1.17$\pm$
      0.23 &      0.39$\pm$     0.07 \\
11&    - &      300.64$\pm$      70.96&      3776.36$\pm$
      562.99&       4.88$^{+1.75}_{-1.73}$&      164.18
$\pm$      4.49&     0.01&      0.53$\pm$      0.12
&  - \\
12&       394.75$\pm$      14.99&      1717.40$\pm$      39.27&      2808.12
$\pm$      33.53&       13.72$^{+1.44}_{-1.21}$&
      168.98$\pm$      9.19&     0.04&       17.45$\pm$
      0.39 &       4.01$\pm$      0.15 \\
13&      57.58$\pm$      13.07&      204.41$\pm$      36.83&      2234.93
$\pm$      192.81&       3.63$^{+0.84}_{-0.81}$&
      190.43$\pm$      9.44&     0.071&       2.81$\pm$
      0.51 &      0.79$\pm$      0.18 \\
14&      477.17$\pm$      68.29&      663.70$\pm$      146.27&      4745.67
$\pm$      583.01&       5.70$^{+1.81}_{-1.78}$&
      155.68$\pm$      2.19&    0.003&      0.27$\pm$
     0.060 &      0.19$\pm$     0.03 \\
16&       - &      174.03$\pm$      30.51&      4815.98$\pm$
      550.28&       4.71$^{+1.42}_{-1.39}$&      123.48
$\pm$      2.73&    0.003&      0.17$\pm$     0.03
   & - \\
17&      ---&      217.72$\pm$      51.11&      4643.30$\pm$
      727.45&       3.89$^{+1.49}_{-1.47}$&      115.05
$\pm$      2.94&    0.004&      0.13$\pm$     0.03
  &---- \\
18&       ---&      221.24$\pm$      44.97&      7632.90$\pm$
      1117.01&       17.40$^{+6.17}_{-6.09}$&      197.54
$\pm$      7.99&   0.0004&      0.36$\pm$     0.07
   &---- \\
19&       - &      278.50$\pm$      24.65&      5868.40$\pm$
      361.53&       25.96$^{+4.80}_{-4.57}$&      150.79
$\pm$      5.04&    0.003&       2.68$\pm$      0.24
       & - \\
20&      ---&      328.33$\pm$      59.00&      6719.39$\pm$
      805.62&       17.52$^{+5.28}_{-5.18}$&      179.99
$\pm$      3.27&    0.003&      0.64$\pm$      0.12
       &--- \\
21&   - &      267.37$\pm$      84.94&      2898.75$\pm$
      469.04&       2.69$^{+1.06}_{-1.05}$&      185.77
$\pm$      2.97&      0.135&      0.47$\pm$      0.15
 & - \\
22&      92.52$\pm$      18.56&      216.55$\pm$      51.74&      4497.51
$\pm$      646.77&       8.56$^{+2.98}_{-2.94}$&
      205.73$\pm$      3.78&    0.008&      0.81$\pm$
      0.19 &     0.35$\pm$     0.07 \\
23&    - &      168.01$\pm$      54.94&      2306.66$\pm$
      303.52&       2.06$^{+0.70}_{-0.69}$&      169.07
$\pm$      6.59&       0.73&      0.73$\pm$      0.24
       & -\\
24&       451.72$\pm$      36.94&      2826.95$\pm$      159.69&      2482.03
$\pm$      84.62&       4.44$^{+0.63}_{-0.58}$&
      136.46$\pm$      3.86&     0.024&       2.72$\pm$
      0.15     & 0.43$\pm$     0.04 \\
25&        ---&      306.51$\pm$      96.78&      6526.64$\pm$
      1356.49&       17.96$^{+8.56}_{-8.50}$&      169.40
$\pm$      4.70&    0.004&      0.77$\pm$      0.24
       &--- \\

\hline
\end{tabular}
 
 \end{table}
\end{landscape}
\end{document}